\documentclass[12pt,reqno]{amsart}

\usepackage{tikz-cd}
\usepackage{array}
\usepackage[mathscr]{euscript}
\usepackage[T2A]{fontenc}
\usepackage[utf8x]{inputenc}
\usepackage{breqn}
\usepackage{todonotes}

\usepackage{slashed}
\usepackage{youngtab}
\usepackage{amsmath}
\usepackage{amssymb}
\usepackage{amsxtra}
\usepackage{color}
\usepackage[margin=1.2in]{geometry}

\usepackage{hyperref}
 \hypersetup{
   linkbordercolor = blue,
   linktocpage = false,
}

\usepackage[numbers,sort&compress]{natbib}
\usepackage{hypernat}

\theoremstyle{definition}

\newtheorem*{proposition}{Proposition}
\theoremstyle{remark}
\newtheorem*{example}{Example}
\newtheorem*{remark}{Remark}

\newcommand{\half} {\frac 12}
\newcommand{\pa} {\partial}

\newcommand {\la} {\left \langle}
\newcommand {\ra} {\right \rangle}

\newcommand{\stimes} {{|}\kern -.05in\times}

\newcommand {\y} {\mathscr Y}

\newcommand {\ac} {\mathfrak{a}}

\newcommand {\qe} {\mathfrak{q}}

\newcommand {\fe} {\mathfrak{f}}

\newcommand {\zb} {{\bar z}}

\newcommand {\CalD} {\mathcal D}
\newcommand {\CalE} {\mathcal E}
\newcommand {\CalF} {\mathcal F}

\newcommand {\CalI} {\mathcal I}

\newcommand {\CalL} {\mathcal L}
\newcommand {\CalM} {\mathcal M}
\newcommand {\CalN} {\mathcal N}
\newcommand {\CalO} {\mathcal O}

\newcommand {\CalR} {\mathcal R}
\newcommand {\CalS} {\mathcal S}

\newcommand {\CalW} {\mathcal W}
\newcommand {\CalZ} {\mathcal Z}

\newcommand {\ec} {\mathscr E}

\newcommand{\pp}{p}

\newcommand {\mv} {{\mathfrak M}}

\newcommand {\pv} {{\mathfrak P}}

\newcommand {\BC}   {\mathbb C}

\newcommand {\BN}   {\mathbb N}

\newcommand {\BR}   {\mathbb R}
\newcommand {\BP}   {\mathbb P}

\newcommand {\bB}   {\mathbf{B}}

\newcommand {\bQ}   {\mathbf{Q}}

\newcommand {\ba}  {\mathbf{a}}
\newcommand {\bb} {\mathbf{b}}

\newcommand {\bn}{ \mathbf{n}}
\newcommand {\bm}{ \mathbf{m}}
\newcommand {\bv}{  \mathbf{v}}
\newcommand {\bk}{  \mathbf{k}}
\newcommand {\bw}{  \mathbf{w}}
\newcommand {\bwt}{ \mathbf{\tilde w}}

\newcommand {\BS}   {\mathbb S}
\newcommand {\BT}   {\mathbb T}

\newcommand {\BZ}   {\mathbb Z}

\newcommand {\CP}   {\mathbb C \mathbb P}

\newcommand {\vt}  {\vartheta}

\newcommand{\g}{\mathfrak{g}}

\renewcommand{\Im} {\mathrm{Im}}

\DeclareMathOperator{\coker}{coker}
\DeclareMathOperator{\codim}{codim}

\DeclareMathOperator{\End}{End}
\DeclareMathOperator{\Hom}{Hom}

\DeclareMathOperator{\tr} {tr}
\DeclareMathOperator{\Tr} {Tr}
\DeclareMathOperator{\str} {str}
\DeclareMathOperator{\rk} {rk}

\DeclareMathOperator{\ind} {ind}

\newcommand{\ch}{\mathrm{ch}}
\newcommand{\td}{\mathrm{td}}

\renewcommand{\hat}{\widehat}

\newcommand{\Gg}{G_{\mathbf{v}}}

\newcommand{\Gf}{G_{\text{\tiny f}}}

\newcommand{\Gammadi}{\boldsymbol{\Gamma}}

\newcommand{\Gq}{{\mathbf{G}}_{\Gamma}}
\newcommand{\Gqp}{{\mathbf{G}}_{\Gamma'}}

\newcommand{\gq}{\mathfrak{g}_{\Gamma}}
\newcommand{\gqp}{\mathfrak{g}_{\Gamma'}}

\newcommand{\Ver}{\mathrm{I}_{\Gamma}}

\newcommand{\Edg}{\mathrm{E}_{\Gamma}}

\newcommand{\ii}{\imath}
\newcommand{\iv}{\bf i}

\newcommand{\Cx}{\mathsf{C}_{\mathsf{x}}}

\newcommand{\mf}{\bm^{\mathrm{f}}}
\newcommand{\mft}{{\tilde \bm}^{\mathrm{f}}}
\newcommand{\mb}{{\bm}^{\mathrm{bf}}}

\newcommand{\Gequiv}{G_{\mathrm{eq}}}
\newcommand{\Gl}{G_{\text{\tiny L}}}
\newcommand{\tequiv}{t_{\mathrm{}}}
\newcommand{\Tequiv}{\mathsf{T}}
\newcommand{\Efix}{\mathsf{E}}
\newcommand{\indfun}{\mathbb{I}}
\newcommand{\Liker}{L}
\newcommand{\sP}{\mathscr P}

\newcommand{\kCS}{\kappa} 

\newcommand{\Braid}{\mathscr{B}}
\DeclareMathOperator{\Li}{Li}

\newcommand{\muf}{\mu^{\mathrm{f}}}
\newcommand{\muft}{\tilde \mu^{\mathrm{f}}}
\newcommand{\RGamma}{R_{\Gamma(\bv,\bw,\bwt)}}

\newcommand{\Hinstk}{ \mathcal{H^{\mathrm{inst}}_{\mathbf{k}}}}
\newcommand{\Minstk}{ \mathcal{M}^{\mathrm{inst}}_{\mathbf{k}}}

\newcommand{\bek}{k}  
\newcommand{\ep}{\epsilon} 

\newcommand{\lfive}{\ell} 

\newcommand{\betir}{\ii \ell } 

\numberwithin{equation}{section}

\newcommand{\Yang}{  \mathbf{Y}}
\newcommand{\Qaff}{ \mathbf{U}^{\mathrm{aff}}}
\newcommand{\Qalg}{\mathbf{U}}
\newcommand{\Qell}{ \mathbf{U}^{\mathrm{ell}}}

\newcommand{\libs}[1]{\href{file://localhost/Users/pestun/Dropbox/lib/spires/#1}{\nolinkurl{#1}}}
\newcommand{\libm}[1]{\href{file://localhost/Users/pestun/Dropbox/lib/math/#1}{\nolinkurl{#1}}}
\newcommand{\libb}[1]{\href{file://localhost/Users/pestun/Dropbox/lib/books/#1}{\nolinkurl{#1}}}
\newcommand{\libr}[1]{\href{file://localhost/Users/pestun/Dropbox/lib/research/#1}{\nolinkurl{#1}}}

\begin{document}

\bibliographystyle{utphys}

\title{Quantum geometry
and quiver gauge theories}

\date{December 23, 2013}
\author{Nikita Nekrasov}

\address{
\hbox{\vbox{Simons Center for Geometry and Physics,
Stony Brook, USA}}
\hbox{\vbox{\qquad \tiny on leave of absence from}} 
\hbox{\vbox{\quad Institut des Hautes Etudes Scientifiques,  France,}} 
\hbox{\vbox{\quad Institute of Theoretical and Experimental Physics,  Russia,}}
\hbox{\vbox{\quad Institute for Information Transmission Problems, Lab. 5, Russia}}}
\email{nikitastring@gmail.com}

\author{Vasily Pestun}
\address{\quad Institute for Advanced Study,  Princeton, USA}

\email{pestun@ias.edu}

\author{Samson Shatashvili}
\address{\hbox{\vbox{School of Mathematics, Trinity College Dublin, Ireland}} 
\hbox{\vbox{\quad Hamilton Mathematics Institute, TCD, Dublin 2, Ireland}}
\hbox{\vbox{\quad Chaire Louis Michel, Institut des Hautes Etudes Scientifiques,  France,}} 
\hbox{\vbox{\qquad \tiny on leave of absence from}}
\hbox{\vbox{\quad Euler International Mathematical Institute, St. Petersburg, Russia}}
\hbox{\vbox{\quad Institute for Information Transmission Problems, Lab. 5, Russia}}}

\email{samson@maths.tcd.ie}
\begin{abstract}
We study macroscopically two dimensional $\mathcal{N}=(2,2)$ supersymmetric gauge theories constructed by compactifying the  quiver gauge theories with eight supercharges on a product $\mathbb{T}^{d} \times \mathbb{R}^{2}_{\epsilon}$ of a $d$-dimensional torus  and a two dimensional cigar with $\Omega$-deformation. 
We compute the universal part of the effective twisted superpotential. In doing so we establish the
correspondence between the gauge theories, quantization 
 of the moduli spaces of instantons on
$\mathbb{R}^{2-d} \times \mathbb{T}^{2+d}$ and singular monopoles on $\mathbb{R}^{2-d} \times
\mathbb{T}^{1+d}$, for $d=0,1,2$,  and the Yangian $\mathbf{Y}_{\epsilon}(\mathfrak{g}_{\Gamma})$, quantum affine
 algebra $\mathbf{U}^{\mathrm{aff}}_q(\mathfrak{g}_{\Gamma})$, or the  quantum elliptic algebra
 $\mathbf{U}^{\mathrm{ell}}_{q,p}(\mathfrak{g}_{\Gamma})$ associated to Kac-Moody algebra $\mathfrak{g}_{\Gamma}$ for quiver $\Gamma$.
 \end{abstract}

 \begin{flushright}
   TCD-MATH-13-17,   HMI-13-02 
 \end{flushright}

\bigskip
\bigskip
\maketitle 
\setcounter{tocdepth}{3} 
\tableofcontents

\section{Introduction}

The Bethe/gauge correspondence between the supersymmetric gauge theories and
quantum integrable systems is a subject of research spanning over a decade \cite{Moore:1997dj,
Gerasimov:2006zt,
Gerasimov:2007ap,
Nekrasov:2009zz,
Nekrasov:2009rc,
Nekrasov:2009ui,
Nekrasov:2009uh,
Nekrasov:2010ka,
Nekrasov:2011bc} and even longer in the context of topological gauge theories \cite{Witten:1989wf,Gorsky:1993dq,Gorsky:1993pe,Gorsky:1994dj}. It can also be viewed as a correspondence between the supersymmetric gauge theories and representation theory of infinite dimensional algebras typical for the two dimensional conformal field theories or integrable deformations thereof. This relation, dubbed the BPS/CFT correspondence in
\cite{Nekrasov:2004sem}, following the prior work in \cite{Nakajima:1994r, Nakajima:1998, Vafa:1994tf, Losev:1995cr, Nekrasov:2002qd, Losev:2003py, Nekrasov:2003rj}, became a subject of intense development after
the seminal work \cite{Alday:2009aq} where the instanton partition functions
of the $\CalS$-class $\CalN=2$ gauge theories in the $\Omega$-background
\cite{Nekrasov:2002qd} were conjectured to be the Liouville (and, more
generally, the $ADE$ Toda) conformal blocks.

\subsection{Gauge theories}

Recently  \cite{NP2012a}  the
Seiberg-Witten geometry, the curves \cite{Seiberg:1994aj,Seiberg:1994rs} and the integrable systems associated to all $\mathcal{N}=2$
(affine) ADE quiver gauge theories, whose gauge group $\Gg = \prod SU({\bv}_{\iv})$ is a product of unitary groups,  were systematically found. The direct application of quantum field theory techniques leads to the
equivariant integration over the instanton moduli on $\BR^4$
\cite{Nekrasov:2002qd,Nekrasov:2003rj,Moore:1997dj,Losev:1997tp, Lossev:1997bz}. The above mentioned geometry emerges in the flat space 
limit  $\ep_1, \ep_2  \to 0$ of the $\Omega$-deformed gauge theory \cite{Nekrasov:2002qd}. 

\subsubsection{Prepotentials and special geometry}

The
prepotential ${\CalF}( {\ac},{\bm}; {\qe})$ of the effective low-energy theory
relates to the gauge theory 
partition $\CalZ$-function \cite{Nekrasov:2002qd} in the flat space limit as follows:
  \begin{equation}
  \label{eq:first}
    \CalF(\ac,\bm; \qe) = - \lim_{\ep_1, \ep_2 \to 0} \ep_1 \ep_2 \log
    {\CalZ} ( \ac,\bm; \qe; \ep_1, \ep_2)
  \end{equation}
where $\qe$ here  denotes the set of gauge coupling constants of the
theory, $\bm$ the set of masses of the hypermultiplets fields, 
and $\ac$ is the set of the flat special coordinates on the moduli space of
vacua $\mv$ on the Coulomb branch of the theory. The latter is identified with the
asymptotics  of the scalar fields of $N=2$ vector
multiplets at infinity of the Euclidean space-time. The geometry of $\mv$ and ${\CalF}(\ac,\bm; \qe)$ are captured by a complex analogue of a classical integrable system, an algebraic integrable system. The phase space $\pv$ of that system admits a partial compactification which is a complex symplectic manifold with the holomorphic symplectic form and 
projection to $\mv$ with Lagrangian fibers, which are, generically, abelian varieties. 
 
The gauge theories studied in \cite{NP2012a} and in this paper are
labeled by the classes of representations $\Gamma(\bv, \bw, \bwt)$  of quivers  which correspond to the
Dynkin diagrams $\Gamma$ of simply laced finite-dimensional or affine
Kac-Moody algebras $\gq$. The dimension vectors ${\bv} = ({\bv}_{\iv}), {\bw} = (\bw_{\iv}), {\bwt} = (\bwt_{\iv})$ encode the number of colors, the number of fundamental flavors and the
number of anti-fundamental flavors charged with respect to the gauge group factor
$U({\bv}_{i})$, respectively. 

For affine Kac-Moody $\gq$, let $\gqp$ be the underlying
finite-dimensional simple Lie algebra.
The corresponding phase spaces $\pv$ turn out to be the moduli spaces of
solutions to the BPS equations,  $\Gq$-monopoles (for finite quivers) on $\BR^{d} \times
\BT^{3 -d}$, 
 or $\Gqp$-instantons (for affine quivers) on $\BR^{d} \times
\BT^{4 -d}$.   The gauge group $\Gq$ ($\Gqp$) for these BPS equations is
 not the original gauge group $\Gg$ of the quiver theory. 
Yet the solution of the quantum $\Gg$-gauge theory is captured by the moduli space of particular classical solutions to the $\Gq$- ($\Gqp$)-gauge theory. 

\subsubsection{Departure from the Seiberg-Witten theory.}

In this paper, we extend the results of \cite{NP2012a} in two ways. 

First of all, we study the effective low-energy theory of the two dimensional ${\CalN}=(2,2)$ supersymmetric theory which is obtained from the original four dimensional 
${\CalN}=2$ theory by $\Omega$-deformation, affecting two out of four space-time dimensions.  Secondly, we focus more on the theories in five dimensions, compactified on a circle. The four dimensional case is obtained by sending the radius of the circle to zero. 

\subsubsection{Effective twisted superpotential.} 

In terms of the general $\Omega$-backgrounds \cite{Nekrasov:2002qd}, we 
send to zero only one equivariant parameter, say $\ep_2=0$,
while keeping the other parameter $\ep_1  = {\ep}$  finite. Thus we work in the limit discussed in \cite{Nekrasov:2009rc} and study the {\it universal part} of \emph{the effective twisted superpotential}
 \begin{equation}
 \label{eq:super}
    {\CalW}(\ac,\bm; \qe, \ep) = - \lim_{\ep_2 \to 0} \ep_2 \log
    {\CalZ} ( \ac,\bm; \qe; \ep_1=\ep, \ep_2)\, , 
  \end{equation}
identified in the Refs. 
\cite{Nekrasov:2009zz}, 
\cite{Nekrasov:2009rc}, 
\cite{Nekrasov:2009ui}, 
\cite{Nekrasov:2009uh}. 

{}Recall that the twisted superpotential is a specific $F$-term in the Lagrangian of a ${\CalN}=(2,2)$ supersymmetric theory in two dimensions. A four dimensional gauge theory can be viewed as a two dimensional theory with an infinite number of degrees of freedom. The two dimensional $\Omega$-deformation of the type we study in this paper breaks the four dimensional $\CalN=2$ supersymmetry down to  two dimensional $\CalN=2$ supersymmetry. However, in order to view the resulting theory as the two dimensional theory with the well-defined effective Lagrangian we need to specify the boundary conditions at infinity. 
  One may view the choice of the boundary conditions at infinity as a choice of a three dimensional supersymmetric theory compactified on a circle, coupled to the original four dimensional theory on the product of a cigar-like two-dimensional geometry and the two dimensional Minkowski world-sheet of the resulting theory. The three dimensional theory at infinity is compactified on a circle because of the asymptotics of the cigar-like geometry, which looks like ${\BR}^{1} \times {\BS}^{1}$ at infinity. The contribution ${\CalW}^{\infty} ({\ac}, {\bm}; {\ep})$ of the three dimensional theory at infinity of the cigar is purely perturbative and is given by a sum of dilogarithm functions
$$
{\CalW}^{\infty} ({\ac}, {\bm}; {\ep}) \sim \ {\ep} \, \sum \text{Li}_{2} \left( e^{\frac{\text{linear} ({\ac}, {\bm})}{\ep} } \right)
$$
According to the Refs. \cite{Nekrasov:2010ka,Nekrasov:2011bc},  the twisted superpotential ${\CalW}^{\text{eff}}({\ac}, {\bm}; {\qe}, {\ep})$ of the effective two-dimensional theory  is the sum:
\begin{equation}
{\CalW}^{\text{eff}} ({\ac}, {\bm}; {\qe}, {\ep}) =  {\CalW}({\ac}, {\bm}; {\qe}, {\ep}) + 
{\CalW}^{\infty} ({\ac}, {\bm}; {\ep})
\label{eq:eff}
\end{equation}
which can be identified with the Yang-Yang function of some quantum integrable system \cite{Nekrasov:2009zz}, \cite{Nekrasov:2009rc}, \cite{Nekrasov:2009ui}, \cite{Nekrasov:2009uh}. The $\Omega$-deformation parameter $\ep$ plays the r\^ole of a Planck constant. It is easy to see by comparing the Eqs. \eqref{eq:super} and \eqref{eq:first} that in the limit ${\ep} \to 0$, the superpotential 
${\CalW}({\ac}, {\bm}; {\qe}, {\ep})$ behaves as
$$
\frac{1}{\ep} {\CalF}( {\ac},{\bm}; {\qe})
$$
and the quantum integrable system approaches the classical one with the phase space $\pv$, and the prepotential ${\CalF}( {\ac},{\bm}; {\qe})$ describing the special geometry of its base $\mv$. 

\subsubsection{Bethe equations and vacua}

{}More precisely, the
Bethe equations of that quantum integrable system read as follows:
\begin{equation}
\label{eq:bethe}
{\exp}\, \frac{\pa\CalW^{\text{eff}}({\ac}, {\bm} ; \qe, \ep)}{\pa {\ac}_{i}} =1\, , \qquad i = 1, \ldots , \text{dim}{\mv}
\end{equation}
The Eq. \eqref{eq:bethe} describes in some specific Darboux coordinates, e.g. \cite{Nekrasov:2011bc}, the intersection of two Lagrangian subvarieties, cf. \cite{Nekrasov:2010ka}, one with the generating function ${\CalW}({\ac})$ and another, with the generating function $-{\CalW}^{\infty}({\ac})$.

\subsubsection{Focus of the paper}

In this paper we explore a particular formalism for the systematic computation of the universal part of the superpotential,  
${\CalW}({\ac}, {\bm}; {\qe}, {\ep})$.
We shall not discuss the choices of boundary conditions and representations of  the noncommutative deformations of the algebra of functions on $\pv$. 

Another aspect in which this paper develops the Ref. \cite{NP2012a} is its focus on the five dimensional version of the theory. The ${\CalN}=2$ gauge theory in four dimensions
can be canonically lifted to the ${\CalN}=1$ theory in five dimensions.  
The five dimensional ${\CalN}=1$ supersymmetric gauge theory
compactified on a circle $\BS^1_{\lfive}$ of circumference $\lfive$ can be viewed \cite{Nekrasov:1996cz}
as a particular deformation of the four dimensional ${\CalN}=2$ theory. One studies the theory with twisted boundary conditions, such that in going around the circle the remaining four dimensional Euclidean space-time is rotated in the two orthogonal two-planes, by the angles $\lfive \ep_1$ and $\lfive \ep_2$, respectively. 
For
\begin{equation}
  q_1 = e^{\ii \lfive \ep_1}, \qquad q_2 =e^{\ii \lfive \ep_2}, \qquad q
  = q_1 q_2 
\end{equation}
 we are interested in the limit 
 \begin{equation}
   q_2  \to 1, \qquad q = q_1
 \end{equation}
and for convergence issues we shall assume in this paper
\begin{equation}
  |q| < 1
\end{equation}
For the quiver which is a Dynkin graph of (affine) ADE Lie algebra $\gq$, we
find that the $\ep$-deformed limit shape partition profile equations of  \cite{Nekrasov:2003rj} can be mapped to the Bethe equations of
a formal trigonometric XXZ $\gq$-spin chain. Despite the  similarity  to the problems
(for finite $A_r$ quivers in four dimensions) studied in
\cite{Poghossian:2010pn,Dorey:2011pa,Chen:2011sj} and also in 
\cite{Bao:2011rc,Mironov:2012ba,Gorsky:1995zq,Gorsky:1996hs,Gorsky:1997jq,Gorsky:1997mw,Gerasimov:2006zt,Gerasimov:2007ap,Nekrasov:2009rc,Bazhanov:1998dq,Dorey:2007zx,Gerasimov:2005qz,Gerasimov:2005,Galakhov:2012hy,Mironov:2009uv,Mironov:2009dv,Teschner:2010je,Muneyuki:2011qu},
there are important conceptual differences which we shall explain
below. In fact, the universal part of the gauge theory twisted superpotential does not correspond to any spin chain. It corresponds to a commutative subalgebra of  noncommutative associative algebra, which is the quantum affine algebra $\Qaff_{q}(\gq)$ associated to $\gq$. One can then study the spectrum of this subalgebra in any representation of $\Qaff_{q}(\gq)$, which would lead to the ordinary Bethe equations.

\subsubsection{Chiral ring generating functions}
In gauge theory, the functions $\y_{\iv} (x)$, where ${\iv} \in \Ver$ runs through
the set $\Ver$ of nodes  of $\Gamma$,  encode the expectation values of
chiral ring observables. For a point $u \in \mv$ on the Coulomb branch moduli
space $\mv$ and a simple gauge group factor $U({\bv}_{\iv})$ we define the generating function,
in the four and five dimensional versions given by 
\begin{equation}
\label{eq:yphys}
  \begin{aligned}
& 4d: \qquad  \y_{\iv}(x - \tfrac \ep 2 ;u) = \exp(\, \la \log \det (x - \phi_{\iv}) \ra_{u})\\
 & 5d: \qquad  \y_{\iv}^{+}(\xi q^{-\frac 1 2};u) = \exp(\, \la \log \det (1 - e^{\ii \lfive (\phi_{\iv}-x)}) \ra_{u}), \qquad \xi = e^{\ii \lfive x}
  \end{aligned}
\end{equation}
respectively, with $\phi_{\iv}$ the complex adjoint scalar in the $SU({\bv}_{\iv})$ gauge vector
multiplet, $\la \ra_{u}$ denotes the expectation value at the point $u
\in \mv$ of the moduli space of vacua, and $\lfive$ is the circumference
of the circle $\mathbb{S}^1_{\lfive}$. We give the geometric definition below, cf. Eqs. \eqref{eq:ygeom4}, \eqref{eq:ygeom5}).

\subsection{Quantum groups}
In this note we will be dealing with quantum  algebras
\begin{equation}
\label{eq:algebras}
\mathbf{U}_{\ep} \gq(\Cx): \qquad \qquad \Yang_{\ep}(\gq), \qquad \Qaff_{q}(\gq),
  \qquad \Qell_{q,p}(\gq)
\end{equation}
associated  to the quiver theories  on
$\BR^4$,  on $\BR^4
\times \BS^1_{\lfive}$ and on $\BR^4 \times \BT^2_{\lfive,
  - \lfive/\tau_p} $, respectively. Here
 $\lfive$ is the \emph{circumference} of the circle
$\BS^1_{\lfive}$ on which we compactify the 5d gauge theory, and
$(\lfive, -\lfive/\tau_p)$ are the periods of the torus
$\BT^2_{\lfive, - \lfive/\tau_p}$ on which we compactify the six dimensional gauge
theory.  The parameters $\ep$, or  $q$ 
\begin{equation}
  q  =e^{\ii \lfive \ep}\, , 
\end{equation}
are the \emph{quantization} parameters (\emph{Planck constant}) of quantum algebras
(\ref{eq:algebras}). 

The algebras (\ref{eq:algebras}) are defined using \emph{second}
Drinfeld realization \cite{Drinfeld:1987}, i.e. quantum $\gq$
currents $(\psi_{\iv}^{\pm}(x), e^{\pm}(x))_{{\iv} \in \Ver}$, see
\ref{se:Drinfeld-currents}. 
The domain $\Cx$ of the additive spectral parameter $x \in \Cx$ in  quantum
$\gq$-currents,  for the four, five and the six dimensional case, is  
\begin{itemize}
\item  a plane  $\Cx= \BR^2 \cong \BC$ for $\gq$-Yangian $\Yang_{\ep}(\gq)$,
\item  a  cylinder 
$\Cx = \BR^2/(\frac{2 \pi}{\lfive} \BZ) \cong  \BR^{1} \times \BS^{1} \cong \BC^{\times}$ for $\gq$-quantum affinization
$\Qaff_{q}(\gq)$,
\item  a torus $\Cx= \BT^2_{\tau_p} \cong \BR^2/( \frac{2
  \pi}{\ell})(\BZ + \tau_p \BZ)  \cong \BC^{\times}/p^{\BZ} =\ec_p$ for $\gq$-quantum elliptic algebra
$\Qell_{q,p}(\gq)$; where the parameter $p = e^{2 \pi \ii \tau_p}$ is the multiplicative elliptic modulus 
\end{itemize}

In the classical limit $\ep = 0$, quantum algebras (\ref{eq:algebras})
reduce to the (universal enveloping of) classical current algebra
$\mathbf{U}\gq(\Cx)$ on the spectral domain $\Cx$, and the results of
this note reduce to the results of  \cite{NP2012a}. Namely, in
\cite{NP2012a}, to each point on the Coulomb moduli space $\mathfrak M$ of the gauge
theory, one associated a \emph{current} $h(x) \in \mathbf{U}\gq(\Cx)$,
defined in terms of the functions $\y_{\iv}(x)$ and gauge theory data. It is shown in \cite{NP2012a} that  $h(x)$ satisfies the equation
\begin{equation}
\label{eq:leash}  \chi_{\iv}( h (x)) = T_{\iv}(x), \qquad i \in \Ver
\end{equation}
where $\chi_{\iv}$ is a (twisted) character of the $i$-th fundamental
$\gq$-module, and $T_{\iv}(x)$ is a polynomial of degree ${\bv}_{\iv}$.

\subsection{Main result}
In this paper we show, the result (\ref{eq:leash})  of \cite{NP2012a} has
natural \emph{quantum} generalization \ref{prop:central-result}:  for $\ep \neq 0$,
the algebra $\mathbf{U}\gq(\Cx)$ is replaced by its quantum version 
$\mathbf{U}_{\ep}\gq(\Cx)$. We construct  a quantum current $h(x) \in
\mathbf{U}_{\ep}\gq(\Cx)$, defined in terms of the gauge theory data
and the generating functions $\y_{\iv}(x)$  such that the character equation (\ref{eq:leash}) still
holds, but $\chi_{\iv}$ are replaced by the characters for quantum algebras known as
$q$-characters \cite{Knight:1995,Frenkel:1998}. 

\subsection{Conventions}
\subsubsection{Quantum parameter}
In this paper, the $q$-numbers are defined by 
\begin{equation}
\label{eq:qnumber}
  [n]_q = \frac{q^{\frac n 2} -q^{-\frac n 2}}{q^{\frac 1 2} -
    q^{-\frac 1 2}}
\end{equation}
Our conventions for the quantum parameter $q = e^{\ii \lfive \ep}$ of the quantum
groups differ from \cite{Frenkel:1998} or \cite{Chari:1994}, so that our
$q$ is the $q^2$ of \cite{Frenkel:1998} or
\cite{Chari:1994}. Our conventions for $q$ agree with e.g. 
\cite{Witten:1988hf,Kirillov:1991ec,Reshetikhin:1991tc,Reshetikhin:1990pr} so that  the $SU(n)$
three dimensional Chern-Simons theory at the level $k$ 
relates to the quantum group $\Qalg_q(\mathfrak{sl}_n)$
 with parameter 
\begin{equation}
\label{eq:qparameter}
  q = \exp \left( \frac{ 2 \pi \ii}{h^{\vee} + k} \right), \quad
  h^{\vee}_{\mathfrak{sl}_n} = n
\end{equation}
and the invariant of the unknot $\bigcirc$ in fundamental representation
$\mathbf{n}$ is $\la
W_{\mathbf{n}}(\bigcirc) \ra = [n]_q$ in convention (\ref{eq:qnumber}). The three dimensional 
Chern-Simons theory has been related to
$\mathcal{N}=4$ SYM on a four-manifold with a boundary in \cite{Witten:2011zz}, and the parameter $q$ in the normalization
(\ref{eq:qparameter}) naturally counts instantons in the $\mathcal{N}=4$
SYM and enjoys Langlands duality, or modular transformation: for a
simply-laced gauge group for $q = e^{\ii \lfive \ep} = e^{2 \pi \ii \tau_{\ep}}$ the Langlands dual $\tau_{\ep}^L = -1/\tau_{\ep}$. 
It would be interesting to explore the modular properties of the $\ep$-`modular' parameter
\begin{equation}
  \tau_{\ep} = \tfrac{ \lfive }{ 2 \pi} \ep
\end{equation}
 in  the context of this paper, dealing with \emph{quantum algebras}
 (\ref{eq:algebras}), in particular in view of the applications to S-duality \cite{Nekrasov:2004js, Nekrasov:2005bb}. We leave this task for the future. 

\subsubsection{Quantum algebras}
There is a variety of conventions for naming the quantum algebras
$\mathbf{U}_{\ep}(\Cx)$  (\ref{eq:algebras}) in the literature. In
this note we follow the naming scheme in which  the quiver $\Gamma = A_1 =\bullet$  that has one node
and no edges, associated to the Lie algebra $\gq = \mathfrak{sl}_2$,
corresponds to
\begin{itemize}
\item 4d:  the $\mathfrak{sl}_2$ Yangian
$\Yang_{\ep}(\mathfrak{sl}_2)$; XXX $\mathfrak{sl}_2$-spinchain for 4d 
\item 5d: the $\mathfrak{sl}_2$ quantum affine algebra
  $\Qaff_q(\mathfrak{sl}_2) \simeq \Qalg_q(\hat{\mathfrak{sl}_2})$; XXZ
  $\mathfrak{sl}_2$-spinchain
\item 6d: the $\mathfrak{sl}_2$ quantum elliptic algebra
  $\Qell_{q,p}(\mathfrak{sl}_2) \cong \mathbf{E}_{\ep,
    \tau_p}(\mathfrak{sl}_2)$; XYZ $\mathfrak{sl}_2$-spinchain
\end{itemize}

It is known that the Yangian $\Yang_{\ep}(\gq)$ or the quantum affinization  $\Qaff_q(\gq)$ can be defined using second Drinfeld
\emph{current} realization \cite{Drinfeld:1987}, see \ref{se:Drinfeld-currents}, for
any generalized symmetrizable affine Kac-Moody algebra $\gq$. In
particular, if $\gq$ is an affine Kac-Moody Lie algebra $\gq = \hat \g$
where $\g$ is a finite dimensional simple Lie algebra, then
$\Qaff_q(\gq)$ 
\begin{equation}
 \Qaff_q(\gq) = \Qaff_q(\hat \g) = \Qalg_q(\hat {\hat \g})
\end{equation}
is often called in the literature $\g$ \emph{quantum toroidal
  algebra} because of double loop $\hat{\hat \g}$
and $\Yang_{\ep}(\gq) = \Yang_{\ep}(\hat \g)$ is called \emph{affine
  Yangian}.  

In this note we will be using consistent naming convention 
\begin{itemize}
\item  $\gq$-Yangian for $\Yang_{\ep}(\gq)$,
\item   $\gq$-quantum affine algebra for $\Qaff_q(\gq)$,
\item $\gq$-quantum elliptic algebra for $\Qell_{q,p}(\gq)$. 
\end{itemize}

 For example, the necklace quiver with $r+1$ nodes, $\gq
=\hat A_r$, is associated in our conventions with $\hat{\mathfrak{gl}_{r+1}}$ \emph{quantum
  affine algebra}, or, equivalently, with $\mathfrak{gl}_{r+1}$ \emph{quantum toroidal
  algebra}, see \cite{Hernandez:2008}. 

In terms of Schur-Weyl duality, quantum deformations of $\g$-Weyl groups
are called  $\g$-Hecke
algebras. If $\g$ is a finite-dimensional simple Lie group, then 
\begin{itemize}
\item $\Qalg_q(\g)$ is Schur-Weyl dual to $\g$-Hecke algebra 
\end{itemize}
Meanwhile, for  the $\Cx = \BC^{\times}$ versions of quantum algebras
(\ref{eq:algebras}), considered in this note, associated to  $\gq = \g$ or
$\gq = \hat \g$ where $\g$ is finite dimensional simple Lie algebra 
\begin{itemize}
\item $\Qaff_q(\g)$ is Schur-Weyl dual to  $\g$-affine Hecke algebra
\item $\Qaff_q(\hat \g)$ is Schur-Weyl dual to  $\g$-double affine Hecke
  algebra (DAHA), see review \cite{Cherednik:2004} 
\end{itemize}

See section  \ref{se:quantum-groups} for more details and literature
review on representation theory of quantum algebras.

\subsection{Acknowledgments}
We are grateful to I.~Cherednik, E.~Frenkel, V.~Kazakov, H.~Nakajima, A.~Okounkov, F.~Smirnov, V.~Smirnov, V.~Tarasov, Z.~Tsuboi and A.~Zayakin for discussions.

Research of NN was supported in part by RFBR grants 12-02-00594, 12-01-00525, by Agence Nationale de Recherche via the grant
ANR 12 BS05 003 02,  by Simons Foundation, and by Stony Brook Foundation.

VP gratefully acknowledges support from Institute for Advanced Study,
NSF grant PHY-1314311, PHYS-1066293, Rogen Dashen Membership, MSERF 14.740.11.0081,
NSh 3349.2012.2, RFBR 10-02-00499, 12-01-00482, 13-02-00478 А and the hospitality of the Aspen Center for Physics.

The work of SSh is supported in part by the Science Foundation Ireland under the RFP program and by the ESF Research Networking Programme ``Low-Dimensional
Topology and Geometry with Mathematical Physics (ITGP)''.

The results of the paper were reported in  2012 at
\cite{Pestun:berlin, Pestun:SCGP, Nikita:talk2012i, Nikita:talk2012ii, Samson:talk2012} and in 2013 at \cite{Samson:talk2013i, Samson:talk2013ii}.  We thank
the organizers of these conferences and the institutions for  the
invitations and the participants for the questions and the comments. 
After we reported on the main results of the present paper 
we received a preprint \cite{Fucito:2012xc} which partially overlaps
 with our section \ref{seq:qSW}.

\section{The partition $\CalZ$-function of quiver gauge theories}

We start with the class of supersymmetric quiver gauge theories studied in
\cite{NP2012a}, that is the 
${\CalN}=2$ supersymmetric four-dimensional gauge theories with a
Lagrangian microscopic description, with a gauge group ${\Gg} = \times_{i} \, SU ({\bv}_{i})$ and matter
hypermultiplets in the fundamental, the bifundamental
or the adjoint representations of $\Gg$, such that the theory is
asymptotically free or conformal in the ultraviolet.

\subsection{The Lagrangian data}

The theory is encoded in the following equipped quiver $\Gamma(\bv, \bw,
\bwt)$,  cf. \cite{NP2012a}:
\begin{itemize}

\item An oriented graph $\Gamma$ whose vertices label gauge
  groups while edges label bifundamental multiplets.

\item The set of vertices of $\Gamma$ is
  denoted by $\Ver$. For our purposes it is sufficient to consider
  non-multiple edges, so the set of edges comes as $\Edg \subset \Ver \times
  \Ver$ with the two natural projection maps $s$ and $t$ on the first and second
  factor, respectively, with $s(e)$ called  the source of the edge $e$ and
  $t(e)$ called the target of the edge $e$.   
  
 \item There are three ${\BZ}$-valued functions on $\Ver$, ${\bv}$, ${\bw}$
 and ${\tilde\bw}$. 
  For the vertex $i$ we denote ${\bv}_{\iv} = {\bv}({\iv}) \in {\BZ}_{> 0}$, ${\bw}_{\iv} = {\bw}({\iv}) \in {\BZ}_{\geq 0}$ and
  ${\tilde\bw}_{\iv} = {\tilde\bw}({\iv}) \in {\BZ}_{\geq 0}$.
  
Then ${\bv}_{\iv}$ is the number of colors in the factor $SU({\bn}_{\iv})$ of the gauge group
  associated to the $\iv$-th vertex,  $\bw_{\iv}$ and  $\bwt_{\iv}$ is the number of fundamental
and anti-fundamental  multiplets charged under the gauge group factor $SU(\bv_{\iv})$, respectively. Note that in the gauge theory on flat space-time there is no distinction between the fundamental and anti-fundamental hypermultiplets. The distinction comes upon the coupling to the $\Omega$-background. 

\item
Each edge $e \in \Edg$ describes a bifundamental hypermultiplet
transforming in $(\bar{ \mathbf{v}}_{s(e)}, \mathbf{v}_{t(e)})$ (or
adjoint hypermultiplet if the edge is a loop  with $s(e) = t(e)$). 
Let $I_{\bf ij}$ be the incidence matrix of the graph $\gamma$ with 
$$
I_{\bf ij} = \# \left( s^{-1}({\iv}) \cap t^{-1}({\bf j}) \right) + \# \left(  s^{-1}({\bf j}) \cap t^{-1}({\iv}) \right)
$$
equal to the number of oriented edges connecting the vertices ${\iv}$ and ${\bf j}$ with either 
orientation. The Cartan matrix of $\Gamma$ is  $a_{\bf ij} = 2 \delta_{\bf ij} - I_{\bf ij}$. The integers 
$(\bv_{\iv}, \bw_{\iv}, \bwt_{\iv})$ should satisfy the inequality
\begin{equation}
  \label{eq:conf-good}
 \sum_{{\bf j}\in \Ver} a_{\bf ij} {\bv}_{\bf j} \geq \bw_{\iv} + \bwt_{\iv}   
\end{equation}
which implies that the four dimensional theory is sensibly defined in the ultraviolet as a quantum field
theory. This implies that $a_{\bf ij}$ is the finite or affine ADE Cartan
matrix and $\Gamma$ is the finite or affine ADE Dynkin diagram. 
If the inequality \eqref{eq:conf-good} is saturated, we say that
$\RGamma$ is in \emph{conformal class} 
\begin{equation}
\label{eq:conformal-class}
 \sum_{{\bf j}\in \Ver} a_{\bf ij} {\bv}_{\bf j} = \bw_{\iv} + \bwt_{\iv}   \quad
 \leftrightarrow\quad  \text{$\RGamma$ is in conformal class}
\end{equation}
All affine type ADE quivers $\Gamma$ saturate the inequality (\ref{eq:conf-good})
at  $\bw_{\iv} = \bwt_{\iv} = 0$ and hence belong to conformal
class automatically. The assigned dimensions $\bv_{\iv}$ 
for affine ADE are uniquely determined by a
single integer $N$ and are given by $\bv_{\iv} = N a_{\iv}$ where $a_{\iv}$ are
Dynkin marks on the vertices of $\Gamma$; so that the imaginary root
$\delta = \sum_{\iv} a_{\iv} \alpha_{\iv}$ where $\alpha_{\iv}$ are simple roots. 
  \item The set ${\boldsymbol\kappa} = ({\kCS}_{\iv})_{{\iv} \in \Ver}$, where $\kCS_{\iv}$ is the level of the five-dimensional Chern-Simons  term for the $\iv$-th
 gauge group factor $SU(\bv_{\iv})$ 
 \begin{equation}
 {\kCS}_{\iv} \int {\tr}_{{\bv}_{\iv}} \left( \frac 13 A \wedge dA \wedge dA +  \frac 14 A^3 \wedge dA  + \frac 15 A^5 \right)
 \end{equation}
Each level $\kCS_{\iv}$ is    
constrained by the inequality 
\begin{equation}
\label{eq:ki-range}
\kCS_{\iv}^{-}
   \leq \kCS_{\iv} \leq  \kCS_{\iv}^{+}
\end{equation}
where
\begin{equation}
\label{eq:kCSplusminus}
  \begin{aligned}
\kCS_{\iv}^{-} =     -  \bv_{\iv} + \bwt_{\iv} + {\sum}_{{\bf j} \in t(s^{-1}({\iv}))} \bv_{\bf j}\\
\kCS_{\iv}^{+} = \bv_{\iv} - \bw_{\iv} - {\sum}_{{\bf j} \in s(t^{-1}({\iv}))} \bv_{\bf j}
  \end{aligned}
\end{equation}
 The range of
allowed Chern-Simons couplings \eqref{eq:ki-range} is non-empty under the conditions
(\ref{eq:conf-good}) when the four dimensional theory with the same quiver is in the
asymptotically free or conformal class.
  \item $\qe_{\iv} \in {\BC}^{\times}$, $0 < |{\qe}_{\iv}| < 1$  is the exponentiated complexified coupling constant of the $SU({\bv}_{\iv})$ gauge factor for the vertex ${\iv} \in \Ver$.
  The coupling constants $\qe$ encode the Yang-Mills coupling
 $g_{\mathrm{YM}}$ and the
$\theta$-parameter for each gauge group factor:
 \begin{equation}
   \qe_{\iv} = \exp( 2 \pi \ii \tau_{\iv}),  \quad \tau = \frac{ 4 \pi
     \ii}{g^2_{\mathrm{YM}}} + \frac{ \theta}{2 \pi}
 \end{equation}
 
  \item $\ac_{{\iv},\ba}$,  ${\iv} \in \Ver, \ba=1,\ldots, \bv_{\iv}$ are the
    eigenvalues
 of the scalars in the vector multiplet serving as the special
 coordinates
 on the classical Coulomb moduli space 
  \item $\mf_{{\iv},\fe}$  ($\mft_{{\iv},\fe}$) are the masses of the fundamental
    (anti-fundamental) matter multiplets  for the vertex $\iv$ with $\fe =
    1,\ldots ,\bw_{\iv}$ ($\fe= 1, \ldots \bwt_{\iv}$)
  \item $\mb_e$ is the mass of the bifundamental matter
    multiplet at edge $e \in \Edg$ \end{itemize}

We will also use exponentiated notations for the order parameters of the
theory  
\begin{equation}
\label{eq:exponents}
  \begin{aligned}
&  w_{{\iv} ,{\ba}} = e^{ \betir {\ac}_{{\iv},{\ba}}} \\
&  q_{1} = e^{ \betir\ep_{1}}, \quad  q_{2} = e^{ \betir\ep_{2}} \\
& \mu_{e} = e^{\betir\mb_e}, \\ 
& \muf_{{\iv},{\fe}}  = e^{\betir\mf_{{\iv},{\fe}} },  \quad 
 \muft_{{\iv},{\fe}} = e^{\betir\mft_{{\iv},{\fe}} }
  \end{aligned}
\end{equation}
Let $\RGamma$ be the representation of quiver $\Gamma(\bv,\bw,\bwt)$, 
that is the representation of the gauge group
\begin{equation}
\Gg = \times_{\iv} SU(\bv_{\iv})
\end{equation}
 for  the matter hypermultiplets in the theory
\begin{equation}
  \RGamma = \bigoplus_{e \in \Edg} (\bar{\bv}_{s(e)} \otimes
  \bv_{t(e)}) \oplus
  \bigoplus_{{\iv} \in \Ver}  (  \bar{\bw}_{\iv}   \otimes \bv_{\iv} )  \oplus
  \bigoplus_{{\iv} \in \Ver}  ( \bar{\bv}_{\iv}  \otimes \tilde{\bw}_{\iv}    ),
\end{equation}
where slightly abusing notations by $\bv_{\iv}$ we denote complex
vector space of dimension $\bv_{\iv}$, etc. It may be useful to think about $\bw_{\iv}$ fundamental matter multiplets for $SU(\bv_{\iv})$
as a bifundamental hypermultiplet with an auxiliary frozen gauge group
$SU(\bw_{\iv})$, and the role of masses $\mf_{{\iv}, {\fe}}$ is played by the frozen scalars
of the $SU(\bw_{\iv})$ vector multiplet. 

Equivalently, 
\begin{equation}
\label{eq:quiver-rep}
\RGamma = \bigoplus_{e \in \Edg} \Hom (\bv_{s(e)} ,   \bv_{t(e)})
 \oplus \bigoplus_{{\iv} \in \Ver} \Hom( \bw_{\iv},  \bv_{\iv}   )
\oplus  \bigoplus_{{\iv} \in \Ver} \Hom( \bv_{\iv}  , \tilde{\bw}_{\iv})
\end{equation}

The hypermultiplet space $\RGamma$ is naturally acted by
flavor symmetry group for the fundamental fields 
\begin{equation}
 G_{\bw,\bwt} = \times_{\iv} U(\bw_{\iv}) \times \times_{\iv} U(\bwt_{\iv}) 
\end{equation}
and  bifundamentals
\begin{equation}
 G_{\mathrm{edge}}=  \times_{e \in \Edg} U(1) 
\end{equation}
The total flavor symmetry group is 
\begin{equation}
\label{eq:flavor}
  \Gf = G_{\mathrm{edge}} \times G_{\bw, \bwt} 
\end{equation}
The theories corresponding to two equipped quivers differing by the orientation of some arrows
(or the choice of the fundamental/anti-fundamental multiplets) are
equivalent  after certain adjustments of the couplings $\kCS_{\iv}$ and $\qe_{\iv}$.

 The rationale for the equation \eqref{eq:conf-good} is to require the non-positive beta-function
$$
{\beta}_{\iv} \leq 0
$$ 
for the four dimensional  running of the $\iv$-th gauge coupling constant: 
$$
\Lambda_{\text{uv}} \frac{d}{d\Lambda_{\text{uv}}} {\tau}_{\iv} =   \beta_{\iv}   = \bw_{\iv} + \bwt_{\iv} - \sum_{\bf j} a_{\bf ij}
  \bv_{\bf j}  \ .
$$
{}The actual reason for this requirement is not so much that the gauge theory is well-defined in the ultraviolet, since we are going to study the five dimensional theory whose definition requires
  some completion at high energies. The class of theories which can be sensibly completed
  at high energy by embedding in some brane configuration in string theory or by geometrical engineering in M-theory or by compactification of the $(0,2)$-theory from six dimensions seems to be much larger. However, there are reasons to suspect that once we go beyond the realm of the theories which are sensible quantum field theories in four dimensions, the microscopic physics of $M$-theory will not decouple. For example, the instanton partition function defined entirely in terms of the gauge theory degrees of freedom will diverge, or will cease to be gauge invariant.  
  
\subsubsection{Six dimensions and anomalies\label{se:6dversion}}  

Formally one can define
and compute the partition ${\CalZ}$-function  even for the six
dimensional theory compactified on the two-torus $\BT^2_{\lfive, -\lfive/\tau_p}$,
  as the equivariant elliptic genus of the four dimensional $\Gg$ instanton moduli spaces \cite{Nekrasov:thesis}, \cite{Baulieu:1997nj}. The Coulomb parameters $\ac$ of the six dimensional theory compactified on $\BT^2_{\lfive, -\lfive/\tau_p}$ parametrize holomorphic ${\Gg}^{\BC}$-bundles on $\BT^2_{\lfive, -\lfive/\tau_p}$. The latter is a quotient of the abelian variety by the action of finite group (the Weyl group of $\Gg$). One has to check that the partition function ${\CalZ}$ is invariant with respect to large gauge transformations on $\BT^2_{\lfive, -\lfive/\tau_p}$. Indeed, we find that if ${\beta}_{i}=0$ then $\CalZ$ is invariant, provided the coupling constants $\qe_{\iv}$ of $SU(\bv_{\iv})$  transform in a suitable way under such large gauge transformations.
This has the following explanation. The six dimensional $\mathcal{N}=(1,0)$ theory for affine ADE
quiver $\Gamma$ with gauge group $\Gg = \times SU(\bv_{\iv})$  where $\bv_{\iv} = N a_{\iv}$
and  $a_{\iv}$ are Dynkin marks can be realized on the stack of N D5 branes
in IIB on ${\BC}^2/\Gammadi$, where $\Gammadi
\subset SU(2)$ is the finite subgroup associated to affine ADE quiver $\Gamma$ by McKay
correspondence. 
   In addition to the vector multiplets of the gauge group $\Gg = \times SU(\bv_{\iv})$ one has  $r$  tensor multiplets coming from the reduction of the IIB potentials on the $r$ cycles generating $H_{2}({\BC}^2/\Gammadi)$.  Now, the theory is not anomalous (see p.11 of \cite{Schwarz:1995zw} in the $A_1$ case and p.8 of \cite{Blum:1997} for the more general case) if the scalar fields of the tensor multiplets are coupled suitably to the gauge fields.  Because of this coupling, indeed, we need to compensate the large gauge transformation, acting by the shift of $\ac$ by the change of $\qe_{\iv}$ when we compute the ${\CalZ}$-function of the six dimensional theory on $\BT^2_{\lfive, -\lfive/\tau_p}$.

We recall the computation from \cite{Blum:1997mm} adopting to our
notations. 
Consider $\mathcal{N}=1$ theory in six dimensions with the gauge group $\Gg$ and the hypermultiplet in the quiver
representation $\RGamma$. The quartic anomaly cancellation condition from \cite{Seiberg:1996qx} requires
\begin{equation}
  \tr_{\mathrm{adj}} F^4 - \tr_{R} F^4 = \sum_{\tilde \iv =1}^{\tilde r} ( \alpha^{\tilde \iv}{}_{\bf j} \tr F_{\bf j})^2
\end{equation}
where $F$ is $\Gg$ curvature, 
index $j$ labels the simple factors in the gauge group $\Gg$ and $\tilde r$ is equal to the number of tensor multiplet that we
need to add to cancel the anomaly in the vector and hypermultiplet.
A useful identity for $SU(\bv_{\iv})$ gauge groups is
\begin{equation}
  \tr_{\mathrm{adj}} F^4_{\iv} = 2 \bv_{\iv} \tr F_{\iv}^4 + 6 (\tr F_{\iv}^2)^2
\end{equation}
Let $x_{{\iv},\alpha}$ be Chern roots of $F_{\iv}$. Each $({\iv},{\bf j})$ bifundamental
hypermultiplet contributes 
\begin{equation}
  -\sum_{\alpha, \beta} (x_{{\iv}\alpha} - x_{{\bf j}\beta})^4 = -(   \bv_{\bf j} \tr F_{\iv}^4
 + 6 \tr F_{\iv}^2 \tr F_{\bf j}^2 + \bv_{\iv} \tr F_{\bf j}^4)
\end{equation}
For a quiver $\Gamma$ with Cartan matrix $a_{ij}$ we find that
 vector multiplet and all bifundamental together contribute 
\begin{equation}
 \sum_{\bf ij} a_{\bf ij} ( \bv_{\iv} \tr F_{\bf j}^4  + 3 \tr F_{\iv}^2 \tr F_{\bf j}^2)
\end{equation}
Adding up the contributions from all the fundamental and the anti-fundamental multiplets we
find, for the total anomaly
\begin{equation}
\sum_{{\iv},{\bf j}} (\bv_{\iv} a_{\bf ij} - \bw_{\bf j}  - \bwt_{\bf j})  \tr F_{\bf j}^4  +
 3 \sum_{{\iv}, {\bf j}}  a_{\bf ij} \tr F_{\iv}^2 \tr F_{\bf j}^2 
\end{equation}
For $\RGamma$ in the conformal class \eqref{eq:conformal-class} the dangerous $\tr F^4$ term vanishes.  To cancel the remaining anomaly by the
$\tilde r$
tensor multiplets we want to present it as a sum of $\tilde r$ squares
$\sum_{\tilde {\iv}=1}^{\tilde r} \left( \alpha^{\tilde {\iv}}{}_{\bf j} \tr F_{\bf j}^2 \right)^2$ with some real coefficients
$\alpha^{\tilde {\iv}}{}_{\bf j}$, so that the two-forms ${\bB}_{\tilde \iv}$  of the tensor multiplets interact with the 
gauge fields by means of the coupling
\begin{equation}
 \alpha^{\tilde {\iv}}{}_{\bf j} {\bB}_{\tilde {\iv}} \wedge \tr F_{\bf j}^2  
\end{equation}
shifting the gauge coupling 
\begin{equation}
  \tau_{\bf j} \to  \tau_{\bf j} + \sum_{\tilde {\iv}} \alpha^{\tilde {\iv}}{}_{\bf j} \int_{T^{2}} {\bB}_{\tilde {\iv}} 
\end{equation}
 In other words, we need to find
$\tilde r$ one-forms $\alpha^{\tilde {\iv}}{}_{\bf j}$, $\tilde {\iv} = 1, \ldots,
\tilde r$  such that 
\begin{equation}
  a_{\bf ij} = \sum_{\tilde {\iv}=1}^{\tilde r} \alpha^{\tilde {\iv}}{}_{\iv} \alpha^{\tilde {\iv}}{}_{\bf j}
\end{equation}
The equation exactly coincides with the definition of the Cartan matrix $a_{\bf ij}$
of quiver $\Gamma$ in terms of simple roots $\alpha_{j} = \sum_{\tilde
  i} \alpha^{\tilde i}{}_j e_{\tilde i}$ expressed in some orthonormal basis
$e_{\tilde i}$. Therefore, $\tilde r = r$ where $r$ denotes the rank of Cartan matrix $a_{ij}$. Recall that the $\Gamma$ of finite type has $r = \# \Ver$ nodes while the $\Gamma$
of affine type has $r = \#\Ver - 1$ nodes. If $\Gamma$ is of affine type, then
$\alpha^{\tilde \iv}_{\bf j}$ for $\tilde {\iv} = 1, \ldots, r$ are the coefficients
of simple roots $\alpha^{\tilde \iv}{}_{\bf j}$ in the expansion  in the basis
$(e_{\tilde {\iv}})_{\tilde {\iv} = 0, \ldots, r}$, where $e_{\tilde 0} = \delta$
is the imaginary root of the affine system \cite{Kac_1990}, and 
$(e_{\tilde {\iv}})|_{\tilde {\iv} = 1, \ldots, r}$ is the orthonormal basis for the root
system for the underlying quiver $\Gamma'$ of finite type obtained by
removing the \emph{affine node '0'} from $\Gamma$.

We conclude that the lift to the six dimensional ${\CalN}=(1,0)$ theories of all of the quiver theories $\RGamma$ of the conformal class
\eqref{eq:conformal-class} is not prohibited by the quartic anomaly
\cite{Seiberg:1996qx}.

\subsection{The $\CalZ$-function}

To the  quiver representation $\RGamma$, the parameters ${\qe, \bm,\ac}$ and two additional parameters
$\ep = (\ep_1,\ep_2)$ introduced in 
 \cite{Moore:1997dj,Losev:1997tp,Lossev:1997bz} we associate the  partition
function ${\CalZ}$ as the formal twisted Witten index of the five dimensional gauge theory  \cite{Nekrasov:2002qd,Nekrasov:2003rj}:
\begin{equation}
  \CalZ(t) = \Tr_{\mathcal{H}} (-1)^{F} e^{\betir H} \tequiv, \qquad \tequiv \in \tilde\Tequiv 
\end{equation}
It is the partition function of the five dimensional theory on the five-dimensional manifold which is the twisted bundle $\widetilde{\BR^{4} {\times} \BS^{1}}_{\la \ep_1, \ep_2 ; \lfive
  \ra}$ mentioned earlier, over the circle  $\BS^{1}_{\la \lfive
  \ra}$ of circumference $\lfive$, with the fibers being ${\BR}^{4}$.  The boundary conditions involve the twists by flavor symmetry, the $R$-symmetry, 
the asymptotic gauge transformation, and the rotation of  $\BR^{4}$. The couplings ${\qe}_{\iv}$ are actually the expectation values of the complexified
Wilson loop of the five dimensional vector multiplet which couples to the conserved topological current
$$
J_{\iv} =\star {\tr}_{{\bv}_{\iv}}\, F \wedge F
$$
Therefore the couplings ${\qe}_{\iv}$ and the couplings ${\mu}_{e}$, ${\mu}_{i, {\fe}}$
are of somewhat similar nature. In fact, in some theories (and most likely in all theories
studied in our paper) the global symmetry of the theory extends to include the transformations generated by the topological charges. The torus $\tilde\Tequiv$ is the maximal torus of the group $\Gequiv \times G_{\mathrm{top}} = \Gg \times \Gf \times \Gl \times G_{\mathrm{top}}$, defined below. The group $G_{\mathrm{top}}$ is generated by the topological charges. For the theories in this paper the it is 
$U(1)^{\Ver}$. 
The factor $\Gg$ is the group of global constant
gauge transformations (or changing of the framing at infinity of the
framed gauge bundle used to describe instantons on $\BR^{4}$), the $\Gf$
is the flavor symmetry group acting on hypermultiplets, the $\Gl =
SO(4)$ is the $\BR^4$ rotation group. 
 The four-dimensional instantons are lifted to
particles in five dimensions. The partition function ${\CalZ}$
factorizes as the product of the tree level, the perturbative and the instanton factors:
\begin{equation}
  {\CalZ} = \CalZ^{\mathrm{tree}} \CalZ^{\mathrm{pert}} \CalZ^{\mathrm{inst}}\ .
\end{equation}
The perturbative part is ambiguous, as we mentioned above, it depends on the choice of boundary conditions. The instanton part is unambiguous and can be defined  mathematically precisely as
the $\Gequiv$-character of the graded vector space $\Hinstk$
\begin{equation}
  \CalZ^{\mathrm{inst}}(\qe; \tequiv) = 
\sum_{\bk} \, \qe^{\bk} \str_{\Hinstk}\, (  \tequiv)
\end{equation}
evaluated on the group element $\tequiv$ of the
maximal torus $\Tequiv$ of the equivariant group $\Gequiv = \Gg \times \Gf \times
\Gl$. The group element $\tequiv \in \Tequiv$ is parametrized by  $(\ac,
\bm, \ep)$.
 The Coulomb  parameters $\ac = ( \ac_{{\iv},\ba} )$ are coordinates in the
Cartan algebra of global constant gauge transformations acting at the
framing of the gauge bundle at infinity
\begin{equation}
\Gg = \prod_{{\iv} \in \Ver} SU(\bv_{\iv})
\end{equation}
 the mass parameters $\bm= (\mb_{e}, \mf_{{\iv},\fe}, \mft_{{\iv},\fe}) $ are the coordinates on the
Cartan subalgebra of the flavor symmetry group (\ref{eq:flavor})
\begin{equation}
\Gf = \times_{e \in \Edg} \, U(1) \, \times_{{\iv} \in \Ver}\, 
U(\bw_{\iv})  \, \times_{{\iv} \in \Ver} \, U(\bwt_{\iv})
\end{equation}
and the epsilon parameters ${\ep} = ({\ep}_1, {\ep}_2)$ are the
coordinates on the Cartan subalgebra of the four dimensional rotation group
\begin{equation}
\Gl  = SO(4)
\end{equation}
acting on $\BR^{4}$ by rotations about a particular point $0$.
The set of integers
$\mathbf{k} \in {\BZ}_{+}^{\Ver} =  \{\bk_{\iv} \in \BZ_{\geq 0} \, |\,  {\iv} \in \Ver\}$ encodes the 
four dimensional instanton charges (the second Chern classes $c_2$) for the gauge group
factors $SU(\bv_{\iv})$ for ${\iv} \in \Ver$. More precisely, 
\begin{equation}
   \Hinstk = H^{\bullet}( \Minstk, \CalE^{R})
\end{equation}
is the sheaf cohomology of the virtual matter bundle $\hat \CalE^{\RGamma} \to
\Minstk$ defined as follows. 
The base space $\Minstk$ is the framed $\Gg$-instanton moduli space on
$\BR^4$. Let $\mathcal{A}_{\Gg}$ be the space of connections on a
principal $\Gg$-bundle over $\BR^{4}_{\ep_1,\ep_2}$ with fixed framing at
infinity and fixed second Chern class $\mathbf{k}$.  Then 
\begin{equation}
  \Minstk = \{ A \in \mathcal{A}_{\Gg} \ |\   F_{A}^{+} =0 \} / \mathcal{G_\infty}
\end{equation}
where $\mathcal{G_\infty}$ denotes the group of framed gauge transformations, i.e. trivial  at
${\infty} = {\BS}^{4} \backslash \BR^{4}_{\ep_1,\ep_2}$. The fiber of the virtual matter bundle
$\CalE^{R}_{A}$ at the point $A \in \Minstk$ is the virtual space of the
zero modes of Dirac operator in representation $\RGamma$ in the background of the instanton 
connection $A$ in the gauge bundle on $\BS^{4}$, that is 
\begin{equation}
\hat  \CalE^{R}_A= \ker \slashed{D}^{R}_{A} - \coker \slashed{D}^{R}_{A} 
\end{equation}

The equivariant partition function ${\CalZ}(\qe, \ac, \bm; \ep)$ is
analytic in the equivariant parameters $(\ac,\bm;\ep)$ which henceforth
are treated  as  complex variables.

\subsection{Index computation}

 In this section we give only the brief review of the gauge theory
 partition ${\CalZ}$-function, mostly in order to set the notations. We
 shall skip the arguments of the partition function $\CalZ$. 
 
 For more details one can consult 
 \cite{Nekrasov:2002qd,Nekrasov:2003rj,Lossev:1997bz,Moore:1997dj,Losev:1997tp,Nakajima:1999,Nakajima:2003Lectures,Nakajima:2003,Shadchin:2005mx, Pestun:2007rz}.
 The $\Tequiv$-equivariant path integral of the supersymmetric
 $\Omega$-deformed gauge theory on ${\BR}^{4}$ localizes to
 the $\Tequiv$-equivariant integral over the instanton moduli space 
$ \bigcup_{\mathbf{k}} \Minstk$, where $\mathbf{k} = ({\bk}_{\iv})_{{\iv} \in \Ver} \in {\BZ}_{+}^{\Ver}$ is the
second Chern class of the anti-self-dual $\Gg$-connection on
$\BR^4$ and
\begin{equation}
 \Minstk = \prod_{{\iv} \in \Ver} \mathcal{M}_{\bv_{\iv},\bk_{\iv}} 
 \label{eq:modinst}
\end{equation}
Then, using the Atiyah-Bott theorem  
 one reduces the integral over $\Minstk$
to the sum over the set of fixed points of the locally defined expressions. The fixed point formula is  similar to the calculation of an integral of a meromorphic form via residues. 
To apply  Atiyah-Bott localization we need to partly compactify  the moduli space of instantons  to the  Gieseker-Nakajima moduli space
$\Minstk$ in which vector bundles are replaced by the torsion free sheaves, or, in differential-geometric language, by the noncommutative instantons \cite{Nekrasov:1998ss}. Thus, in particular, 
${\CalM}_{\bv_{\iv},\bk_{\iv}}$ is the moduli space of framed torsion free rank $\bv_{\iv}$
sheaves $E_{\iv}$ on ${\BC\BP}^{2}$, trivialized over a complex line
${\BC\BP}^{1}_{\infty}$ at infinity, with the second Chern characters
$\text{ch}_{2}(E_{\iv}) = \bk_{\iv}$. 

Let $\CalE^{\iv}$ denote the universal torsion
free sheaf over ${\CalM}_{\bv,\bk} \times {\BC\BP}^{2}$ in the $\iv$-th factor, and let $\pi:
\CalM_{\bv,\bk} \times \CP^{2} \to \CalM_{\bv,\bk}$ be the projection.
For a quiver representation $\RGamma$ the matter sheaf cf. \eqref{eq:quiver-rep}) over
$\CalM_{\bv,\bk} \times \CP^2$ is 
\begin{equation}
\mathcal{E}^{\RGamma} = \bigoplus_{e \in \Edg} \Hom (\CalE_{s(e)} ,  \CalE_{t(e)})
 \oplus \bigoplus_{{\iv} \in \Ver} \Hom( \bw_{\iv},  \CalE_{\iv} )
\oplus  \bigoplus_{{\iv} \in \Ver} \Hom( \CalE_{\iv} , \tilde{\bw}_{\iv})
\label{eq:mattsh}
\end{equation}
Let
 \[
{\hat \CalE}^{\RGamma} = R {\pi}_{*} {\CalE}^{\RGamma}
\]
 be the derived
pushforward of the matter sheaf to $\CalM_{\bv,\bk}$, which means that 
the virtual fiber $\hat \CalE^{\RGamma}|_{m}$ at a point $m \in
\CalM_{\bv, \bk}$ is the graded space of sheaf cohomology
$H^{\bullet}(\CP^2, \CalE^{\RGamma}|_m)$.

\subsection{Four dimensions}

The instanton part of the four dimensional gauge theory partition function is 
\begin{equation}
4d: \qquad   \CalZ^{\text{inst}} = \sum_{\bk=\boldsymbol{0}}^{\boldsymbol{\infty}} \qe^{\bk}
  \int_{\CalM_{\bv, \bk}} e_{\Tequiv} (\hat \CalE^{\RGamma})
\end{equation}
where equivariant integration $\int_{\CalM_{\bv,\bk}}$ denotes the
pushforward of the $\Tequiv$-cohomologies of $\CalM_{\bv, \bk}$ by 
 projection map from  $\CalM_{\bv, \bk}$ to a point, $e_{\Tequiv}$
 is the $\Tequiv$-equivariant Euler class, 
 and
 \begin{equation}
 {\qe}^{\bk} = \prod_{{\iv} \in \Ver} {\qe}_{\iv}^{{\bk}_{\iv}}
 \label{eq:qebk}
 \end{equation}
 
 \subsubsection{The ${\y}$-operators in four dimensions}
 
The geometric definition of generating functions $\y_{\iv}(x)$
(\ref{eq:yphys}) is conveniently given  in terms of the spectral Chern
class $c(x,E)$. For $x_I$ the virtual Chern roots of a vector bundle $E$, define the
rational $x$-spectral Chern class $c(x,E)$ as 
\begin{equation}
  c(x,E) = \prod_{I} (x - x_I)
\end{equation}
so that $c(1,E)= c(E^{\vee})$ is the ordinary  total Chern class. Then
\begin{equation}
  \label{eq:ygeom4}
4d: \qquad   \y_{\iv}(x - \tfrac \ep 2) = \frac { \sum_{\bk=\boldsymbol{0}}^{\boldsymbol{\infty}}   {\qe}^{\bk} \int_{\CalM_{\bv,
      \bk}}  c_{\Tequiv} (x,\mathring{\CalE}_{\iv})  e_{\Tequiv} (\hat
  \CalE^{\RGamma})}
{\sum_{\bk=\boldsymbol{0}}^{\boldsymbol{\infty}}  {\qe}^{\bk}  \int_{\CalM_{\bv,
      \bk}}   e_{\Tequiv} (\hat \CalE^{\RGamma})}
  \end{equation}
and $\mathring{\CalE}_{\iv}$ is the restriction of the sheaf $\CalE_{\iv}$ at the
$\Tequiv$-fixed point $0 \in \BC^2$. The restriction is defined for sheaves
using the Koszul resolution complex and derived pushforward map
\begin{equation}
  \mathring{\CalE}_{\iv} = R \pi_{*} \CalE_{\iv} \otimes ( \Lambda^2
  T^{*}_{\BC^2} \to T^{*}_{\BC^2} \to \mathcal{O})
\end{equation}

\subsection{Five dimensions}

The five dimensional, K-theoretic, version of the partition function is 
\begin{equation}
5d: \qquad     \CalZ^{\text{inst}} = \sum_{\bk} \qe^{k}
\ch_{\Tequiv} H_{\Tequiv}^{\bullet}(\CalM_{\bv,\bk}, \mathbb{I}(\hat \CalE^{\RGamma}) )
\end{equation}
where $\mathbb{I}$ is the index functor defined on sheaves by 
\begin{equation}
  \mathbb{I}: \CalE \to \sum_{l=0}^{\infty} (-1)^{l} \Lambda^{l} {\CalE} 
\end{equation}
which satisfies  
\begin{equation}
  \indfun\left[ \CalE_1 \oplus \CalE_2 \right]  = \indfun\left[\CalE_1\right] \indfun\left[\CalE_2\right] \\
\end{equation}
and
\begin{equation}
\indfun\left[ {\CalE}^{\vee} \right] = \indfun\left[ {\CalE} \right]^{\vee} = (-1)^{\mathrm{rk}{\CalE}}
\left( \Lambda^{\mathrm{rk}{\CalE}} {\CalE} \right)^{-1} \otimes \indfun\left[ {\CalE} \right]
\label{eq:serrd}
\end{equation}
At the level of Chern characters, with $x_I$ the virtual Chern roots,
\begin{equation}
\label{eq:indfun}
    \indfun\left[\sum_{I} e^{x_I}\right]  = \left[\prod_{I} (1 - e^{x_I})\right]
\end{equation}
Remember, that in our theories we have the mass parameters, i.e. the equivariant parameters for the symmetries, which rescale the fibers of various summands in the matter sheaf \eqref{eq:mattsh}. Thus the sum in \eqref{eq:indfun} is more refined then it may appear. 

\subsubsection{The ${\y}$-operators in five dimensions}

The five-dimensional, K-theoretic geometric definition of the $\y_{\iv}(\xi)$ function is 
\begin{equation}
\label{eq:ygeom5}
  \y_{\iv}(\xi q^{-\frac 12}) =  \frac{ \sum_{\bk=\boldsymbol{0}}^{\boldsymbol{\infty}}
    \qe^{\bk} 
\int_{\CalM_{\bv,\bk}} \td_{\Tequiv} (T{\CalM_{\bv,\bk}} ) \hat
c_{\Tequiv}(\xi,\mathring{\CalE}_{\iv})  \hat c_{\Tequiv}(1,\hat
\CalE^{\RGamma})
}{\sum_{\bk=\boldsymbol{0}}^{\boldsymbol{\infty}}
    \qe^{\bk} 
\int_{\CalM_{\bv,\bk}} \td_{\Tequiv} ( T{\CalM_{\bv,\bk}} ) \hat c_{\Tequiv} (1,\hat
\CalE^{\RGamma})
}
\end{equation}
where the $K$-theoretic trigonometric $\xi$-spectral characteristic class $\hat c(\xi, E)$ is defined in
terms of the $\Tequiv$-equivariant Chern roots $x_I$ of the sheaf $E$ by 
\begin{equation}
\hat c_{\Tequiv}(\xi, E) = \prod_{I} (1 - \xi_I/\xi)  \qquad \xi =
e^{\ii  \lfive x}, \quad \xi_I = e^{\ii \lfive x_I} 
\end{equation}
where $\lfive$ is the circumference of the 5d compactification circle
$\mathbb{S}^1_{\lfive}$. The class ${\hat c}_{\Tequiv}({\xi}, E)$ has the following transformation property:
\begin{equation}
{\hat c}_{\Tequiv} ({\xi}, E^{\vee}) = e^{-c_{1}(E)} (-{\xi})^{-\mathrm{rk}E} {\hat c}_{\Tequiv} ({\xi}^{-1}, E)
\label{eq:hcte}
\end{equation}

\subsubsection{Asymptotics of the $\y$-operators}

Using the definitions above, and the fact that $\rk \CalE_{\iv} = {\bv}_{\iv}$, 
we get: 
\begin{equation}
\label{eq:yfun1}
  \begin{aligned}
&    4d:\qquad \y_{\iv}(x) \to x^{\bv_{\iv}}, \qquad x \to \infty\\
&    5d:\qquad
    \begin{cases}
      \y_{\iv}(\xi) \to 1, \qquad \xi \to \infty \\
      \y_{\iv}(\xi) \to  \xi^{-\bv_{\iv}} \prod_{\ba=1}^{\bv_{\iv}} (-w_{{\iv}, \ba}q^{-\frac 12}),
      \qquad \xi \to 0
    \end{cases}
  \end{aligned}
\end{equation}
In the limit $\qe = 0$ the functions $\y_{\iv}(x), \y_{\iv}(\xi)$ are (Laurent) polynomials
\begin{equation}
  \begin{aligned}
  4d: \qquad \lim_{\qe \to 0}  \y_{\iv}(x- \tfrac \ep 2) = \prod_{\ba=1}^{\bv_{\iv}} ( x -
  \ac_{{\iv},\ba})\\
 5d: \qquad \lim_{\qe \to 0} \y_{\iv}(\xi q^{-\frac 1 2}) = \prod_{\ba=1}^{\bv_{\iv}} ( 1 -
  w_{{\iv},\ba}/\xi)
  \end{aligned}
\end{equation}
The shifts $x \mapsto x - \tfrac \ep 2$, ${\xi} \mapsto {\xi} q^{-\frac 12}$ are introduced here for a later convenience of 
relating $\y_{\iv}$ to the representation theory of quantum groups.

\subsection{Partition function is a sum over partitions}

The equivariant integration over $\CalM_{\bv,\bk}$ reduces to the sum
over the space  $\CalM_{\bv,\bk}^{\Tequiv}$ of $\Tequiv$-fixed points by
Atiyah-Bott localization
\begin{equation}
\label{eq:Z-sum}
  \CalZ^{\text{inst}} = \sum_{\bk} \sum_{{\lambda} \in \CalM_{\bv, \bk}^{\Tequiv}} Z_{\lambda}
\end{equation}
Note that $Z_{\lambda}$ in \eqref{eq:Z-sum} contains the factor 
\[
{\qe}^{\bek} = \prod_{{\iv} \in \Ver} {\qe}_{\iv}^{{\bek}_{\iv}} \] 
which helps to make the instanton sum a convergent series, at least for $| {\qe}_{\iv} | \ll 1$. 
 The $\Tequiv$-fixed points $\lambda$ are the sheaves which split as the
 sum of the $\Tequiv$-equivariant rank one torsion free sheaves. The
 latter are the finite codimension monomial ideals ${\CalI}$ in the ring
 of polynomials in two variables $\BC[z_1,z_2]$. The latter are in
 one-to-one correspondence with partitions of the size $\codim \CalI$ . The use of torsion free sheaves in lieu of the honest bundles reduces
the rotation symmetry group from $SO(4)$ to $U(2)$, since it requires
a choice of complex structure on ${\BR}^{4} \approx {\BC}^{2}$. Thus the symmetry
group for the ${\CalM}_{\bv_{\iv},\bk_{\iv}}$ moduli space is $U(\bv_{\iv}) \times U(2)$. The maximal torus
$\Tequiv$ is unchanged by this reduction.

The set of fixed points $\CalM_{\bv, \bk}^{\Tequiv}$ is a two indexed set of partitions labeled by the
quiver node ${\iv} \in \Ver$ and by the color $\ba = 1, \ldots , {\bv}_{i}$ of the $SU(\bv_{\iv})$-factor in $\Gg$:
\begin{equation}
 \CalM_{\bv, \bk}^{\Tequiv} = \{ \lambda_{{\iv}, {\ba}} \ | \  {\iv} \in \Ver, \, {\ba} =1,
 \ldots, {\bv}_{\iv} \ \},
 \label{eq:flam}
\end{equation}
where  each partition $\lambda_{{\iv},{\ba}}$ is a non-increasing sequence of
non-negative integers 
\begin{equation}
  \lambda_{{\iv}, \ba } = (\lambda_{{\iv}, \ba, k})_{k \in {\BN}} = (\lambda_{i, \ba, 1}
  \geq \lambda_{{\iv}, {\ba}, 2} \geq
  \dots \geq 0 = 0 = \ldots ) 
\end{equation}
whose size is defined as:
\[ |{\lambda}_{{\iv}, {\ba}}| = \sum_{k=1}^{\infty} {\lambda}_{{\iv}, {\ba}, k}
\]
Let us denote by ${\lambda}_{\iv} = ({\lambda}_{{\iv}, {\ba}})_{{\ba}=1}^{{\bv}_{\iv}}$ the colored partition corresponding to $U({\bv}_{\iv})$ (cf. \cite{Nekrasov:2003rj}), and
\[ | {\lambda}_{\iv} | = \sum_{{\ba}=1}^{{\bv}_{\iv}} | {\lambda}_{{\iv}, {\ba}}| 
\]
The fixed point ${\lambda} \in \CalM_{\bv, \bk}$ corresponding to ${\lambda} = ( \lambda_{\iv} )_{{\iv} \in \Ver}$ (we thus use the same letter $\lambda$ to denote both the fixed point in the moduli space of sheaves and the quiver-colored partition it corresponds to) is the collection
$({\Efix}_{{\iv}, {\lambda}_{\iv}})_{{\iv} \in \Ver}$ of the sheaves which are the direct sums of the monomial ideals corresponding to ${\lambda}_{{\iv}, {\ba}}$. The topology of the sheaf determines the sizes of the colored partitions:
\begin{equation}
| {\lambda}_{\iv} |  = {\bk}_{\iv} 
\label{eq:instchkl}
\end{equation}
Let $\CalE_{{\iv},\lambda_{\iv}}$ be the
$U({\bv}_{\iv}) \times U(2)$ equivariant Chern character of ${\Efix}_{{\iv},\lambda_{\iv}}$ restricted at $0 \in {\BC}^{2}$. Let $(\ac_{i, \ba})$, $\ep=
(\ep_1,\ep_2)$ be coordinates in the Cartan algebra of $SU(\bv_{\iv})
\times SO(4)$. From  the ADHM \cite{Atiyah:1978ri} construction one derives (cf. \cite{Nakajima:1999})
\begin{equation}
  \mathcal{E}_{{\iv},\lambda_{\iv}} = W_{\iv} - (1 - q_1) (1 - q_2) V_{{\iv},{\lambda}_{\iv}}
\end{equation}
where $q_{1},q_{2}$ are defined in \eqref{eq:exponents}, 
\begin{equation}
  \begin{aligned}
  W_{\iv} &= \sum_{\ba=1}^{\bv_{\iv}} e^{ \betir{\ac}_{\iv, \ba }} \\
  V_{{\iv}, {\lambda}_\iv} &= \sum_{\ba=1}^{\bv_{\iv}} \sum_{ s \in \lambda_{\iv, \ba}} 
  e^{ {\betir}  c_{\iv, \ba,  s} }
  \end{aligned}
\end{equation}
where $s \in \lambda_{\iv, \ba}$ labels boxes in the partition
$\lambda_{\iv, \ba}$ so that $s=(s_1,s_2)$ is a pair of integers with $s_{1} = 1, 2, \ldots, $ and $s_{2} = 1, \ldots, \lambda_{\iv, \ba, s_1}$, and
$c_{\iv, \ba, s}$ is the $\ac_{\iv, \ba}$-shifted content \cite{Nekrasov:2003rj} of the box $s$
\begin{equation}
  c_{\iv,\ba, s} = \ac_{\iv, \ba } + (s_1 - 1) \ep_1 + (s_2 - 1) \ep_2
\end{equation} 
The contribution of the fixed point $\lambda$ to the partition function 
is a product of factors labeled by the vector and hypermultiplets
\begin{equation}
\label{eq:Zfull}
  Z_{\lambda} =  \left( \prod_{{\iv} \in \Ver}
  Z^{\text{cs}}_{{\iv},\lambda}Z_{{\iv},\lambda}^{\text{top}}
  Z_{{\iv},\lambda}^{\text{vec}} Z_{{\iv},\lambda}^{\text{f}} Z_{{\iv} ,\lambda}^{\text{af}} \right) \left(  
\prod_{e \in \Edg} Z_{e,{\lambda}}^{\text{bf}} \right)
\end{equation}

\subsubsection{Plethystic exponents and the asymptotics ${\ep}_{2} \to 0$}

The contributions to the five dimensional index $Z_{\lambda}$ are found by applying the plethystic exponent (index functor) $\indfun$ to the corresponding characters ${\chi}_{\lambda}$.
Applying $\indfun$ to $\chi$ is equivalent to 
using Atiyah-Singer formula to compute the $\Tequiv$-equivariant
index of the Dolbeault operator $\bar \partial$
for the complex of $(0, \bullet)$-forms on  (virtual) vector
space which is $\Tequiv$-module with Chern character $\chi$.

First, we introduce the $q$-analogue of the $\log \Gamma$ function
\begin{equation}
  \gamma_{q}(\xi) = \log \indfun \left[ \frac{\xi}{1 -q} \right] =  \log \prod_{ n \geq
    0} (1 - \xi q^{n})
\end{equation}

To compute $\CalZ$ in the limit $\ep_2 \to 0$ we will need
the limit of $\gamma_{q_2}(x)$ function for $q_2 = e^{{\betir}
\ep_2} \to 1$. We find
\begin{equation}
  \gamma_{q_2}(\xi)  =  \sum_{m=1}^{\infty} \frac 1m \frac{{\xi}^{m}}{1- q_{2}^{m}} \stackrel{\ep_2 \to 0} {\approx}\  \exp \left( -\frac 1{\betir\ep_2} \Li_2 (\xi) \right)
\end{equation}

{}We now proceed with the definitions of each factor in \eqref{eq:Zfull}.

\subsubsection{The bi-fundamental contribution}
First, we describe explicitly the bi-fundamental factor 
$Z_{e,\lambda}^{\text{bf}}$.  
Let
\begin{equation}
\chi_{e,\lambda} = \ind \slashed{D}_{e,\lambda}  
\end{equation}
 be the index of the Dirac operator
on $\BR^{4}$
in the bifundamental representation $ e= (\mathbf{v}_{t(e)}), \bar{\mathbf{v}}_{s(e)})$ evaluated in the background of the connection associated to the fixed
point $\lambda \in \Minstk$. 
From the equivariant Atiyah-Singer theorem we find the character
\begin{equation}
  \chi_{e,\lambda} = - \mu_{e}^{-1} \frac{  ( q_1
     q_2)^{\frac 12}}{ (1 - q_1) (1-q_2)} 
 \CalE_{t(e),\lambda_{t(e)}}\CalE_{s(e),\lambda_{s(e)}}^{\vee}
 \label{eq:bifch}
\end{equation}
which contains the overall factor ${\mu}_e^{-1}$ (\ref{eq:exponents}) accounting 
for the bi-fundamental mass. It is the $U(1)_{e}$ flavor
symmetry group element of the bi-fundamental hypermultiplet corresponding to the edge $e$. By $\CalE^{\vee}$ we
denote the equivariant Chern character of the dual bundle $\Efix^{\vee}$ in which all weights are
the negatives of the weights in $\CalE$.

The character ${\CalE}_{{\iv}, {\lambda}}$ of the universal bundle $\Efix_{{\iv}, \lambda}$ can be
represented by the infinite sum
\begin{equation}
  {\CalE}_{{\iv}, \lambda} = \sum_{\ba=1}^{\infty}  \sum_{k=1}^{\infty} (1-q_1) e^{ 
    {\betir}  x_{{\iv}, \ba, k} } 
\end{equation}
where $x_{{\iv}, \ba , \bek }$ 
\begin{equation}
  x_{{\iv}, \ba, k} = {\ac}_{{\iv}, \ba} + \ep_1 (k - 1)  + \ep_2  \lambda_{{\iv}, \ba, k}  
\end{equation}
denotes the ${\ac}_{{\iv}, \ba}$-shifted content of the box $s = (s_1,s_2) = (k, \lambda_{{\iv}, \ba, k} +1)$ (which is right outside the Young diagram of the partition $\lambda_{{\iv}, \ba}$). 
The  profile of the partition $\lambda_{{\iv}, {\ba}}$ can be parametrized by
the infinite sequence ${\xi}_{{\iv}, {\ba}} = \left( {\xi}_{{\iv}, \ba,  1}, {\xi}_{{\iv}, \ba, 2}, \ldots \right)$
\begin{equation}
  \xi_{{\iv}, {\ba}, k} = e^{ {\betir} x_{{\iv}, {\ba}, k}}
\end{equation}
Let
\begin{equation}
\label{eq:setXi}
  {\Xi}_{\iv} = ( {\xi}_{{\iv}, \ba, k} )_{{\ba}=1, \ldots, {\bv}_{\iv} ; k = 1, 2 , \ldots}
\end{equation}
denote the set of all ${\xi}_{{\iv}, \ba, k}$ with fixed ${\iv} \in \Ver$ and
$1 \leq \ba \leq {\bv}_{\iv}$, $k  \in {\BN}$. 
Then
\begin{equation}
\label{eq:chi-bif}
  \chi_{e, {\lambda}} =   \frac{q_{1}^{1/2}  - q_{1}^{-1/2}} { q_{2}^{1/2}  - q_{2}^{-1/2}} \sum_{ (\xi_t, \xi_s) \in \Xi_{t(e)} \times \Xi_{s(e)}} \frac{\xi_{t(e)}}{{\mu}_{e}\xi_{s(e)}}
\end{equation}
The contribution $Z^{\text{bf}}_{e,\lambda}$ of the bi-fundamental hypermultiplets, computed 
from 
the Chern character  $\chi_{e,\lambda}$ \eqref{eq:chi-bif},
asymptotes to
\begin{equation}
\label{eq:bifund}
  Z^{\text{bf}}_{e,\lambda} \sim \exp  \left ( - \frac 1{\betir\ep_2} \sum_{ (\xi_t, \xi_s)
    \in \Xi_{t(e)} \times \Xi_{s(e)}} \Liker \left( \frac{\xi_{t}}{{\mu}_{e}\xi_{s}} \right) \right)
\end{equation}
where
\begin{equation}
\Liker(\xi)  = \Li_2 ( q_1^{\frac 1 2} \xi) - \Li_2 (q_1^{-\frac 12} \xi )
\end{equation}
Actually, both \eqref{eq:chi-bif} and \eqref{eq:bifund} are formal expressions at this stage, as the convergence of the series for ${\CalE}_{{\iv}, {\lambda}_{\iv}}$ requires $|q_{1}|< 1$.
Of course, the original formula  \eqref{eq:bifch} makes sense, as it operates with finite expressions, so that the correct formula
is 
\begin{multline}
{\chi}_{e, \lambda} = - \mu_{e}^{-1} \frac{  ( q_1
     q_2)^{\frac 12}}{ (1 - q_1) (1-q_2)} \left( 
W_{t(e),\lambda_{t(e)}}W_{s(e),\lambda_{s(e)}}^{\vee} \right) + \\
\mu_{e}^{-1} \left( ( q_1
     q_2)^{\frac 12}  W_{t(e),\lambda_{t(e)}} V_{s(e),\lambda_{s(e)}}^{\vee}
     +    ( q_1
     q_2)^{-\frac 12} V_{t(e),\lambda_{t(e)}} W_{s(e),\lambda_{s(e)}}^{\vee} \right) - \\
     - {\mu}^{-1}_{e} (q_{1}^{\frac 12} - q_{1}^{-\frac 12})( q_{2}^{\frac 12} - q_{2}^{-\frac 12})
     V_{t(e),\lambda_{t(e)}} V_{s(e),\lambda_{s(e)}}^{\vee} 
     \label{eq:chi-bifco}
\end{multline}
One can actually make sense out of  \eqref{eq:chi-bif} and \eqref{eq:bifund} using a prescription where one sums over ${\xi}_{s}$ first, viewing it as a telescopic sum. The telescopic sum is actually finite. The remaining sum over $\xi_{t}$ is convergent. One should use a similar prescription for the evaluation of $Z^{\text{vec}}_{{\iv},{\lambda}}$ below. 

One can actually avoid using the formal expressions whatsoever, by working with the finite formulae like \eqref{eq:chi-bifco}. Below we write the formal expressions for the sake of brevity. 

\subsubsection{The fundamental and the anti-fundamental hypermultiplet contributions}
The contribution of the fundamentals $Z^{\text{f}}_{{\iv},\lambda}$
is obtained by applying the $ \indfun$-functor to the simpler version of the  character \eqref{eq:bifch} \begin{equation}
   \chi_{\iv}^{\text{f}} = - \sum_{{\fe} = 1}^{\bw_{\iv}} \frac{ ( q_1
     q_2)^{\frac 12}}
{ ( 1- q_1) (1-q_2) } \CalE_{\iv}  \mu_{{\iv},{\fe}}^{-1}
 \end{equation}
Thus, in the ${\ep}_{2} \to 0$ limit, 
\begin{equation}
  Z^{\text{f}}_{{\iv},\lambda} \sim \exp \left( -\frac 1{\betir \ep_2} 
   \sum_{ (\xi_t, \xi_s)
    \in \Xi_{\iv} \times \{ \mu_{{\iv},\fe}\} } \Liker(\xi_{t}/\xi_{s}) \right)
\end{equation}
and, similarly, the anti-fundamental contribution is
\begin{equation}
  Z^{\text{af}}_{{\iv},\lambda} = \exp \left( - \frac 1{\betir \ep_2}  \sum_{ (\xi_t, \xi_s)
    \in \{ \tilde \mu_{{\iv},\fe}\} \times \Xi_{\iv}  } \Liker(\xi_{t}/\xi_{s}) \right)
\end{equation}

\subsubsection{The vector multiplet contribution}
To compute the vector multiplet contribution
$Z^{\text{vec}}_{{\iv},{\lambda}}$ we need to find the
$\Gequiv$-character
of the tangent space $T_{\lambda}\Minstk$ to the moduli space at the 
fixed point $\lambda \in \Minstk$. From the deformation theory, the
character of the tangent space is dual to the 
index of Dirac operator in the adjoint
representation twisted by the square root  of the canonical bundle. Thus, 
\begin{equation}
\label{eq:chi-vec}
  \chi_{\iv}^{\text{vec}} = \frac{ q_1 q_2} { (1 - q_1)(1-q_2)} \CalE_{\iv} \CalE_{\iv}^{\vee}
\end{equation}
Therefore, in the ${\ep}_{2} \to 0$ limit, 
\begin{equation}
  Z^{\text{vec}}_{{\iv},{\lambda}} = \exp  \left(  \frac 1{\betir \ep_2}
\sum_{ (\xi_t, \xi_s)
    \in \Xi_{\iv} \times \Xi_{\iv} } \Liker( q_1^{\frac 12} \xi_{t}/\xi_{s}) \right) 
\end{equation}
\subsubsection{The topological contributions}
The factor $Z^{\text{top}}_{{\iv},\lambda}$ accounts for the
topological instanton charge given by the second Chern class $\bk_{\iv}$ for the
$\iv$-th gauge group factor $U({\bv}_{\iv})$
\begin{equation}
\label{eq:Ztop}
  Z^{\text{top}}_{{\iv},\lambda} = \qe_{\iv}^{\bk_{\iv}} = 
\exp \, \frac{2 \pi \ii \tau_{\iv}}{\ii \lfive \ep_2} \left(
  \sum_{k=1}^{\infty} \sum_{{\ba}=1}^{{\bv}_{\iv}} \log 
  \frac{\xi_{{\iv}, \ba, k}}{\mathring{\xi}_{{\iv}, \ba, k}}
\right)
\end{equation}
where 
\begin{equation}
\label{eq:Xi0}
\mathring{\Xi}_{\iv} = \{  \mathring{\xi}_{{\iv},{\ba},k} =w_{{\iv},{\ba}} q_1^{k-1} \}
\end{equation}
corresponds to the empty partition $({\lambda}_{{\iv},{\ba},k}) = (0,0,\dots, )$.
 The  second Chern class of $\Efix_{{\iv}, {\lambda}_{\iv}}$ is the size $| {\lambda}_{\iv}|$ of the colored partition
 \begin{equation}
   \bk_{\iv} = \sum_{\ba=1}^{\bv_{\iv}} \sum_{k=1}^{\infty}  \lambda_{{\iv},{\ba}, k}
 \end{equation}
Finally, the factor $Z^{\text{cs}}_{{\iv},\lambda}$ accounts for the contribution from the
Chern-Simons coupling. It is the descendant of the cubic term in
the tree level prepotential
\begin{equation}
  \CalL_{\text{cs}, \iv} = -\kCS_{\iv} \int d^{5}x d^{4}{\vartheta}\, \tr {\bf\Phi}_{\iv}^3\sim \frac{ \betir \kCS_{\iv}}{ 3! \ep_1 \ep_2}
  \tr \Phi_{\iv}^3
\end{equation}
The fixed point contribution of the cubic term $\frac{1}{6} \tr \Phi_{\iv}^3$ is equal to the third component of the Chern
character of the universal bundle, restricted at the fixed point ${\lambda} \times 0 \in {\CalM}^{\text{inst}}_{\bk} \times {\BC}^{2}$:
\begin{equation}
 \frac 1{3!} \tr \Phi_{\iv}^3 \Biggr\vert_{{\lambda} \times \{ 0 \} } 
  =  \sum_\ba \left( \frac{{\ac}_{{\iv},{\ba}}^3}{6}  - \frac{{\ep}_{1}{\ep}_{2}({\ep}_{1}+{\ep}_{2})}{2} |{\lambda}_{{\iv},{\ba}} | - {\ep}_1 {\ep}_2 \sum_{s \in
  \lambda_{{\iv},{\ba}}} c_{s}  \right) 
\end{equation}
which gives, in terms of the $\xi_{{\iv},{\ba},k}$ variables
\begin{equation}
  Z^{\text{cs}}_{{\iv},\lambda} = \exp  \frac{\kCS_{\iv}}{2\ep_2} 
 \sum_{ \xi \in \Xi_{\iv}}   {\log}( {\xi}/{\mathring{\xi}} ) \left( \frac{{\log} ( {\xi}{\mathring{\xi}} )}{\betir} + 
 \ep_1 +  \ep_2   \right)
\end{equation}

\subsubsection{Six dimensional case}
For a six dimensional theory compactified on the torus 
\begin{equation}
\BT_{p} = {\BC}/\left( {\lfive}{\BZ}  \oplus {-\lfive/\tau_p}{\BZ} \right)
\label{eq:ptor}
\end{equation}
 with the metric
 \begin{equation}
 ds_{p}^{2} = \frac{{\lfive}^{2}}{|{\tau_p}|^{2}} \left( {\Im}{\tau_p} dz d{\zb} \right), \qquad z \sim z + \frac{\lfive}{\tau_p} \left( n_{1} + n_{2} {\tau_p} \right)
 \end{equation}
and the nome $\pp = e^{2 \pi {\ii} \tau_p}$, the elliptic analogue
of the index functor \eqref{eq:indfun} is 
\begin{equation}
    \indfun_\pp \left[\sum_{I} e^{x_I}\right]  = \prod_{I} \theta_1(e^{x_I}; \pp)
\end{equation}
where 
\begin{equation}
\label{eq:theta}
  \begin{aligned}
&    \theta_1(\xi; \pp) =  -{\ii} \sum_{r \in \BZ + \frac 12} 
 (-1)^{r-\frac 12} \xi^{r} {\pp}^{\frac{r^2}{2}} = {\ii} p^{\frac 1 8} \xi^{-\frac 1 2} \theta(\xi; p) \\
&  \theta(\xi;\pp) = 
 \sum_{n \in \BZ} (-1)^n \xi^n {\pp}^{ \frac 12 n(n-1)} 
 =(\xi,p)_{\infty} (p,p)_{\infty} (p \xi^{-1}, p)_\infty
  \end{aligned}
\end{equation}
In terms of the additive arguments 
\begin{equation}
    \xi = e^{\betir x}, \qquad p = e^{ 2 \pi \ii \tau_p}
\end{equation}
\begin{equation}
\theta_1(\xi; \pp) = {\vartheta}_{1}(x; {\tau_p})
\end{equation}
Under the shifts
\begin{equation}
  x \mapsto  {\tilde x} = x + \frac{2 \pi}{\lfive}( n_1 + n_2 \tau_p), \qquad n_1, n_2 \in \BZ
\end{equation}
the function $\vartheta_1(x;\tau_p)$ transforms as
\begin{equation}
  \vartheta_1(x; \tau_p) \mapsto  e^{- (\pi \ii \tau_p n_{2}^2 + \pi \ii (n_{1}+n_{2}) +\ii \lfive
    n_{2} x)} \vartheta_1(x; \tau_p) 
\end{equation}
For future use, let us rewrite the multiplier for the ${\vartheta}_{1}$ transformation 

{}under the shifts by $(n_{1}, n_{2}) = (n,0)$ in the form:
\begin{equation}
\label{eq:thetri}
T_{1}^n: e^{ -\ii {\pi}n}  = e^{ - \frac{{\ii}{\ell}}{2} ({\tilde x} - x )} \end{equation}
{}under the shifts by $(n_{1}, n_{2}) = (0,n)$ in the form:
\begin{equation}
\label{eq:thetrii}
T_{2}^n: e^{- (\pi \ii \tau_p n^2 +\ii \lfive
    n x+ \ii {\pi}n)}  = e^{- \frac{{\ii}{\ell}^{2}}{4\pi\tau_p} \left({\tilde x}^2 - x^2\right) - \frac{{\ii}{\ell}}{2\tau_{p}} ({\tilde x} - x )} \end{equation}
Under the modular transformations of the parameter ${\tau}_{p}$:
\begin{equation}
\label{eq:modular}
  \vartheta_1 \left( \frac{x}{\tau_p}; -\frac{1}{\tau_p} \right) = e^{-\frac{3}{4}  \pi \ii } \tau_p^{\frac 1 2} e^{\frac   {  \ii \lfive^2 x^2}{4 \pi \tau_p} } \vartheta_1(x; \tau_p)\qquad  \vartheta_1(x, \tau_p + 1)= e^{\frac 1 4 \pi \ii} \vartheta_1(x, \tau_p)
\end{equation}
\subsubsection{The gauge and modular anomalies}
Define 
\[ \left[ \sum_{I}  e^{ \betir w_I} \right]_{d} := \sum_{I} w_I^d\]
 The transformation of $\indfun_p[\chi]$ under the large gauge transformations, which act on the Coulomb moduli ${\ac}_{{\iv}, {\ba}}$ by the shifts:
\begin{equation}
{\ac}_{{\iv}, {\ba}} \mapsto {\tilde\ac}_{{\iv}, {\ba}} = {\ac}_{{\iv}, {\ba}} + \frac{2 \pi}{\lfive}( n_{{\iv}, {\ba}} + m_{{\iv}, {\ba}} \tau_p), \qquad n_{{\iv}, {\ba}}, m_{{\iv}, {\ba}} \in {\BZ}
\label{eq:lgt}\end{equation}
is given by $\indfun_p[\chi] \mapsto \indfun_p[\tilde\chi]$, cf. Eqs. \eqref{eq:thetri}, \eqref{eq:thetrii}
\begin{equation}
\begin{aligned}
& T_{1}^n :  \indfun_p[\tilde\chi] - \indfun_p[\chi]  = - \frac{\ii \lfive}{2}
\left( \left[ {\tilde\chi} \right]_{1} - \left[ {\chi} \right]_{1} \right) \\
& 
T_{2}^n : \indfun_p[\tilde\chi] - \indfun_p[\chi] = - \frac{{\ii}{\lfive}^2}{4\pi \tau_p} \left( \left[ {\tilde\chi} \right]_2 - \left[ {\chi} \right]_{2} \right) - \frac{\ii \lfive}{2\tau_p}
\left( \left[ {\tilde\chi} \right]_{1} - \left[ {\chi} \right]_{1} \right) \end{aligned}
\label{eq:gaugean}
\end{equation}
\proposition{The six dimensional theory compactified on a torus is gauge invariant, if the right hand side of \eqref{eq:gaugean} vanishes for all $(n_{\iv, \ba}, m_{\iv, \ba})$, or if the right hand side of \eqref{eq:gaugean} can be compensated by some consistent transformation of the coupling constants, i.e. ${\qe}_{\iv}$'s. For that to be the case, the right hand side of \eqref{eq:gaugean} must be proportional to ${\bk}_{\iv}$ and should not depend on the individual colored partitions $\lambda_{\iv}$. }
 
The transformation of $\indfun_p[\chi]$  under $\tau_p \to -1/\tau_p$, i.e. the modular anomaly,  is proportional to
the exponent of $[\chi]_2$.

\proposition{For the theory compactified on the torus to make sense on its own, the partition function must be modular invariant, perhaps at the expense of making the gauge couplings ${\qe}_{\iv}$ transform under the modular group. The latter means that $[ {\chi} ]_{2}$ can be
linear in ${\bk}_{\iv}$, but should not depend on any moduli such as ${\ac}_{{\iv}, {\ba}}$.}

\subsubsection{Example: the $\hat A_{0}$ theory in six dimensions}
For the $\hat A_0 = ({\CalN}=2^{*})$ theory the set of vertices consists of one element, so we skip the index ${\iv}$ in what follows. We have 
\begin{equation}
  \chi^{\text{vec}} = \frac{ q_1 q_2} { (1 - q_1)(1-q_2)} \CalE \CalE^{\vee}
\end{equation}
and
\begin{equation}
  \chi^{\mathrm{adj}} = - \frac{  ( q_1
     q_2)^{\frac 12}  \mu^{-1} }{ (1 - q_1) (1-q_2)} 
 \CalE \CalE^{\vee}
 \label{eq:adj}
\end{equation}
Then
  \begin{equation}
 \chi^{\mathrm{vec, inst}} =   -W V^{\vee}  - q_1 q_2 W_{}^{\vee} V_{} +
  (1 - q_1)(1-q_2) V_{}\, V_{}^{\vee}
\end{equation}
For the $\hat A_0$ theory we find, in the $k$-instanton sector
\begin{multline}
\label{eq:quadraticA0}
 \left[\chi^{\mathrm{vec}}_{\lambda} + \chi^{\mathrm{adj}}_{\lambda} \right]_2  =  
  - \sum_{{\ba}, {\bb}, s \in
    {\lambda}_{\bb} } \left( ({\ac}_{\ba} - c_{{\bb}, s})^2 + ( 2 \ep_{+} +
  c_{{\bb},s} - \ac_{\ba})^2 \right)  + \\
  + \sum_{{\ba}, {\bb}, s \in
    {\lambda}_{\bb}} \left( ({\ac}_{\ba} - c_{{\bb},s} -{\ep}_{+} - m)^2 + ( {\ep}_{+} +
  c_{{\bb},s} - {\ac}_{\ba} - m )^2 \right) = \\
= 2 k {\bv}  ( m + {\ep}_{+})(m - {\ep}_{+})
\end{multline}
Since this is ${\ac}$-independent, there is no gauge anomaly. 
In addition, using~(\ref{eq:quadraticA0}) we see that under $\tau_p  \to
-1/\tau_p$
\begin{equation}
  Z_{\lambda}  \mapsto \exp \left( |{\lambda}| \frac{  \ii \lfive^2  {\bv} (m^2 - \ep_{+}^2)} { 2 \pi \tau_p} \right) Z_{\lambda} 
\end{equation}
The instanton partition function of the ${\hat A}_{0}$ theory in six dimensions can be rendered modular invariant by requiring the gauge coupling to transform 
\begin{equation}
  \qe \mapsto \exp\left(- \frac{  \ii \lfive^2  {\bv} (m^2 - \ep_+^2)}{ 2 \pi \tau_p} \right) {\qe}
\end{equation}
under the $\tau_p \to -1/\tau_p$ transformation.  

The factor $m^2 - \ep_+^2$ should be related to the anomaly cancellation condition 
\begin{equation}
  dH = F^2 - R^2 \propto m^2 - \ep_+^2 
\end{equation}
where $F$ and $R$ are the equivariant curvatures of the
R-symmetry connection and spin-connection, respectively, on the compactification torus
$\BT_{\lfive, -\lfive/\tau_p}$. 

\subsubsection{Example: the $A_{1}$ theory in six dimensions}

Here is an example of potentially sick theory. Again, we skip the $\iv$ index below. 
The $SU({\bv})$ theory with ${\bv}$ fundamental hypermultiplets and ${\bv}$ anti-fundamental hypermultiplets has, in the instanton sector:
\begin{multline}
\label{eq:quadraticA1}
 \left[\chi^{\mathrm{vec}}_{\lambda} + \chi^{\mathrm{f}}_{\lambda} +
   \chi^{\mathrm{af}}_{\lambda} \right]_2  =  2 {\ep}_{1}{\ep}_{2} k^{2} 
  - \sum_{\ba, {\bb}, s \in
    \lambda_{\bb}} \left( (\ac_{\ba} - c_{{\bb},s})^2 + ( 2 \ep_{+} +
  c_{{\bb},s} - \ac_{\ba})^2 \right)  + \\
  + \sum_{\bb, {\fe},  s \in
    \lambda_{\bb}} \left( (c_{{\bb},s}+{\ep}_{+} -m_{\fe})^2 + ( {\ep}_{+} + {\tilde m}_{\fe} - 
  c_{{\bb},s} )^2 \right) = \\
= 2 {\ep}_{1}{\ep}_{2} k^{2} 
+2 \left( \sum_{{\bb}, s \in {\lambda}_{\bb}} c_{{\bb}, s} \right) \left( \sum_{{\ba}=1}^{\bv} \left( 2 {\ac}_{\ba}   - 2{\ep}_{+} - m_{\ba} - {\tilde m}_{\ba} \right)\right) \\
+ k \sum_{{\ba}=1}^{\bv} \left( - 2 ( {\ac}_{\ba}-  {\ep}_{+})^{2} + m_{\ba}^2 + {\tilde m}_{\ba}^{2} + 2 {\ep}_{+} ( {\tilde m}_{\ba} - m_{\ba} ) \right) 
\end{multline}
To make the right hand side gauge invariant we need to insist on the following \emph{traceless}
constraint:
\begin{equation}
\sum_{{\ba}=1}^{\bv} \left( 2 {\ac}_{\ba}   - 2{\ep}_{+} - m_{\ba} - {\tilde m}_{\ba} \right) = 0
\end{equation}
which means that only the $SU({\bv})$ part of the gauge group is possibly anomaly free, hardly a surprise. In addition, we must make the gauge coupling $\qe$ transform under the gauge transformations, so that the combination
\begin{equation}
{\qe}\ {\exp}  - \frac{{\ii}{\ell}^{2}}{4\pi \tau_p} \sum_{{\ba}=1}^{\bv} \left( - 2 ( {\ac}_{\ba}-  {\ep}_{+})^{2} + m_{\ba}^2 + {\tilde m}_{\ba}^{2} + 2 {\ep}_{+} ( {\tilde m}_{\ba} - m_{\ba} ) \right) 
\end{equation}
is invariant. 
However the modular invariance cannot be fixed because of the first term $\propto k^2$. 
\section{The twisted superpotential $\CalW$ of quiver theories}

We now wish to evaluate the $\CalZ$-partition function in the limit ${\ep}_{2} \to 0$ with ${\ep}_{1} = {\ep}$ fixed. On the physical grounds (the extensitivity of the partition function of the instanton gas) we expect the following behavior:
\begin{equation}
{\CalZ} ( {\ac}, {\bm}; {\qe}, {\ep}_{1} = {\ep}, {\ep}_{2} \to 0) \sim {\exp} \, \left( \, - \frac{{\CalW} ({\ac}, {\bm}; {\qe}, {\ep})
}{{\ep}_{2}} \, \right) \label{eq:fztw}
\end{equation}
Our task is to demonstrate the validity of the asymptotics \eqref{eq:fztw} and to evaluate $\CalW$. 

\subsection{The limit shape}

We shall now investigate the limit ${\ep}_{2} \to 0$. 
Observe that each contribution to the full partition function
\eqref{eq:Zfull}, for small ${\ep}_{2}$, if expressed through the quantities $x_{i, {\ba}, {\bek}}$, contains the factor $\tfrac{1}{\ep_2}$ in
the exponential. This suggests to look for a dominant contribution to \eqref{eq:Zfull} among the configurations with finite $(x_{i, {\ba}, {\bek}})$, for all $i \in \Ver$, ${\ba}= 1, \ldots , {\bv}_{i}$, ${\bek} = 1, 2, \ldots $.   
Recall that for finite sums 
\begin{equation}
\lim_{\ep_2 \to 0 } {\ep}_{2} \, {\log} \left( \sum_{\lambda \in {\Lambda}}
  {\exp} \, {S_{\lambda}\over {\ep}_{2}} \right) = S_{\lambda_{*}} 
\label{eq:trop}
\end{equation}
where $\lambda_{*}$ is the element of the set $\Lambda$ on which 
\begin{equation}
{\rm Re}\left(\frac{ S_{\lambda_{*}}}{\ep_{2}} \right) \quad \text{is maximal}
\end{equation}
under the assumption that the phases ${\rm Im} \left(\frac{ S_{\lambda}}{\ep_{2}} \right)$ of the contributions of the configurations $\lambda$ with ${\rm Re}\left( \frac{S_{\lambda}}{{\ep}_{2}} \right)$ close the maximal one, are also close to each other. 
A typical example is the evaluation of the series
$$
\sum_{n=0}^{\infty} \frac{({\qe}/{\ep}_{2})^{n}}{n!}
$$
for ${\rm Re}({\qe}/{\ep}_{2}) > 0$. The dominant contribution comes from $n \sim {\qe}/{\ep}_{2}$, i.e. from $x = {\ep}_{2}n \sim {\qe}$. To make the similar evaluation for all values of ${\qe}/{\ep}_{2}$ we use the analyticity of the series (assuming it converges, which in the present example is easy to establish), and then evaluate it with an arbitrary accuracy for ${\qe}/{\ep}_{2} \in {\BR}_{+}$.

In the case at hand we need to show that there is a domain in the space of parameters ${\ac}, {\bm}, {\ep}_{1}, {\ep}_{2}$ such that the phases of all contributions to the partition function in the vicinity of the dominant one are aligned. We shall make the argument in the four dimensional case, with ${\tilde\bw}_{\iv} = 0$, leaving the general five dimensional case to a curious student.

It is convenient, for this purpose, to choose all the parameters in the problem: ${\ac}_{{\iv}, {\ba}}$, ${\ep}_{1,2}$, ${\bm}_{\fe}$, ${\mb}_{e}$, ${\qe}_{\iv}$ etc., to be real. 
We shall also assume that ${\ep}_{1} = {\ep} \gg -{\ep}_{2} > 0$, and that ${\ac}_{{\iv}, {\ba}} - {\ac}_{{\iv}, {\bb}} \gg {\ep}$ for ${\ba} \neq {\bb}$ for all $\iv$. We shall also assume that all masses are real, and much larger then $|{\ac}_{{\iv}, {\ba}}|$ for all $({\iv}, {\ba})$.

The contribution \eqref{eq:Zfull} $Z_{\lambda}$ of a set ${\lambda} = ({\lambda}_{{\iv}, {\ba}})$ of quiver colored partitions to the partition function $\CalZ$ is then also real. We only need to make sure it is positive for all $\lambda$'s close to the dominant one.

Let us now compare the contributions of two close configurations, namely $\lambda$ and $\lambda'$ obtained from $\lambda$ by adding one square to the ${\lambda}_{{\iv}, {\ba}}$ partition. The ratio can be easily evaluated to be  (cf. the Eq. \eqref{eq:dlogz} below) 
\begin{multline}
Z_{\lambda'}/Z_{\lambda} = (-1)^{{\bv}_{\iv} - 1  + \sum_{e \in s^{-1}({\iv})} {\bv}_{t(e)}} \times \\
{\qe}_{\iv} P_{\iv}(x) \frac{({\ep}_{1}+{\ep}_{2})}{{\ep}_{1}{\ep}_{2}} \times \\ \frac{\prod_{e \in t^{-1}({\iv})} {\y}_{s(e)} ( x + m_{e} + {\ep}_{1} + {\ep}_{2} ) \prod_{e \in s^{-1}(i)} {\y}_{t(e)}(x - m_{e})}{{\y}_{\iv}^{\prime}(x) {\y}_{\iv} (x + {\ep}_{1} + {\ep}_{2})} \\
\times \prod_{e \in s^{-1}({\iv}) \cap t^{-1}({\iv})} \frac{(m_{e} + {\ep}_{1})(m_{e} + {\ep}_{2})}{m_{e}(m_{e} + {\ep}_{1} + {\ep}_{2})} 
\label{eq:zzpr}
\end{multline}  
where $x$ is one of the zeroes of ${\y}_{\iv}$. It determines the position of the box one can add to ${\lambda}_{{\iv}, {\ba}}$, in fact, $x$ is equal to the ${\ac}_{{\iv}, {\ba}}$-shifted content of that box. 
By taking $|m_{e}| \gg {\ep}$ we can neglect the last line in \eqref{eq:zzpr}. Also, by taking the masses of the fundamentals to be much larger then the Coulomb moduli, we can drop the fundamental polynomial $P_{\iv} (x)$, replacing it by $(-1)^{{\bw}_{\iv}}$. 

Using the fact that the zeroes and the poles of ${\y}_{\iv}$ interlace for ${\ep}_{1}/{\ep}_{2} < 0$, 
and assuming $|{\ac}_{{\iv},{\ba}} - {\ac}_{{\iv}, {\bb}} | \gg {\ep}$, the sign of the denominator of \eqref{eq:zzpr} is the same as the sign of 
${\ep}_{1} + {\ep}_{2}$. Now, the final adjustment is the choice of the chamber where for all $e \in \Edg$, ${\ba} = 1, \ldots , {\bv}_{s(e)}$, ${\bb} = 1, \ldots , {\bv}_{t(e)}$: 
\begin{equation}
a_{t(e), {\bb}} - a_{s(e), {\ba}} + m_{e} \gg 0
\end{equation}
It implies that the period of the $1$-cochain  $m_{e}$ on any closed $1$-cycle in the quiver is non-trivial. The sign of the numerator in the third line of \eqref{eq:zzpr} is then equal to
$$
\prod_{e \in t^{-1}({\iv})} (-1)^{{\bv}_{s(e)}}
$$
Collecting all the signs and using the equality in \eqref{eq:conformal-class}, we obtain:
\begin{equation}
{\rm sign} \frac{Z_{\lambda'}}{Z_{\lambda}} = {\rm sign} ({\qe}_{\iv}) (-1)^{{\bv}_{\iv} - 1}
\label{eq:qisi}
\end{equation}
Thus, we shall assume ${\qe}_{\iv} \in (-1)^{{\bv}_{\iv} - 1} \cdot {\BR}_{+}$. 

Therefore, in the limit $\ep_2 \to 0$ the sum over the
set of all fixed points can be computed semi-classically by finding the
dominant configuration, as was originally done in
\cite{Nekrasov:2003rj}. The important difference is that now, for finite $\ep_1$,
 the profile of the dominant partition  cannot be assumed to be a smooth
 function. Instead, the profile of partition $\lambda_{{\iv}, \ba}$ shall
be described by an infinite series of continuous variables $(\xi_{{\iv},{\ba},k})$. This approach was suggested in \cite{Poghossian:2010pn,Dorey:2011pa,Chen:2011sj}. An alternative approach is based on the Mayer expansion  \cite{Losev:2003py} and the ${\ep}_{2} \to 0$ analysis of the asymptotics of the gauge partition function expressed via the contour integral form of the ADHM integrals 
\cite{Moore:1997dj}, leading to the TBA-like integral equations
\cite{Nekrasov:2009rc,Kozlowski:2010tv}. It will be discussed elsewhere.

\subsection{Entropy estimates}

Let us now explain why finding the dominant configuration suffices for the evaluation of ${\CalW}({\ba}, {\bm}; {\qe}, {\ep})$.

Indeed, the argument similar to \eqref{eq:trop} may fail if the number ${\CalN}({\ep}_{2})$ of configurations $\lambda$ whose actions $S_{\lambda}$ fall within an order ${\ep}_2$ error in the vicinity of $S_{\lambda_{*}}$, grows as $e^{C/{\ep}_{2}}$ for some constant $C$, for ${\ep}_{2} \to 0$. This is how, for example, one may understand the phase transition in the Ising model, starting with the high temperature expansion, leading to the counting of random walks on the lattice \cite{Polyakov:1987ez}.

Let us now explain why in our problem the entropy factor ${\CalN}({\ep}_{2})$ grows at most in a power-like fashion with $1/{\ep}_{2}$. 

Recall that the configurations $\lambda$ in our problem are the multi-sets
${\lambda}_{i, {\ba}}$ of partitions of integers, indexed by the vertices $i \in \Ver$ of the quiver, and the colors ${\ba} = 1 , \ldots , {\bv}_{i}$ of the corresponding gauge group factor.

Let us estimate the number of partitions, close to the given set $\lambda = ({\lambda}_{{\iv}, {\ba}})$ of partitions ${\lambda}_{{\iv}, {\ba}}$, such that the action $S_{\lambda}$ is close to $S_{\lambda_{*}}$. This number is equal to the number of boxes one can add, or remove from the Young diagrams of the all these partitions ${\lambda}_{{\iv}, {\ba}}$. 
For each row $j = 1, \ldots , {\lambda}^{t}_{1}$ we
can add one box, ${\lambda}_{j} \mapsto {\lambda}_{j} +1$ if ${\lambda}_{j-1} > {\lambda}_{j}$. Analogously we can remove one box ${\lambda}_{j} \mapsto {\lambda}_{j} -1$ if ${\lambda}_{j} > {\lambda}_{j+1}$. At any rate, the number of such modifications of a single partition is bounded above by $2^{{\lambda}_{1}^{t}}$, and for the whole configuration $\lambda$ it is bounded above by
\begin{equation}
e^{S_{\lambda}^{\rm entropy}} \leq  {\exp} \, K'\, \sum_{{\iv}, {\ba}} ({\lambda}_{{\iv}, {\ba}})^{t}_{1} 
\label{eq:entro}
\end{equation}
for some constant $K'$. 
Now let us estimate $({\lambda}_{{\iv}, {\ba}})^{t}_{1}$. We shall show below, using the analysis of the limit shape equations in the form of the Eq. \eqref{eq:SWq}, that 
\begin{equation}
{\ep}_{2} {\lambda}_{{\iv}, {\ba}, k} \sim \mathscr{C}_{{\iv}, {\ba}, k} {\qe}_{\iv}^{k}, \qquad {\rm for} \qquad
k \longrightarrow \infty
\label{eq:asxi}
\end{equation}
where
\begin{multline}
\label{eq:civbak}
\mathscr{C}_{{\iv}, {\ba}, k} = \frac{{\mathscr A}({\ac}_{{\iv}, {\ba}} + {\ep}k)}{{\mathscr A}^{\prime}({\ac}_{{\iv}, {\ba}})} \times \\
\prod_{j=1}^{k} \left( \frac{ P_{\iv} ( {\ac}_{{\iv}, {\ba}} + {\ep}(j-1) )}{{\mathscr A}_{\iv}( {\ac}_{{\iv}, {\ba}} + {\ep}j )^2} \prod_{e \in t^{-1}({\iv})} {\mathscr A}_{s(e)}( {\ac}_{{\iv}, {\ba}} + {\ep}(j-1) - m_{e}) \prod_{e \in s^{-1}({\iv})} {\mathscr A}_{t(e)} ( {\ac}_{{\iv}, {\ba}} + {\ep} j + m_{e} )  \right) 
\end{multline}
where
\[ {\mathscr A}_{\iv} (x) = \prod_{{\ba}=1}^{{\bv}_{\iv}} ( x - {\ac}_{{\iv}, {\ba}} )
\]
Since for large $k$,  $\mathscr{C}_{{\iv}, {\ba}, k} \propto k^{K''}$ for some constant $K''$ which depends only on $(\ac, \bm, \ep)$, we can conclude that
 $({\lambda}_{{\iv}, {\ba}})^{t}_{1} \sim \frac{{\rm log} {\ep}_{2}}{{\rm log}{\qe}_{\iv}}$
so that 
\begin{equation}
e^{S_{\lambda}^{\rm entropy}} \leq {\ep}_{2}^{-K}
\label{eq:epk}
\end{equation}
for some uniformly bounded constant $K$. Thus, the entropy factor is subleading compared to the exponential of the critical action ${\exp} ( S_{\lambda_{*}}/{\ep}_{2} )$. 

\remark{Note that the generating function of the leading asymptotics \eqref{eq:civbak}
\begin{equation}
J_{{\iv}, {\ba}} ({\qe}) = \sum_{k=1}^{\infty} {\mathscr{C}}_{{\iv}, {\ba}, k} {\qe}^{k}
\end{equation}
is a generalized hypergeometric function, and can be identified with a vortex partition function \cite{Shadchin:2005hp} of some gauged linear sigma model (and the $J$-function of \cite{Givental:1997}). It would be interesting to understand the physics of this coincidence.}

\subsection{Analysis of the limit shape equations}
In the limit $q_2 = 1$ the  functions $\y_{\iv}(\xi)$ (\ref{eq:ygeom5}) can be
represented as the infinite product over the set $\Xi_{\iv}$ (\ref{eq:setXi}) as
\begin{equation}
\label{eq:Yi}
\y_{\iv}(\xi)=  \y_{\iv}^{+}(\xi) =    \frac{ Q_{\iv}^{+}(q^{\frac 12} \xi)} {Q_{{\iv}}^{+}(q^{-\frac 12} \xi)} 
\end{equation}
where the \emph{Baxter} function
\begin{equation}
\label{eq:Baxter}
 Q_{{\iv}}^+(\xi) = \prod_{\xi' \in \Xi_{\iv}} (1 - \xi'/\xi).
\end{equation}
The product \eqref{eq:Baxter} converges because of the set $\Xi_{\iv}$ (\ref{eq:setXi}) is asymptotic to ${\Xi}_{\iv}^{\circ}$:
\begin{equation}
\label{eq:Bethe-as}
  \xi_{{\iv}, \ba, k}  \to {\xi}_{{\iv}, {\ba}, k}^{\circ},  \quad k \to \infty
\end{equation}
where $\mathring{\xi}_{{\iv}, \ba, k} =  w_{{\iv}, \ba} q^{k-1}$ by
\eqref{eq:Xi0}. 

Explicitly
\begin{equation}
\label{eq:Yivee}
  \y_{\iv}^{+}(\xi) :=\prod_{\ba=1}^{\bv_{\iv}} \left(( 1- q^{-\frac 1 2} \xi_{{\iv},\ba, 1}/\xi)
\prod_{k=1}^{\infty}  \frac {  1 - q^{-\frac 1 2} \xi_{{\iv}, \ba, k+1}/\xi }
{ 1 - q^{\frac 12} \xi_{{\iv}, \ba, k}/\xi}\right)
\end{equation}
It is convenient to introduce the notation $\y_{\iv}^{-}(\xi)$ for 
\begin{equation}
\label{eq:Yiminus}
  \y_{\iv}^{-}(\xi) :=  \y_{\iv}^{+}(\xi)  \prod_{\ba=1}^{\bv_{\iv}}\left(  -\frac{ q^{\frac 12}
      \xi }{  w_{{\iv},\ba}} \right)
\end{equation}
and from the asymptotics ~(\ref{eq:Bethe-as}),(\ref{eq:Xi0}) we
see\footnote{This product formula follows from 
\begin{equation}
     \y_{\iv}^{+}(\xi) = \y_{\iv}^{-}(\xi) \left( \prod_{\ba=1}^{\bv_{\iv}} \frac { 1 - q^{-\frac 12} \xi_{{\iv},\ba, 1}/\xi} 
{ 1 - q^{\frac 1 2} \xi/\xi_{{\iv},\ba, 1}} \prod_{k=1}^{\infty} 
\frac{\xi_{{\iv},\ba,  k+1}}
{ q \xi_{{\iv},\ba, k}} \right)
\end{equation}
}
\begin{equation}
  \y_{\iv}^{-}(\xi) = \prod_{\ba =1}^{\bv_{\iv}} \left( ( 1- q^{\frac 1 2 } \xi /\xi_{i, \ba, 1})
\prod_{\bek=1}^{\infty}   \frac { 1 - q^{\frac 1 2} \xi/ \xi_{i, \ba, \bek+1}}
{ 1 - q^{- \frac 1 2} \xi/ \xi_{i,\ba, \bek}} \right)
\end{equation}
The asymptotics of $\y_{\iv}^{\pm}(\xi)$ functions are (c.f. (\ref{eq:yfun1})
\begin{equation}
\label{eq:yasympt}
  \begin{aligned}
 &   \xi \to 0: \quad &  & \y_{\iv}^{+}(\xi) \to  \xi^{-\bv_{\iv}} \prod_{\ba=1}^{\bv_{\iv}}
 \left( - w_{i, \ba} q^{-\frac
     1 2} \right), \quad & & \y_{\iv}^{-}(\xi) \to 1 \\
 &   \xi \to \infty: \quad &  & \y_{\iv}^{+}(\xi) \to 1,   \quad & & 
\y_{\iv}^{-}(\xi) \to \xi^{\bv_{\iv}} \prod_{\ba=1}^{\bv_{\iv}}  \left( - w_{i, \ba}^{-1} q^{\frac
     1 2} \right)
  \end{aligned}
\end{equation}
Let 
\begin{equation}
\label{eq:fund}
  P_i^{-}({\xi}) = \prod_{{\fe}=1}^{w_i} (1 - \xi / \muf_{i,\fe}),
\quad
 {\tilde P}_i^{+}({\xi}) = \prod_{{\fe}=1}^{\tilde w_i} (1 - \muft_{i,\fe}/\xi)
\end{equation}
be the fundamental (anti-fundamental) matter polynomials.

The critical point equations are obtained by a simple variation of the
exponents in the constituent factors of (\ref{eq:Zfull}). Using
\begin{equation}
  \partial_{ \log \xi} \Li_2 (\xi) = \Li_1(\xi) = -\log(1 - \xi)
\end{equation}
we find the critical point equations on the dominant fixed point
$\lambda_{*}$ encoded by  the sets ${\Xi}_{\iv}$ to be 
\begin{equation}
 \exp ( \betir \ep_2 \partial_{\log \xi_{{\iv}, \ba, k}} \log Z_{\lambda_{*}}) = 1
\end{equation}
for each $\xi_{{\iv}, \ba, k}$ denoted below as $\xi$. Explicitly, we find
\begin{equation}
\label{eq:dlogz}
\begin{aligned}
 &   \exp (  \betir  \ep_2 \partial_{\log \xi} \log Z_{\lambda}) = \\
& \qquad\qquad - \qe_{\iv} ( q^{\frac 1 2} \xi)^{\kCS_{\iv}} 
\frac{P_{\iv}^{-}(\xi) \tilde P_{\iv}^{+}(\xi)}{\y_{\iv}^{-}(q^{\frac 1 2} \xi) \y_{\iv}^{+} (q^{-\frac 1 2} \xi)}
\\
&  \qquad\qquad\qquad\qquad \times
  \prod_{e \in t^{-1}({\iv})} \y_{s(e)}^{-}(\mu_e^{-1} \xi) \prod_{e \in s^{-1}({\iv})}
  \y^{+}_{t(e)}(\mu_e \xi), \quad \xi \in \Xi_{\iv}
\end{aligned}
\end{equation}
Therefore, the critical profiles
(\ref{eq:Z-sum}) are the solutions to the following infinite
system of difference equations in the variables $\xi_{{\iv},\ba,k}$:
For each ${\iv} \in \Ver$ and for each $\xi \in \Xi_{\iv}$, 
\begin{equation}
\label{eq:Bethe}
\begin{aligned}
&    \y_{\iv}^{-}(q^{\frac 1 2} \xi) \y_{\iv}^{+} (q^{-\frac 1 2} \xi)  = \\
& \qquad\qquad
- \qe_{\iv} ( q^{\frac 1 2} \xi)^{\kCS_{\iv}} P_{\iv}^{-}(\xi) \tilde P_{\iv}^{+}(\xi)  \prod_{e \in t^{-1}({\iv})}
\y_{s(e)}^{-}(\mu_{e}^{-1} \xi) \prod_{e \in s^{-1}({\iv})} \y^{+}_{t(e)}(\mu_{e}
\xi),\qquad \xi \in \Xi_{\iv}
\end{aligned}
\end{equation} 
It would appear that the left hand side of the equations \eqref{eq:Bethe} cannot be evaluated
 at $\xi = \xi_{{\iv},\ba, k}$  because of the first order pole in $\y^{-}_{\iv} (q^{1/2} \xi)$ at $ \xi \to
\xi_{{\iv},\ba, k}$. However, $\y^{+}_{\iv} ( q^{-\frac 12}\xi)$ has a first order zero at
$\xi \to \xi_{{\iv}, \ba, k}$. Therefore the product $ \y_{\iv}^{-}(q^{\frac 1 2} \xi) \y_{\iv}^{+} (q^{-\frac 1 2} \xi)$ is non-singular in the limit $\xi \to \xi_{{\iv},{\ba}, k}$.
The equations (\ref{eq:Bethe}) are formally equivalent to the 
algebraic trigonometric Bethe ansatzn equations for the $\gq$-spin chain,
with an infinite number of Bethe roots $\xi_{{\iv}, {\ba}, k}$, see
e.g. \cite{Bazhanov:1989yk,Kirillov:1987zz,Kuniba:1994na,Kuniba:2010ir}.
These Bethe roots
come in the Regge-like trajectories, or Bethe strings. The strings are labeled
by the quiver nodes $\iv$ and the individual colors $\ba = 1,\ldots, \bv_{\iv}$.
Each string contains an
infinite number of Bethe roots $\xi_{{\iv},\ba, k}$ labeled by $k =
1,\ldots, \infty$. Recall, that  the sequences $(\xi_{{\iv},\ba, k})$
parametrize the asymptotic profiles of the colored partitions $(\lambda_{{\iv}, \ba, k})$.

We summarize: \emph{the critical profile equations of
  \cite{Nekrasov:2003rj,NP2012a} for the five dimensional $\gq$-quiver theory  with 
$\ep_1 = \ep$ finite and $\ep_2 = 0$ are the Bethe equations
for a formal $\gq$ XXZ spin chain.  The  number of Bethe roots is infinite with
prescribed regular asymptotics at infinity: they have the slope given by $\ep$ and the intercepts equal to the Coulomb moduli $\ac$ of the theory.}

\subsubsection{On-shell versus off-shell}

Note that these Bethe equations should not be confused with those in \eqref{eq:bethe} where one extremizes with respect to the Coulomb moduli $\ac$.  Here we do no such minimization, so we are \emph{off-shell} in that sense. In other words, in order to construct the \emph{off-shell} superpotential, more precisely the universal twisted superpotential ${\CalW}({\ac}, {\bm}, {\qe}; {\ep})$ we solve the Bethe equation for the $U_{q}({\hat\gq})$ XXZ spin chain; in order to find the states in a \emph{proper} Hilbert space, i.e. to go \emph{on-shell} we need to solve another Bethe equation which follows from the extremization of $\CalW^{\text{eff}}$ with expect to the Coulomb moduli $\ac$.

For $A_{r}$ theories the proposition was formulated with some restrictions 
in \cite{Poghossian:2010pn,Dorey:2011pa,Chen:2011sj,Fucito:2011pn}. The general ADE case was presented in  \cite{Pestun:berlin, Pestun:SCGP}, the formal construction in the four dimensional case can also be found  in \cite{Fucito:2012xc}.

\begin{remark}
Since $\y_{\iv}^{+}(\xi)$ and $\y_{\iv}^{-}(\xi)$ as well as $P_i^{-}(\xi)$
and $P_i^{+}(\xi)$ differ only by a power of $\xi$ and a constant
factor, one can change an
orientation of an edge in the quiver, or replace the fundamental by anti-fundamental
matter and adjust the 
Chern-Simons level $\kCS_i$ so that it compensates the change of the
orientation in the equations
\eqref{eq:Bethe}; in this way such quiver theories are isomorphic. 
\end{remark}

Using \eqref{eq:Yiminus} we can write the equations \eqref{eq:Bethe}
only in terms of $\y_{\iv}^{-}(\xi)$ as
\begin{equation}
\label{eq:Betheminus}
  \y_{\iv}^{-}(q^{\frac 1 2} \xi) \y_{\iv}^{-}(q^{-\frac 1 2}\xi)  = 
-\sP_i^{-}(\xi) \prod_{e \in t^{-1}(i)} \y_{s(e)}^{-}(\mu_e^{-1} \xi) \prod_{e \in s^{-1}(i)}
\y_{t(e)}^{-}(\mu_e\xi), \quad \xi \in \Xi_{i} 
\end{equation}
where 
\begin{equation}
\label{eq:P-source}
  \sP_{i}^{-}(\xi) = -\qe_{\iv} (q^{\frac 1 2} \xi)^{\kCS_i} 
 \prod_{\ba=1}^{\bv_{\iv}} \left ( - \frac \xi {w_{i,\ba}} \right) 
P_i^{-}(\xi)  \tilde
  P_i^{+}(\xi)  \left(
 \prod_{e \in s^{-1}(i)} \prod_{\ba=1}^{\bv_{t(e)}} \left ( - \frac { w_{t(e),
        \ba}} {q^{\frac 12} \mu_e \xi } \right)  \right)
\end{equation}
or, equivalently, in term of $\y_{\iv}^{+}(\xi)$ as
\begin{equation}
\label{eq:Betheplus}
  \y_{\iv}^{+}(q^{\frac 1 2} \xi) \y_{\iv}^{+}(q^{-\frac 1 2}\xi)  = 
-\sP_i^+(\xi) \prod_{e \in t^{-1}(i)} \y_{s(e)}^{+}(\mu_e^{-1} \xi) \prod_{e \in s^{-1}(i)}
\y_{t(e)}^{+}(\mu_e\xi), \quad \xi \in \Xi_{i} 
\end{equation}
where 
\begin{equation}
\label{eq:Pplus-source}
  \sP_{i}^{+}(\xi) = -\qe_{\iv} (q^{\frac 12} \xi)^{\kCS_i} 
 \prod_{\ba=1}^{\bv_{\iv}} \left ( - \frac  {w_{i,\ba}} {q \xi} \right) 
 P_i^{-}(\xi)  \tilde
  P_i^{+}(\xi)\left(
 \prod_{e\in t^{-1}(i)} \prod_{\ba=1}^{\bv_{s(e)}} \left ( - \frac  {q^{\frac 12} \xi } {\mu_e w_{t(e),
        \ba}} \right)\right)
\end{equation}

The \emph{matter polynomials} $\sP_i^{-}(\xi) \in {\BC} [\xi,\xi^{-1}]$ are Laurent polynomials in
$\xi$, 
 with the following asymptotics at  $\xi \to 0, \infty$
\begin{equation}
\label{eq:P-asymptotics-minus}
  \begin{aligned}
    \xi \to 0: \quad \sP_i^{-}(\xi)  \sim  \xi^{ -  \bwt_{\iv} + \kCS_i + \bv_{\iv} - \sum_{j \in
        t(i)} \bv_j}  \\
    \xi \to \infty: \quad \sP_i^{-}(\xi) \sim  \xi^{  \bw_{\iv} + \kCS_i + \bv_{\iv} - \sum_{j \in
        t(i)} \bv_j} \\
  \end{aligned}
\end{equation}
Now pick the CS levels $\kCS_i$ in such a way to set 
asymptotics $\sP_i^{-}(\xi) \sim \xi^{0}$ at $\xi \to 0$
\begin{equation}
  \kCS_{i}^{-} = -\bv_{\iv} + \bwt_{\iv} +  \sum_{j \in t(i)} \bv_j 
\end{equation}
Then
\begin{equation}
  \kCS_i = \kCS_i^{-}: \qquad 
  \begin{cases}
    \xi \to 0: \quad \sP_i^{-}(\xi)  \sim \xi^{0}  \\
    \xi \to \infty: \quad \sP_i^{-}(\xi) \sim  \xi^{  \bw_{\iv} + \bwt_{\iv}} \\
  \end{cases}
\end{equation}
Similarly, the matter polynomials $\sP_i^{+}(\xi) \in {\BC}
[\xi,\xi^{-1}]$ have the following asymptotics at  $\xi \to 0, \infty$
\begin{equation}
\label{eq:P-asymptotics-plus}
  \begin{aligned}
    \xi \to 0: \quad \sP_i^{+}(\xi)  \sim  \xi^{ -  \bwt_{\iv} + \kCS_i - \bv_{\iv} + \sum_{j \in
        s(i)} \bv_j}  \\
    \xi \to \infty: \quad \sP_i^{+}(\xi) \sim  \xi^{  \bw_{\iv} + \kCS_i - \bv_{\iv} + \sum_{j \in
        s(i)} \bv_j} \\
  \end{aligned}
\end{equation}
and we can pick the CS levels $\kCS_i$ in such a way to set asymptotics
 $\sP_i^{+}(\xi) \sim \xi^{0}$ at $\xi \to \infty$
\begin{equation}
  \kCS_{i}^{+} = \bv_{\iv} - \bw_{\iv} - \sum_{j \in s(i)} \bv_j 
\end{equation}
then 
\begin{equation}
  \kCS_i = \kCS_i^{+}: \qquad 
  \begin{cases}
    \xi \to 0: \quad \sP_i^{+}(\xi)  \sim \xi^{-\bwt_{\iv} - \bw_{\iv}} \\
    \xi \to \infty: \quad \sP_i^{+}(\xi) \sim  \xi^{0} 
  \end{cases}
\end{equation}
Notice that $\kCS_{i}^{\pm}$ are precisely the boundary of the allowed
range of CS couplings (\ref{eq:ki-range}), and $\kCS_{i}^- =
\kCS_{i}^{+}$ for the quiver gauge theories in the conformal class
(\ref{eq:conformal-class}).

If the set of bifundamental masses $\mu_e$ is a trivial cocycle, i.e. if for any $1$-cycle 
$$
c = \sum_{e} c_{e} [e] \in C_{1}({\gamma}) , \qquad {\partial}c = \sum_{e} c_{e} ( [t(e)] - [s(e)] ) = 0 \in C_{0}({\gamma})  
$$ 
\begin{equation}
\label{eq:coboundary}
\prod_{e \in \text{cycle in $\gamma$}} {\mu}_{e}^{c_{e}} = 1,
\end{equation}
then we can find the set $(\tilde \mu_{i})_{i \in \Ver}$ of compensators
${\tilde \mu}_{i} \in {\BC}^{\times}$, such that
\begin{equation}
  \label{eq:coboundary2}
  \mu_e = \tilde \mu_{s(e)}/\tilde \mu_{t(e)}
\end{equation}
This is possible for all acyclic quivers, that is for all cases except
affine $A$-quiver.  Then, if we set 
\newcommand{\yt}{\tilde \y}
\newcommand{\ytpm}{\tilde \y^{\pm}}
\begin{equation}
\label{eq:mass-shift}
 \ytpm_i(\xi)  =\y^{\pm}_i(\xi/\tilde \mu_i), \quad \tilde
 \sP_i^{\pm}(\xi) = \sP_i^{\pm}(\xi/{\tilde \mu_i})
\end{equation}
the equations (\ref{eq:Betheminus})~(\ref{eq:Betheplus}) convert to 
\begin{equation}
  \label{eq:BetheMain}
\boxed{  \yt_i(q^{\frac 1 2} \xi) \yt_i(q^{-\frac 1 2}\xi)  = 
-\tilde \sP_i(\xi) \prod_{j \in s(t^{-1}(i)) \cup t(s^{-1}(i))} \yt_{j}(\xi), \quad \xi \in \Xi_i.}
\end{equation}
where $\yt_i(\xi),\tilde \sP(\xi)$ denotes $+$ or $-$ form of
$\yt_i^{\pm}(\xi), \tilde \sP_i^{\pm}(\xi)$. 

The equations~(\ref{eq:BetheMain}) are the $q$-version of the cross-cut
equations of \cite{NP2012a}.

For $\hat A_{r}$ quiver we consider its universal covering, the $A_{\infty}$ quiver, and the
infinite set of functions $Y_i(\xi)$ for $i \in \BZ$  with twisted periodicity
mod $r+1$. 

Let $\Gamma$ be $\hat A_{r}$ quiver with $r+1$ nodes on a cycle, labeled 
$0, \dots, r$, 
and $( \mu_{e})_{e \in \Edg}$ representing a non-trivial class in $H^1(\gamma,\BC)$. Let us define ${\mu} \in {\BC}^{\times}$ by
\begin{equation}
\prod_{e} {\mu}_{e} = {\mu}^{r+1}
\label{eq:mum}
\end{equation}
Then $({\mu}_{e}/{\mu})_{e}$ represents a trivial cocycle, for which one can find the compensators, and define $Y^{\prime}_{i}$ as above. Now 
the equations (\ref{eq:Betheminus})(\ref{eq:Betheplus}) convert to
\begin{equation}
\label{eq:Bethe4}
  \yt_{i}(q^{\frac 1 2} \xi) \yt_{i}(q^{-\frac 1 2} \xi) = -\tilde \sP_i \yt_{i-1}(\mu^{-1}
  \xi) \yt_{i+1}(\mu \xi),\quad i \in \BZ/(r+1)\BZ, \quad \xi \in \Xi_i'
\end{equation}
Let us now define functions $\tilde \yt_i(\xi)$ labeled by the vertices $i \in \BZ$
 on the universal cover $\tilde \Gamma$ of $\gamma$
\begin{equation}
 \tilde \yt_i(  \xi) =  \yt_{i\, \mathrm{mod}\, r+1}'(\mu^{i} \xi) \quad i \in \BZ
\end{equation}
In terms of functions $\tilde \yt_i(\xi)$ the equations (\ref{eq:Bethe4}) take
the canonical form of equations (\ref{eq:BetheMain}) for $A_{\infty}$ quiver
\begin{equation}
\label{eq:AinftyBethe}
  \tilde \yt_i(q^{\frac 1 2} \xi) \tilde \yt_i(q^{-\frac 1 2} \xi) = -
  \tilde {\tilde \sP}_i \tilde \yt_{i-1}(\xi) 
\tilde \yt_{i+1}(\xi), \quad i \in \BZ, \qquad  \xi \in \Xi_{i}
\end{equation}

In any case, the limit shape equations are 
(\ref{eq:BetheMain}) associated to the universal cover $\tilde\Gamma$ 
of the original quiver $\Gamma$, with $\tilde\Gamma = \Gamma$
if $\Gamma$ is of the finite $ADE$ or affine $\hat D\hat E$ type, 
and $\tilde\Gamma = A_\infty$ 
if $\Gamma= \hat A_{r}$. In the $\hat A_{r}$ case the functions
$\y_{\iv}(\xi)$ are lifted to the cover with the twisted
$r+1$-periodicity. In the language of quantum groups, such lifting
corresponds to the relation between  $\Qaff_{q}(^{\mu}\hat{\mathfrak{gl}}_{r+1})$ 
and  $\Qalg_q(\mathfrak{gl}_{\infty})$ see \cite{Hernandez:2011}, \cite{Frenkel:2002}. 

In the subsequent sections  we drop the symbol tilde on functions
$\yt_i(\xi)$ in the canonical form (\ref{eq:BetheMain}),  keeping in mind the change of
variables  (\ref{eq:mass-shift}).

\section{Quantum geometry \label{seq:qSW}}

The  set of algebraic equations on the set of Bethe
roots ($\xi_{i,\ba,\bek}$)  in the theory of  $\gq$  spin chains (known 
as the nested algebraic Bethe Ansatz equations in the case $\gq =\mathfrak{sl}_r$),
one equation per each Bethe root,  can be converted to a set of functional $q$-difference
 equations on functions $\y_{\iv}(\xi)$  (for
 elementary introduction to algebraic Bethe Ansatz see
 \cite{Faddeev:1996iy},  for most 
generic case see \cite{Frenkel:1998,Frenkel:2013dh}, and for solvable
lattice models of  statistical mechanics see \cite{Baxter:1985}). The zeroes and poles of the functions $\y_{\iv}(\xi)$ encode
 information about Bethe roots.
 In
 the limit $q \to 1$ the functional $q$-difference equations on $\y_{\iv}(\xi)$
reduces to the ordinary character equations determining 
the  Seiberg-Witten curve in \cite{NP2012a}. The functions $\y_{\iv}(x)$ of \cite{NP2012a}
provide the classical limit of their more complicated $q$-relatives
$\y_{\iv}(x)$ of the present paper.

For illustration we first consider the quiver $\gq = A_1= \mathfrak{sl}_2$. The Bethe
Ansatz equations are simply (we can use here either \emph{plus} or
\emph{minus} form of functions, $\y_{\iv}^{\pm}(\xi), \sP_i^{\pm}(\xi)$)
\begin{equation}
\label{eq:BetheA1}
  \y_1(q^{\frac 1 2} \xi) \y_{1} (q^{-\frac 1 2} \xi) = -\sP_{1}(\xi),
  \quad \xi \in \Xi_1
\end{equation}
with $\Xi_1$ being the set of poles of $\y_1(q^{\frac 1 2} \xi)$, and we
remind that $\xi \in \Xi_1$ should be understood in the limit sense (see
remark after (\ref{eq:Bethe})).
Now consider the functional
\begin{equation}
\label{eq:chi1A1}
  \chi_1[\y_1(\xi)] = \y_1 (\xi) + \frac { \sP_1(q^{- \frac 1 2 } \xi)}{ \y_{1}
    (q^{-1} \xi)}
\end{equation}
The function $\chi_1(\xi)$ is regular function everywhere on the
 multiplicative plane $\BC^{\times}_{\la \xi \ra}$ provided the equations
(\ref{eq:BetheA1}). Indeed, the 
poles which appear in the first term, $Y_1(\xi)$, are precisely
canceled by the second term due (\ref{eq:BetheA1}). Therefore,
$\chi_1(\xi)$ is Laurent polynomial in $\BC[\xi, \xi^{-1}]$, i.e.
a holomorphic function on  the cylinder $\BC^{\times}_{\la x \ra }$. Now consider the asymptotics of $\chi_1(\xi)$ at
$\xi \to 0$ and $\xi \to \infty$.

 We find that the degree in $\xi$
 of the first term and second term in $\chi_1(\xi)$ of (\ref{eq:chi1A1})
 is respectively (using $\y_1^{-}(\xi), \sP_1^{-}(\xi)$ form)
 \begin{equation}
   \begin{aligned}
   \xi \to 0:      &\quad [\xi^{0}],  & \quad    [\xi^{0}]\\
   \xi \to \infty: &\quad [\xi^{\bv_{\iv}}], & \quad   [\xi^{\bw_{\iv} + \bwt_{\iv}
     - \bv_{\iv} }]
   \end{aligned}
 \end{equation}
Therefore, at $\xi \to \infty$  the first term dominates the second term
by the positivity condition (\ref{eq:conf-good}). 
We conclude that
\begin{equation}
  \label{eq:SW-curve1}
  \chi_1[\y_1(\xi), \sP_1(\xi)] = T_1(\xi)
\end{equation}
where $T_1(\xi)$ is polynomial in $\xi$ of degree $\bv_1$, in which
the highest and lowest coefficients in $\xi$ are fixed from the
known asymptotics of  $\y_{\iv}(\xi)$ and $\sP_i(\xi)$, i.e. in terms of
masses and coupling constants. The functional $q$-difference equation
(\ref{eq:SW-curve1}) is quantum version of Seiberg-Witten  for the $A_1$ quiver
theory, that is $SU(\bv_1)$ gauge theory with $\bw_1$ and $\bwt_1$
fundamentals and anti-fundamental. 
 The Laurent  polynomial $\chi_1$ in $\BC[Y(\xi q^{\frac 1 2
  \BZ})^{\pm}]$ given by (\ref{eq:chi1A1}) is the
Frenkel-Reshetikhin $q$-character for $\Qaff_q(A_1)$ \cite{Frenkel:1998}.
The  $\bv_1-1$ middle term coefficients in \emph{gauge polynomial} $T_1(\xi)$ play the role of the
Coulomb moduli of the theory and are implicitly defined in terms of the 
$SU(\bv_1)$ Coulomb parameters ${\ac}_{1, \ba}$.

Note that in the four dimensional $\lfive \to 0$ limit the Eq. \eqref{eq:SW-curve1}  
formally becomes identical to the Baxter equation for the $A_1$ XXX spin chain. 
It is not yet a
vacuum equation of the four dimensional $\mathcal{N}=2$ theory in the $\Omega_{\epsilon}$-background. As we discussed in the Introduction, the
latter requires the second minimization with respect to the Coulomb
moduli $\ac$ and this corresponds to finding the \emph{proper}
solution of Baxter equation. We know the
definition of \emph{proper}  from the integrable model side only in a few cases ---
in particular in the cases considered in
\cite{Nekrasov:2009rc}. Similar interpretation holds for the five dimensional $A_1$ case
(and XXZ spin chain), as well as for the arbitrary $ADE$ or affine $\hat
A\hat D \hat E$ quivers both in five and four dimensions.

The cancellation of poles between the two terms in (\ref{eq:chi1A1}) is
the $q$-analogue of the invariance of the classical character used in
\cite{NP2012a} under crossing the cuts supporting the charge densities
used to define the classical potentials $\mathscr{Y}_i(\xi)$. In the
classical limit $q \to 1$ the strings of Bethe roots condense on a certain finite 
intervals $I_{i,\ba}$, which are interpreted as cuts of the analytic
functions $\mathscr{Y}_i(\xi)$.  The remaining Bethe roots  
form the semi-infinite strings  $(\xi_{i,\ba,\bek} | \bek > \bek^*_{i,\ba}
)$ for certain  $\bek^*_{i,\ba}$, and the classical limit their slope approaches a constant 
so that their contribution to ${\y}_{i}(\xi)$
  defined  by the product formula (\ref{eq:Yi}) cancels. 

Below we will use short-hand notations 
\begin{equation}
  \begin{aligned}
\y_{i,b} \equiv \y_{\iv}(q^{-b} \xi) \\
\sP_{i,b} \equiv \sP_{i}(q^{-b} \xi)  
  \end{aligned}
  \label{eq:nota}
\end{equation}

For any (affine) simply-laced $\gq$ one can find
 recursively the $q$-characters $\chi_{i}$
 \cite{Frenkel:1998,Frenkel:2001, Hernandez:2004} 
for $i$-th fundamental evaluation representation of $\Qaff_q(\gq)$ as follows.
Start from the highest weight
\begin{equation}
  \tilde \chi_{i} = \y_{i,0} + \dots
\end{equation}
add a term to cancel the poles in the highest weight term 
\begin{equation}
\label{eq:chi-character}
  \tilde \chi_{i} = \y_{i,0} + \frac{ \sP_{i,\frac 1 2} \prod_{j \in s(t^{-1}(i)) \cup t(s^{-1}(i))}
    \y_{j,\frac 1 2}} {\y_{i,1}} + \dots 
\end{equation}
This term introduces new poles from the functions $\y_{j}(\xi)$ 
of the nodes $j$ linked with $i$. 
 To cancel these poles, we add new monomials  $\y_{j,b}(\xi)$ to (\ref{eq:chi-character}) until
all poles are canceled. The precise algorithm to compute 
the $q$-characters (\ref{eq:chi-character}) for $\Qaff_q(\gq)$ is given
by Frenkel-Mukhin \cite{Frenkel:2001} for all finite dimensional 
 Lie algebras $\gq$. It is generalized for the general
 symmetrizable  Kac-Moody Lie algebra $\gq$ in  \cite{Hernandez:2004}.

 The symbols
 $Y_{i,b}(\xi)$ in the $q$-character are the \emph{current-weights} encoding the
 (generalized) eigenvalues of an element of the maximal commuting subalgebra of
 $\Qaff_q(\gq)$ on a (generalized) eigenspace in a representation of
 $\Qaff_q(\gq)$, see \cite{Frenkel:1998, Frenkel:2013dh} and appendix
 for details. We give explicit formula for the current $h(x) \in
 \Qaff_{q}(\gq)$ in (\ref{eq:helement}), (\ref{eq:ginh}).

\newcommand{\TT}{\mathrm{T}}
The $q$-analogue of the \emph{iWeyl group} $W_{\gq}$ of \cite{NP2012a}
is  the braid group $\Braid(W_{\gq})$, see e.g. \cite{Etingof:2000}.
We remind that as $\xi$ crosses a classical cut of $\y_{\iv}(\xi)$ the $\y_{\iv}(\xi)$ is
transformed by the reflection in $W_{\gq}$ generated by $i$-th simple
root \cite{NP2012a}. 
 The braid group $\Braid(W_{\gq})$ is freely build on
generators $\TT_{i}, i \in \Ver$ with the relations (for simply laced $\gq$)
\begin{equation}
  \begin{aligned}
&    \TT_{i} \TT_{j} = \TT_{j} \TT_{i}, \quad a_{ij} = 0    \\
&    \TT_{i} \TT_{j} \TT_{i} = \TT_{j} \TT_{i} \TT_{j}, \quad a_{ij} = -1.
  \end{aligned}
\end{equation}
The braid group $\Braid(W_{\gq})$ action on the weight space of $\Qaff_q(\gq)$ generated
by fundamental weights $Y_i$ is the natural analogue of the Weyl
reflection
\begin{equation}
\TT_i:  \y_{i,0} \mapsto \frac{ \sP_{i,\frac 1 2} \prod_{j \in s(t^{-1}(i)) \cup t(s^{-1}(i))}
    \y_{j,\frac 1 2}} {\y_{i,1}}; \quad \y_{j} \mapsto \y_{j},  \quad j \neq i.
\end{equation}

The central result of the present work is the following statement, which could
be seen as the $q$-version of the cameral Seiberg-Witten curve of \cite{NP2012a} for
$\gq$-quiver gauge theories.

\begin{proposition}\label{prop:central-result}
The $\mathcal{N}=2$ gauge theory functions $(\y_{\iv}(\xi))_{i \in \Ver}$ of $\gq$-quiver gauge
theory on $\BR^{4}_{\ep,0} \times \BS^{1}_{\lfive}$ (the $\Omega$-background of
\cite{Nekrasov:2009rc}) defined by \eqref{eq:ygeom5} satisfy a system
of functional $q=e^{\ii \lfive \ep}$ difference equations
 \begin{equation}
\label{eq:SWq}
   \{ \tilde \chi_{i}\left[ \{\y_j(q^{\BZ} \xi), \sP_j(q^{\BZ/2}\xi)\}_{i
     \in \Ver} \right] = T_i(\xi) \  | \  i \in \Ver \ \}
 \end{equation}
where $\tilde \chi_i$ is the twisted  $q$-characters for the $i$-th
fundamental module of $\Qaff_q(\gq)$ (see
\cite{Frenkel:1998,Hernandez:2004} and Appendix), and
 $T_i(\xi)$  is a polynomial of degree ${\bv}_i$ with the highest and
 lowest coefficient fixed by the masses and
coupling constants of the theory, and the middle $\bv_{\iv} - 1$
coefficients parametrizing the Coulomb branch.
\end{proposition}

\subsection{The slope estimates}

It follows from \eqref{eq:SWq} that $\xi_{{\iv},\ba,k}$ approach $\xi_{{\iv},\ba, k}^{\circ}$
as $k \to \infty$ exponentially fast (for $|{\qe}_{\iv}| < 1$):
\begin{equation}
\xi_{{\iv},{\ba},k } = \xi_{i,\ba,k}^{\circ} + {\mathscr C}_{{\iv}, \ba,k } {\qe}_{\iv}^{k}, \qquad {k} \longrightarrow \infty
\label{eq:expas}
\end{equation}
where ${\mathscr C}_{{\iv}, \ba, k}$ all have a finite limit as ${\qe} \to 0$, and behave at most power-like with $k$ for large $k$. 
The idea is to study the small $\qe_{\iv}$ limit of \eqref{eq:SWq} evaluated at ${\xi} = {\xi}^{\circ}_{{\iv}, {\ba}, k} = w_{{\iv}, {\ba}} q^{k-1}$ in five dimensions, or, at $x = {\ac}_{{\iv}, {\ba}} + {\ep}(k-1)$ in four dimensions. Let us stick with four dimensions.   It is easy to see that 
\begin{multline}
{\y}_{\iv} ( {\ac}_{{\iv}, {\ba}} + {\ep}(k-1)  ) \sim {\qe}_{\iv} {\mathscr A}_{\iv} ( {\ac}_{{\iv}, {\ba}} + {\ep}(k-1)  ) \frac{{\mathscr C}_{{\iv}, \ba, k}}{{\mathscr C}_{{\iv}, \ba, k-1}} , \qquad
 k > 1 \\
 {\y}_{\iv} ( {\ac}_{{\iv}, {\ba}}  ) \sim {\qe}_{\iv} {\mathscr A}_{\iv}^{\prime} ( {\ac}_{{\iv}, {\ba}} ) \frac{{\mathscr C}_{{\iv}, \ba, 1}} , \qquad
 k > 1 \\
{\y}_{\bf j} ( {\ac}_{{\iv}, {\ba}} + {\ep}(k-1)  ) \sim {\mathscr A}_{\bf j} ( {\ac}_{{\iv}, {\ba}} + {\ep}(k-1)  ) , \qquad {\bf j} \in \Ver, \qquad {\bf j} \neq {\iv}\\
\label{eq:yiasym}
\end{multline}
so that in the $q$-character ${\chi}_{\iv}$ it is the term proportional to ${\qe}_{\iv}$ will survive the $\qe \to 0$ limit, all other terms being suppressed either by the explicit powers of ${\qe}$'s, or by the asymptotics of ${\y}_{\iv}$ itself, cf. \eqref{eq:yiasym}. From this the Eq. \eqref{eq:civbak} follows. 

\subsection{The convergence of $q$-characters}
Because of the constraints  (\ref{eq:ki-range}) on the Chern-Simons couplings the
highest term in (\ref{eq:chi-character}) dominates the other terms
in the limit $\xi \to \infty$ (using $-$ version of $\sP_i(\xi),
\y_{\iv}(\xi)$ ) and $\xi \to 0$ (using $+$ version of $\sP_i(\xi),
\y_{\iv}(\xi)$). 
For $\gq$-quiver theories in the conformal class
(\ref{eq:conformal-class}) all terms have the same degree in
$\xi$.  For affine $\gq$ quiver, the $q$-character $\tilde \chi_i$ is
infinite convergent series in coupling constants $(\qe_j)_{j \in \Ver}$ under the assumption $(|\qe_{\iv} | < 1)_{i \in \Ver}$.

\subsection{The observables}

In this section we record the formulae for the observables of the gauge
theory which are encoded in the functions $\y_{\iv}^{-}(\xi)$ obeying
 (\ref{eq:SWq}). For the practical purposes, the
expansion of functions $\y_{\iv}^{-}(\xi)$
 in the instanton parameters $\qe_{\iv}$ can be conveniently obtained from
 (\ref{eq:SWq}) by representing the solution $\y_{\iv}^{-}(\xi)$ in the continuous fraction
 form in terms of the polynomials $T_i(\xi)$ and $\sP_i(\xi)$ at $\xi \to
 \infty$ starting from the zeroth order approximation (\ref{eq:yasympt})
 \begin{equation}
\y_{\iv}^{-}(\xi) = \xi^{\bv_{\iv}} \prod_{\ba=1}^{\bv_{\iv}}  \left( - w_{i, \ba}^{-1} q^{\frac
     1 2} \right) + \dots \quad \xi \to \infty
 \end{equation}
The Chern character of the universal bundle ${\CalE}_{i}$ over the instanton
moduli space corresponds to the observable
\begin{equation}
\label{eq:ObO}
\CalO_{i,n} =    {\tr}_{{\bv}_{i}} \, e^{  \betir n \phi_{i}} \,  
\end{equation}
in the gauge theory where $\phi_i$ is the adjoint complex scalar of
the $\mathcal{N}=2$ supersymmetric $SU(\bv_{\iv})$ gauge vector multiplet. 
The relation between (\ref{eq:yphys}) and (\ref{eq:ygeom5}) follows from
the expansion of $\y_{\iv}(q^{-\frac 1 2} \xi)$ at $\xi = \infty$ 
\begin{equation}
\label{eq:Ovalue}
   {\tr}_{{\bv}_{i}} \, e^{  \betir n \phi_{i}}    = \oint_{\xi = \infty} \xi^{n} d \log \y_{\iv}^{-} \, (q^{-\frac 1 2 }\xi) 
\end{equation}
where the contour runs around $\xi = \infty$ 
and  encloses all zeroes and poles of $Y_i$.
Therefore, the gauge theoretic definition of
the functions $\y_{\iv}^{\pm}(\xi)$ is:
\begin{multline}
  \y_{\iv}^{-}(q^{-\frac 12} \xi)  =  \xi^{\bv_{\iv}} \prod_{\ba=1}^{\bv_{\iv}}\left(-
     w^{-1}_{i,\ba} \right)  \y_{\iv}^{+}(q^{-\frac 1 2} \xi) = \\
=\xi^{\bv_{\iv}} \prod_{\ba=1}^{\bv_{\iv}}
   \left(- w^{-1}_{i,\ba} \right)  
 \exp \, \left( \, -\sum_{i=1}^{\infty} \frac{\xi^{-n}}{n} \la \, \tr_{{\bv}_{i}} e^{\betir n \phi_{i}} \, \ra  \, \right) \\
     = \prod_{\ba=1}^{\bv_{\iv}}\left(
   - w^{-1}_{i,\ba} \right)  
\exp \left( \la\ {\log} \, {\det}_{{\bv}_{i}} \left( \xi -  e^{\ii
      \lfive \phi_i} \right)  \ra\right)
    \label{eq:yifived}
\end{multline}
and
\begin{equation}
  \y_{\iv}^{+}(q^{-\frac 1 2} \xi) = \exp ( \langle {\det}_{\bv_{\iv}} (1 - e^{\ii
    \lfive \phi_i}/\xi) \rangle)
\end{equation}

\subsection{Superpotential}

Here we compute the $\qe$-derivatives of the twisted superpotential
\begin{equation}
    {\CalW}(\ac,\bm; \qe, \ep) = - \lim_{\ep_2 \to 0} \ep_2 \log
    {\CalZ} ( \ac,\bm; \qe; \ep_1=\ep, \ep_2)
\end{equation}
From (\ref{eq:fztw}),(\ref{eq:Ztop}) we find
\begin{equation}
{\CalW}_i := \partial_{\log \qe_{\iv}} {\CalW} : = - \sum_{\xi \in \Xi_i} \log \left. \frac{
    \xi}{\mathring \xi}\right|_{\text{crit}}
\end{equation}
Then, using the Baxter function $Q^{+}_i(\xi; \Xi )$ (\ref{eq:Baxter}) we find
\begin{equation}
  \CalW_{i} = -\frac{1}{ 2 \pi i} \oint_{C_{i}} \log {\xi}  \, d  \,
  {\log}\frac{Q^{+}_{i}(\xi; \Xi_i) }{ Q^{+}_{i}(\xi; \mathring{\Xi_i})}
\end{equation}
in terms of the sets of Bethe roots
$\Xi_i=\{\xi_{i,\ba,\bek}\}$ (\ref{eq:setXi}) and
$\mathring{\Xi}_i=\{\xi_{i,\ba,\bek}\}$~(\ref{eq:Xi0}), 
where the contour $C_i$ encloses all points in $\Xi_i$ and $\mathring{\Xi}_i$.
In terms of the functions $\y_{\iv}^{+}(\xi)$ we find
\begin{equation}
   \frac{ Q_{i}^{+}(\xi) }{\mathring{ Q}_{i}^{+}(\xi)}  =
   \prod_{j=0}^{\infty}   \frac{ \y_{\iv}^{+}(q^{-j} \xi)
   }{\mathring{\y}_i^{+}(q^{-j} \xi)}
 \end{equation}
For practical purposes of computation ${\CalW}_{i}$ in the $\qe$-expansion it is convenient to integrate by parts and
represent the contour $C_{i}$ as the difference of contours around $ \xi =
\infty$ and $\xi = 0$. Then we  get
\begin{equation}
    {\CalW}_{\iv} =  \frac{1}{ 2 \pi \ii} \oint_{C_{\infty} - C_{0}} \frac{d\xi}{\xi} \ {\log}
\prod_{j=0}^{\infty}   \frac{ \y_{\iv}^{+}(q^{-j} \xi) }{\mathring{ \y}^{+}_{\iv}(q^{-j}\xi)}
\end{equation}
Using the continuous fraction expression of  $\y_{\iv}^{+}(\xi)$ from the system 
of  $q$-characters the integrand can be expanded in the series in $\qe$
with coefficients in the rational functions of $\xi$ after which the contour integral is computed
by taking the difference of coefficients at $\xi^{-1}$ in the series 
expansion at $\xi=0$ and at $\xi= \infty$. 

\subsection{Dual periods}

Finally, we give the formula for the partial derivatives 
\begin{equation}
{\ac}_{{\iv},{\ba}}^{D} = \pa_{\ac_{{\iv},{\ba}}} {\CalW} 
\label{eq:wpadw}
\end{equation}
which can be obtained from the variational limit shape
problem
\begin{equation}
  {\exp}\, \left( {\ac}_{{\iv},\ba}^{D}  \right) \ =
\prod_{k=1}^{\infty} \frac{\prod_{e \in t^{-1}({\iv})} \y^{-}_{s(e)}(\mu_e^{-1} \xi_{{\iv}, \ba, k }) \prod_{e \in s^{-1}({\iv})}
\y^{+}_{t(e)}(\mu_e \xi_{{\iv}, {\ba}, k})}
{\prod_{ \bb \neq \ba}
 \y^{-}_{{\iv},\bb}(q^{\frac 1 2} \xi_{{\iv}
      ,\ba, k})
\y_{{\iv},\bb}^{+}(q^{-\frac 1 2} \xi_{{\iv}, \ba, k})}\end{equation}
where we using the \emph{colored} functions $\y_{{\iv},\ba}^{+}(\xi)$
defined as factors in the product over $\ba$ in (\ref{eq:Yivee}) and 
\begin{equation}
\label{eq:Yib}
\begin{aligned}
  \y^{-}_{{\iv},\ba}(\xi) := ( 1- q^{\frac 1 2 } \xi /\xi_{{\iv}, \ba, 1})
\prod_{k=1}^{\infty}   \frac { 1 - q^{\frac 1 2} \xi/ \xi_{{\iv}, \ba, k+1}}
{ 1 - q^{- \frac 1 2} \xi/ \xi_{{\iv},\ba, k}} \end{aligned}
\end{equation}
 so that (c.f. \ref{eq:Yi}) 
 \begin{equation}
   \y_{{\iv},\ba}^{-} ({\xi}) = (-q^{\frac 1 2} \xi/w_{{\iv},\ba}) \y_{{\iv},\ba}^{+}(\xi)
 \end{equation}
It would be interesting to compare the above computation of $\ac_{{\iv},\ba}^{D}$ with 
the approach of \cite{Smirnov:2003}, \cite{Mironov:2009dv}.

\section{Quantum groups\label{se:quantum-groups}}

\subsection{Introduction}
In  \cite{Drinfeld:1985, Jimbo:1985} Drinfeld and Jimbo introduced the
notion of Quantum Groups
generalizing the algebraic structures underlying  quantum integrable
systems known at that time after the work of Sklyanin, Takhtajan,
Faddeev, Kulish, Reshetikhin and others \cite{Sklyanin:1978pj, 
Sklyanin:1979, Takhtajan:1979iv, Sklyanin:1980ij, Sklyanin:1982tf,
Faddeev:1987ih, Kulish:1983md}. Namely, for any symmetrizable Kac-Moody Lie algebra $\g$ and a
complex parameter $q \in \BC^{\times}$  the works \cite{Drinfeld:1985,
  Jimbo:1985} associated  the Hopf algebra
$\Qalg_q(\g)$, also often called {\it the quantum group}, by $q$-deforming the
Serre relations of $\g$, this construction is often referred as
\emph{first Drinfeld realization} \cite{Drinfeld:1985, Jimbo:1985}.  

Later Drinfeld
found \cite{Drinfeld:1987} that for an affine Kac-Moody Lie algebra $\hat \g$, the algebra
 $\Qalg_q(\hat \g)$  can be obtained by a certain canonical
 \emph{quantum affinization} process applied to finite-dimensional
 simple Lie algebra $\g$, with the result being quantum affine algebra
$\Qaff_q(\g) \cong \Qalg_q(\hat \g)$ \cite{Beck:1994,Chari:1991,Chari:1994}. The
construction \cite{Drinfeld:1987} is known as \emph{second Drinfeld
  realization} of quantum affine algebras, or \emph{Drinfeld loop
  realization}, or \emph{Drinfeld currents realization}.

In fact, \emph{Drinfeld current realization} of $\Qaff_{q}(\gq)$ is
defined for any symmetrizable  Kac-Moody algebra $\gq$
\cite{Jing:1998,Nakajima:2001q}. If $\gq = \hat \g$ is an affine
Kac-Moody Lie algebra, then the $\gq$-\emph{quantum affine algebra}
$\Qaff_{q}(\gq) =\Qaff_{q}(\hat \g) = \Qalg_q(\hat{\hat{\g}})$  is often
called the $\g$-\emph{quantum toroidal algebra} \cite{Ginzburg:1995b,Hernandez:2008},
because of the presence of two loops inside. 
The $\gq$-quantum affine algebra $\Qaff_{q}(\gq)$ prominently appears in
Nakajima's work \cite{Nakajima:2000,Nakajima:2001q,Nakajima:2002b} on
equivariant $K$-theory of quiver varieties associated to $\Gamma$.

\subsubsection{Five dimensional version}
In our work, $\gq$-quantum affine algebra $\Qaff_q(\gq)$, where $\gq$
is  ADE or affine ADE,  appears in the study of the 5d $\gq$-quiver gauge
theory on $\BS^1_{\lfive} \times \BR^{4}_{\ep_1, \ep_2}$  in
the limit $\ep_1 = \ep, \ep_2 = 0$ \cite{Nekrasov:2009rc} with
quantization parameter
\begin{equation}
  q = e^{\ii \lfive \ep}.
\end{equation}

\subsubsection{Four dimensional version} In the \emph{four-dimensional limit} $\gq$-quantum affine algebra
$\Qaff_q(\gq)$ contracts to $\gq$-Yangian  $\Yang_{\ep}(\gq)$
\cite{Drinfeld:1986,Drinfeld:1987,Gautam:2013,Guay:2012}. 
The formulae in the present note are presented in terms of the multiplicative spectral parameter 
\begin{equation}
  \xi = e^{ \betir  x}
\end{equation}
on $\Cx \cong \BC^{\times}$ for the 5d case of $\Qaff_q(\gq)$  but are
easily adapted to the 4d case of $\Yang_{\ep}(\gq)$. 
It is sufficient to take the limit $\lfive \to 0$, with 
the multiplicative variables such as $\xi=e^{ i \lfive x}$   and
$q=e^{i \lfive \ep}$ sent to $1$, but keeping additive variables  $x$ and
$\ep$ finite. At the level of the quantum integrable systems, 4d limit
of the 5d theory is the isotropic limit in which the trigonometric $\gq$
XXZ spin chain associated to the 5d theory becomes the
rational $\gq$ XXX spin chain associated to the 4d theory; for classical Seiberg-Witten for $\gq = \mathfrak{sl}_r$ see \cite{Gorsky:1997jq}.

\subsubsection{Six dimensional version} 
The lift of  the gauge theory to \emph{six-dimensional space-time} $\BR^{4}_{\ep,0}\times \BT^{2}_{\lfive,
  -\lfive/\tau_p}$, where $(\lfive, -\lfive/\tau_p)$ denote the periods of
the compactification torus $\BT^2_{\lfive, -\lfive/\tau_p}$,  promotes $\Qaff_q(\gq)$ to the $\gq$-\emph{quantum elliptic
algebra} $\Qell_{q,p}(\gq)$ \cite{Felder:1996,Felder:1995,Jimbo:1999}
with $p$ denoting the multiplicative modulus of the compactification
torus $\BT^2_{\lfive, -\lfive/\tau_p}$ 
\begin{equation}
  p = e^{2 \pi i \tau_p}
\end{equation}
This is done perturbatively by summing over Kaluza-Klein modes \cite{Nekrasov:1996cz,Marshakov:1997cj} and non-perturbatively by the study of elliptic genera of instanton moduli spaces \cite{Nekrasov:thesis,Losev:1997tp,Baulieu:1997nj, Hollowood:2003cv} replacing the quantum affine group
$\Qaff_q( \gq)$ by the
$\gq$-quantum elliptic algebra $\Qell_{q,p}(\gq)$ \cite{Jimbo:1999,Konno:2009}. 
The spectral additive parameter $x$ lives on the elliptic curve
\begin{equation}
  \Cx = \BC/(\BZ \oplus \tau_p \BZ) \cong \BC^{\times}/\pp^{\BZ} = \ec_p
\end{equation}
dual to the torus 
$\BT^2_{\lfive, -\lfive/\tau_p}$ and hence  gets two periods 
\begin{equation}
  x \to x  + \frac{2\pi}{\lfive}, \qquad x \to x + \frac{2\pi}{\lfive} \tau_p
\end{equation}
 see \ref{se:6dversion} for gauge theory discussion. The quantum affine algebra $\Qaff_q(\gq)$ is
promoted to quantum elliptic algebra $\Qell_{q,p}(\gq)$ 
\cite{Jimbo:1999,Konno:2009}. (See \cite{Felder:1995,Felder:1996}
on earlier definitions, \cite{Felder:1995iv,Felder:1997} for elliptic KZ
equation and \cite{Felder:1999} on the defining elliptic gamma function, 
\cite{Feigin:1996} for
connection to elliptic $\mathcal{W}_{q,p}(\gq)$ algebra.) 
For generalized Kac-Moody algebra $\gq$ with symmetric Cartan matrix the elliptic algebra $\Qell_{q,p}(\gq)$ is
defined using an \emph{elliptic version of  Drinfeld currents
  realization} in \ref{se:Drinfeld-currents}. 
 For $\gq$ of finite-dimensional ADE type see definition in
 \cite{Jimbo:1999}, appendix A; keeping in mind that our quantization
 parameter $q$ corresponds to $q^2$ in \cite{Jimbo:1999}. 
Elliptic version, as usual, is achieved by promoting factors 
\begin{equation}
   \xi^{\frac 1 2} - \xi^{-\frac 1 2}  \to \theta_1(\xi;p) 
\end{equation}
where $\theta_1(\xi;p)$ is (\ref{eq:theta}). 

In the world of quantum
integrable systems the lift to the six dimensional theory corresponds to the deformation
of the $\gq$ XXZ spin chain to the anisotropic $\gq$ XYZ spin chain with elliptic
$R$-matrix.  The classical limit $\ep=0$ of the associated integrable
system \cite{NP2012a} for $\gq$ of finite
type is the $\gq$-monopoles on $\BS^1 \times \ec_p \cong \BT^3$. 
The classical SW theory for $\gq = \mathfrak{sl}_r$ quiver, was
studied in \cite{Gorsky:1997mw}.   The classical limit $\ep = 0$ of the
associated integrable system \cite{NP2012a} for
affine $\gq = \hat \gqp$, where $\gqp$ is finite-dimensional simple Lie
algebra, is the moduli space of $\Gqp$
instantons on $T^4 = \ec_{p} \times \ec_{\qe}$ for $\qe = \prod_{i}
\qe_{i}^{a_i}$, where $a_i$ are Dynkin marks on $\Gamma$.

\subsubsection{Dimension uniform description}
To say uniformly, we deal with a \emph{quantum current algebra}
$\mathbf{U}_{\ep}\gq(\Cx)$, where complex one-dimensional curve $\Cx$ is the domain of the additive
spectral parameter $x \in \Cx$, or \emph{loop} variable in the
definition of $\mathbf{U}_{\ep}\gq(\Cx)$ by Drinfeld
currents $(\psi^{\pm}_{i}(x), e^{\pm}_{i}(x))|_{i \in \Ver}$, see \ref{se:Drinfeld-currents}, so that
\begin{equation}
  \mathbf{U}_{\ep}\gq(\Cx) =
  \begin{cases}
    \Yang_{\ep}(\gq), \quad\,\,\,\, \Cx = \BC\\
    \Qaff_{q}(\gq), \quad \Cx = \BC/\BZ \cong \BC^{\times}\\
    \Qell_{q,p}(\gq), \quad \Cx = \BC/(\BZ \oplus \tau_p \BZ) \cong \BC^{\times}/\pp^{\BZ}
  \end{cases}
\end{equation}

The representation theory of quantum algebras $\mathbf{U}_{\ep}\gq(\Cx)$
is  the topic of  active research with a vast literature.
We shall mention only some facts of direct relevance to what follows.  The basic result of Chari and
Presley \cite{Chari:1991,Chari:1994} is that the irreducible finite dimensional
representations of $\Qaff_q(\gq)$ for finite-dimensional simple Lie algebra $\gq$ of rank $r$, are classified by the $r$-tuples of
polynomials, called Drinfeld polynomials. These $\Qaff_q( \gq)$-modules can be
described by the \emph{$q$-characters} of Frenkel and Reshetikhin
\cite{Frenkel:1998}, such that the tensor product of representations
corresponds to the product of $q$-characters. In the rational limit the $q$-characters reduce to the characters of
the Yangian $\Yang_{\ep}(\gq)$ found in \cite{Knight:1995},  see also
\cite{Chari:1996}. 
Frenkel and Mukhin \cite{Frenkel:2001} gave combinatorial algorithm for the
$q$-characters of $\Qaff_q( \gq)$.
 Nakajima \cite{Nakajima:2001q}
realized geometrically representation of $\Qaff_q( \gq)$
 on equivariant $K$-homology group of the quiver variety and in \cite{Nakajima:2001,Nakajima:2004} he 
defined non-commutative $t$-deformation
of the $q$-character building on the geometrical 
 methods of quiver varieties
 \cite{Nakajima:1994,Nakajima:1994r,Nakajima:1998} and computed
 explicitly $(q,t)$-characters of $\Qaff_q(\g)$ for the $\g = A_{r},D_{r}$ series 
in \cite{Nakajima:2003} and for $\g = E_{6}, E_{7}, E_{8}$ in
\cite{Nakajima:2010}.  Hernandez 
 \cite{Hernandez:2004} constructed $(q,t)$-characters for 
$\gq$-quantum affine algebra $\Qaff_q(\gq)$ of generic symmetrizable Kac-Moody Lie algebra $\gq$, developed
 representation theory in \cite{Herandez:2003} and constructed quantum
 fusion tensor category in \cite{Hernandez:2005a};
 Chari et al \cite{Chari:2001,Chari:2004} constructed fundamental $\Qaff_q(\g)$
 $q$-characters  for classical series $\g = A,B,C,D$ using
 the action of the braid group $\Braid(W_{\g})$ corresponding to the
 Weyl group $W_{\g}$.

 More general category of
infinite-dimensional representations for the Borel subalgebra $\Qaff_q(
\gq)$ was studied  by
Hernandez and Jimbo \cite{Hernandez:2011b} following
the work \cite{Bazhanov:1998dq, Bazhanov:1996dr} on the case $\gq =
\mathfrak{sl}_2$.  The recent paper \cite{Frenkel:2013dh}  interprets the Baxter functions for certain infinite-dimensional representations.

\subsubsection{The twist mass for the $A$-type affine quivers}
Recall that if $\gq = \hat A_{r}$ the $\gq$-theory has mass
parameter $m$ associated to the $H^1(\Gamma)$ that can not be removed
by the shifts of scalars in the $U(1)$ vector multiplets \cite{NP2012a}.
The additional equivariant parameter $\mu=e^{i \lfive m}$ appears in H.~Nakajima's work \cite{Nakajima:2002b} on $K$-theory of the quiver variety ${\mathfrak M}_{\Gamma}$ for $\Gamma$
of type $\hat A_{r}$. In his picture the multiplicative group
$\BC^{\times}_{q}$, with defining character $q$,  acts in the K-theory of
moduli space of sheaves on  $\BC^2_{\langle z_3, z_4 \rangle}/\boldsymbol{\Gamma}$ by\footnote{With the identification
  $q_{\text{present}}^{1/2} = q$ of \cite{Nakajima:2002b}} via the scaling symmetry of the base space: $(z_3, z_{4} ) \mapsto ( z_3 q^{1/2},  z_4 q^{1/2})$. If $\boldsymbol{\Gamma} \subset SU(2)$ is cyclic
$\boldsymbol{\Gamma}  = \BZ_{r+1}$, then there is another $\BC^{\times}_{\mu}$
action on $\BC^2$ commuting with $\boldsymbol{\Gamma}$, namely $(z_3, z_{4})  \to ( \mu z_3 ,  \mu^{-1} z_4)$. Hence, for cyclic $\boldsymbol{\Gamma}$ the
equivariant K-theory of moduli space of sheaves on $\BC^2/\boldsymbol{\Gamma}$ depends on two
parameters $(q,\mu)$ for $\BC^{\times} \times \BC^{\times}$ action on
$\BC^2$. Naturally, we expect that there exists  a  $\mu$-twisted version of
$\Qaff_q(\hat{\mathfrak{gl}}_{r+1})$.  Indeed, such $\mu$-twisted
quantum toroidal algebra $\Qaff_{q}(^{\mu}\hat{\mathfrak{gl}}_{r+1})$ was constructed
by Varagnolo and Vasserot \cite{Varagnolo:1999} and it has two
parameters $q$ and $\mu$.   Our conventions correspond to the conventions
of \cite{Varagnolo:1999} as follows: $\mu$-twisted quantum toroidal algebra $\Qaff_{q}(^{\mu}\hat{\mathfrak{gl}}_{r+1})$ is $\mathbf{U}_{q,t}(\BZ/n\BZ)$ of \cite{Varagnolo:1999} for
$n_{\text{\cite{Varagnolo:1999}}} = r+1$, 
$q_{\text{\cite{Varagnolo:1999}}} = q^{\frac 1 2} \mu$ and
$t_{\text{\cite{Varagnolo:1999}}} = q^{\frac 1 2}\mu^{-1}$.

The affine Yangian $\Yang_{\epsilon} (^m \hat{ \mathfrak{gl}}_{r+1}) $
associated to $\hat A_{r}$ quiver is a rational limit for $\mu =
e^{\betir m}$ of
quantum toroidal algebra $\Qaff_{q}(^{\mu}\hat{\mathfrak{gl}}_{r+1})$, 
which admits additional deformation parameter that we identify with the
twist mass $m$ coupled to the cycle in  $H_1(\Gamma)$, see more on
affine Yangian in \cite{Varagnolo:2005,Guay:2005,Guay:2007,Guay:2009a}.

The twist mass parameter $m$ in $H^1(\Gamma)$ for $\Gamma = \hat A_{r}$ corresponds to
the parameter of the non-commutative deformation of $\Cx \times \ec_{\qe}$. Recall that the phase space of the associated integrable
system is the moduli space of $U(r+1)$-instantons on the non-commutative $\Cx \times \ec_{\qe}$  \cite{NP2012a}. The six dimensional case of $\gq = \hat A_{0}^{*}$ theory (i.e. $N=2^{*}$ theory), the phase space of classical integrable system is the moduli
space of non-commutative $U(1)$ instantons on $T^4$ was
studied in \cite{Braden:2001yc, Braden:2003gv, Hollowood:2003cv}. Our
work implies that the quantization of the integrable system for the six dimensional theory is provided by the  elliptic version
of the $\mathfrak{gl}_1$-toroidal algebra, that we call
$\Qell_{q,p}(^\mu\hat{\mathfrak{gl}}_1)$.

\subsubsection{Quantum toroidal algebras}
Quantum $\g$-toroidal algebra  was introduced in 
\cite{Ginzburg:1995b} for all simple Lie algebras $\g$  and 
in \cite{Varagnolo:1996b} it was shown that quantum
$\mathfrak{gl}_{r+1}$-toroidal algebra is Schur-Weyl dual to double affine Hecke algebra
\cite{Cherednik:1992,Cherednik:2004}. The vertex representation at level
1 for $\mathfrak{sl}_{r+1}$-toroidal was constructed in
\cite{Saito:1996b}.  Representation theory of quantum toroidal algebras is based on
non-symmetric MacDonald polynomials  \cite{Cherednik:1995m}. 

 In
\cite{Nakajima:2002b} for $\g = ADE$ and  \cite{Varagnolo:1999} for
$\g = A$ was shown that 
 equivariant
$K$-theory of Nakajima's quiver variety \cite{Nakajima:1998}
realizes representation of quantum $\g$ toroidal algebra. Another construction of representations of quantum toroidal algebra was
given using Schur-Weyl duality to Dunkl-Cherednik representation
\cite{Cherednik:1995} of DAHA \cite{Cherednik:1995}. In \cite{Varagnolo:1996} \cite{Saito:1998} was
constructed representation of  quantum
$\mathfrak{gl}_{r+1}$-toroidal algebra in the q-Fock space (see
\cite{Kashiwara:1995} for quantum $\mathfrak{gl}_{r+1}$-affine
version). The isomorphism between the two constructions has been
shown in \cite{Nagao:2007}. The braid group and automorphisms of quantum
$\mathfrak{gl}_{r+1}$ toroidal algebra was studied in \cite{Miki:1999}.

\newcommand{\torus}[2]{(\mathbb{C}^{\times})^{#1}_{#2}}
 Consider
 the algebra $\torus{d}{\mu}$, called the \emph{non-commutative multiplicative
 $d$-torus}, generated by the variables $x_1^{\pm 1}, \dots, x_d^{\pm 1}$
 subject to the relations 
 \begin{equation}
 x_i x_j = \mu_{ij} x_j x_i 
 \label{eq:mdtor}
 \end{equation} where $\mu_{ij} =
\mu_{ji}^{-1}$ is the matrix of the non-commutativity $c$-valued parameters. 
For example, for $d=2$,
the algebra $\torus{2}{\mu}$ is the non-commutative $2$-torus, also known
as algebra of the $\mu$-difference operators on $\BC^{\times}$, generated by the two
variables $x_1^{\pm 1}, x_2^{\pm 1}$ modulo relations $x_1 x_2 = \mu x_2
x_1$.  Let $\g$ be a finite-dimensional simple Lie algebra. 
The paper \cite{Berman:1996} defined possibly $\mu$-non-commutative $\torus{d}{\mu}$-toroidal
$\g$-algebras $\g[\torus{d}{\mu}]$ and their central extensions. In particular, it was shown
in \cite{Berman:1996} that only the $A$-type $d$-toroidal
$\g$-algebras allow non-trivial non-commutative deformation by parameter
$\mu$. 

The toroidal algebra $\Qaff_{q}(\hat \g)$ associated to the affine
quiver gauge theory $\gq = \hat \g$ is isomorphic to $\mathbf{U}_{q}
\g[\torus{2}{\mu}]$, that is to the $q$-deformation of the universal
enveloping algebra of the $d=2$ $\mu$-noncommutative toroidal algebra
$\g[\torus{2}{\mu}]$ of \cite{Berman:1996}; moreover, the non-commutativity
parameter $\mu$ in $H^1(\Gamma)$ is possibly non-trivial only for the
affine $A$-series, in agreement with
our gauge theory considerations and the constructions of H.~Nakajima. 

In this way, the  simplest, $\mathfrak{gl}_1$ $d=2$ toroidal algebra,  $\Qaff_{q}(^\mu\hat{\mathfrak{gl}}_1)$, with
quantization parameter $q$ and non-commutativity parameter $\mu$ was
constructed and studied by Miki in \cite{Miki:2007} (this $\mathfrak{gl}_1$-toroidal algebra is called there
$\mathcal{U}_{q,\gamma}$ with the parameter $\gamma$ related to our $\mu$,
cf. Eqs. (3.18) - (3.24) of \cite{Miki:2007}).

Morover, in \cite{Miki:2007} Miki has shown that the tensor product of $n$
level $1$ modules of $\mathfrak{gl}_1$-toroidal algebra relates to the $q$-deformed $\mathcal{W}$-algebra
$\mathcal{W}_q(\mathfrak{sl}_n)$ of \cite{Awata:1996,Shiraishi:1996}. 
 On the other hand, from Nakajima
 \cite{Nakajima:1995,Nakajima:1994,Nakajima:1998,Nakajima:2002b} and
 related works
 \cite{Ginzburg:1995b,Ginzburg:1993,Grojnowski:1995,Baranovsky:2000,Varagnolo:1999,Varagnolo:1996,
Varagnolo:1996b,Varagnolo:2005,Schiffmann:2009,Carlsson:2008,Carlsson:2013}, 
 $\mathfrak{gl}_1$-toroidal algebra
$\Qaff_{q}(^\mu\hat{\mathfrak{gl}}_1)$ at level $n$ acts in the equivariant $K$-theory of
 the moduli space of framed torsion free rank $n$ sheaves on $\BC^2_{q_1, q_2}$ 
with two equivariant parameters  $q_3 = q^{\frac 1 2}  \mu$ and $q_4
= q^{\frac 1 2} \mu^{-1}$. Hence, in \cite{Miki:2007}, by relating the tensor product of $n$
level $1$ modules of $\mathfrak{gl}_1$-toroidal algebra to the
$q$-deformed $\mathfrak{sl}_n$ quantum $\mathcal{W}$-algebra
\cite{Awata:1996,Shiraishi:1996,Feigin:1996}  Miki 
has shown (see also remark A.7 in \cite{Feigin:2009com}) a version of correspondence between equivariant rank $n$ gauge
theory on $\BC^2_{q_3, q_4}$ \cite{Nekrasov:2002qd} and
$q$-deformed $\mathfrak{sl}_n$-Toda theory that appeared later as an AGT
conjecture \cite{Alday:2009aq,Wyllard:2009hg,Awata:2009ur,Fateev:2011hq}.

In the 4d, rational limit,
the non-commutative 2-torus $\torus{2}{\mu}$, or the algebra of $\mu$-difference operators, contracts to
the Lie algebra of differential operators of one-variable, called
$\mathfrak{d}$. Its central extension $\hat{\mathfrak{d}}$ is also known as
$\mathcal{W}_{1+\infty}$ algebra \cite{Miki:2007}; and its quantization
would produce the rational limit of $\mathfrak{gl}_1$-toroidal algebra,
which is $\mathfrak{gl}_1$ affine Yangian
$\Yang_{\ep}(^{m}\hat{\mathfrak{gl}}_1)$. For explicit $R$-matrix for $\Yang_{\epsilon} (^m \hat{
  \mathfrak{gl}}_{1})$ see \cite{Smirnov:2013}.
The tensor product of $n$
level $1$ modules of $\mathfrak{gl}_1$ affine Yangian relates to the
ordinary $\mathcal{W}(\mathfrak{sl}_n)$-algebra, the symmetry of the 2d conformal
$\mathfrak{sl}_n$-Toda field theory; and since in the rational limit
equivariant $K$-theory contracts to the equivariant cohomology, the
ordinary $\mathcal{W}(\mathfrak{sl}_n)$ algebra relates to the equivariant rank $n$ gauge
theory on $\BC^2_{\ep_3, \ep_4}$  (AGT conjecture
\cite{Alday:2009aq,Wyllard:2009hg}) that has been proved in this form in
\cite{Awata:2011dc, Maulik:2012,Schiffmann:2012b} after \cite{Alba:2010qc,Fateev:2011hq}. 

The $\mathfrak{gl}_1$ toroidal algebra
$\Qaff_{q}(^\mu\hat{\mathfrak{gl}}_1)$ appeared under different names in
the literature. It was called \emph{quantum continuous
$\mathfrak{gl}_\infty$}, or deformation of universal enveloping of $q$-difference operators in
one-variable  in \cite{Feigin:2010a,Feigin:2010b};  \emph{spherical elliptic
Hall algebra} or $\mathfrak{gl}_\infty$ DAHA in \cite{Burban:2005} and
\cite{Schiffmann:2009,Schiffmann:2010}. In \cite{Feigin:2009b} the same algebra without Serre relations is
called \emph{Ding-Iohara algebra} \cite{Ding:1996mq} or trigonometric
limit of  Feigin-Odesskii
\cite{Feigin:1997} \emph{elliptic shuffle algebra} (see also
\cite{Enriquez:1998}). 
The  elliptic deformation $\Qell_{q,p}(^\mu\hat{\mathfrak{gl}}_1)$ of the
$\mathfrak{gl}_1$ toroidal algebra $\Qaff_{q}(^\mu\hat{\mathfrak{gl}}_1)$ is called $\mathcal{A}(q_1, q_2,
q_3,p)$ in \cite{Feigin:2009com} with $q_1 = q^{\frac 1 2} \mu, q_2 =
q^{\frac 1 2} \mu^{-1}, q_3 = q^{-1}$ and is equivalent without Serre
relations to the Feigin-Odesskii elliptic shuffle algebra
\cite{Feigin:1997}.  For details in equivalence of $\mathfrak{gl}_n$
toroidal algebra and shuffle algebra see \cite{Negut:2013}. The elliptic
deformation of $\mathfrak{gl}_1$ toroidal algebra in a certain limit has been related to the
quantum cohomology of Hilbert scheme of points on $\BC^2$
\cite{Okounkov:2004}  in \cite{Feigin:2009com}, and the relation to elliptic
Macdonald operator was further studied in
\cite{Saito:2013,Saito:2013b}.

The characters of quantum $\mathfrak{gl}_1$ toroidal algebra are computed as a
sum over plane partitions \cite{Feigin:2011, Feigin:2013m}; and
representation theory of quantum $\mathfrak{gl}_r$ toroidal algebra is
discussed in \cite{Feigin:2012, Feigin:2013fga}.

\subsubsection{Other topics}
The representation theory of quantum affine and toroidal algebras
 $\Qaff_q(\g)$ is connected to many
 exciting topics in geometry, algebra, and mathematical physics,
 e.g. the cluster algebras, the $Y$- and the $T$-systems, 
 affine and double affine Hecke algebras \cite{Cherednik:2004}, 
knots and three dimensional theories \cite{Fuji:2012nx}, 
quantum Knizhnik-Zamolodchikov-Smirnov
 equation \cite{Smirnov:1991me, Smirnov:1992vz, Frenkel:1992},  affine Toda integrable field theories and many other
 integrable systems, $W$-algebras \cite{Feigin:1994in, Feigin:1996}, 
 the putative five dimensional version of the AGT \cite{Alday:2009aq} correspondence, 
 Stokes multipliers for
  ordinary differential equations and CFT correlation functions \cite{Bazhanov:1994ft, Bazhanov:1996dr, Bazhanov:1998dq, Dorey:2007zx}, 
holomorphic anomaly equation in the limit of \cite{Nekrasov:2009rc} (see, e.g. \cite{Huang:2012kn}), 
geometric Langlands correspondence \cite{Chervov:2006xk}, 
  spectrum of anomalous dimensions in ${\CalN}=4$ super Yang-Mills. 
  We do not consider in the present note
  the above topics and the web of their interrelation leaving the discussion to the future.

\subsection{Quantum group interpretation of the gauge theory results}

There are two ways to think about the Coulomb moduli space $\mv$.

In the \emph{first interpretation} the expectation values of the observables
(\ref{eq:ObO}) parametrize an element $h(\xi) \in \Qaff_q(\gq)$ (in 
the maximal commutative subalgebra) which morally
speaking is similar to picking a conjugacy class in a group (if $\Qaff_q( \gq)$
were a group).
 For gauge groups of finite rank $\bv_{\iv}$, this 
parametrization is clearly redundant because of the relations between
${\CalO}_{i,n}$ for high enough $n$. Conceptually, the elements $h(\xi)$
in the maximal
commutative subalgebra of $\Qaff_q( \gq)$ parametrize the union of Coulomb
moduli spaces over all ranks $\bv_{\iv}$ of the gauge groups and  couplings
${\qe}_{i}$. 
This interpretation in the classical limit $q \to 1$ reduces to the
construction in \cite{NP2012a}. 

In the \emph{second interpretation} the  polynomial $T_{\iv}(\xi)$ 
is an eigenvalue of the transfer matrix operator $t_{V_{\iv}} \in U_q(\hat
\g)$
 (where $V_{\iv}$ is a fundamental $\Qaff_q(\g)$ module  usually called
\emph{an auxiliary space} in the algebraic Bethe Ansatz) on an eigenstate $| w
\rangle \in W$ where another
 $\Qaff_q(\gq)$-module $W$ is the \emph{physical space of states of $\gq$-spinchain}.
 This $\Qaff_q(\gq)$-module $W$, defined by masses of the gauge theory,
 is usually not finite-dimensional, but
 the spectral problem in the $\Qaff_q(\gq)$-module $W$ can be converted
 to the spectral problem in a finite-dimensional $\Qaff_q(\gq)$-module
 $\tilde W$  for special discrete choices of Coulomb parameter $\ac_{i,\ba}$ as for example 
in \cite{Dorey:2011pa,Chen:2011sj} (\emph{special electric} $\ac_{i, \ba}$) or
\cite{Nekrasov:2009rc} (special magnetic $\ac_{i,\ba}$). 
For example, for $\gq = \mathfrak{sl}_2$ with $\bw_1 = 2 \bv_1$, as was
advocated by Dorey et al \cite{Dorey:2011pa,Chen:2011sj}, one can split
the set of masses $M = \{ \mu_{1,\fe}\}$ into two disjoint sets $M^{-} = \{ \mu_{1,\ba}^{-}\}$
and $M^{+} =\{\mu_{1,\ba}^{+}\}$ with $\ba = 1, \dots, \bv$ and then
choose Coulomb parameters $w_{1,\ba} =  q^{n_{\ba}} \mu_{1, \ba}^{-}$
for some integers  $n_{\ba} \in \BZ_{\geq 0}$. Then $\Qaff_q({
  \mathfrak{sl}_2})$-module $\tilde W$ is 
the tensor product $\tilde W = W_{s_1, \zeta_1} \otimes W_{s_2, \zeta_2}
\otimes \dots \otimes W_{s_n, \zeta_n}$ of the
$\Qaff_q({\mathfrak{sl}_2})$ evaluation Verma modules $W_{s_{\ba}, \zeta_{\ba}}$ at spectral
parameter $\zeta_{\ba} = \sqrt{ \mu_{\ba}^{-} \mu_{\ba}^{+}}$  and
highest weight  $s_{\ba} = \log_{q}(\mu_{\ba}^{+}/\mu_{\ba}^{-})$.

\subsubsection{First interpretation of the gauge theory $q$-character equations}

The conventions for the quantum affine algebras are summarized in the appendix
\ref{se:quantum-affine}. 

First, we find an element $\mathring{\psi}(\xi|\Xi) \in \Qaff_q( \gq)$ in the maximal commutative
subalgebra of $\Qaff_q( \gq)$ such that its evaluation  in a
finite-dimensional $\Qaff_q( \gq)$-module $V$ has the generalized eigenvalues 
(the symbols $Y_{{\iv},\zeta}$) equal to the
gauge theory function $\y_{\iv}(\xi/\zeta)$
\begin{equation}
  \mathrm{ev}_{\mathring{\psi}(\xi|\Xi)}  Y_{{\iv},\zeta}= \y_{\iv}^{+} (\xi/\zeta) 
\end{equation}
Explicitly, comparing the evaluation definition (\ref{eq:evaluation})
and the definition of the $Y$-functions in the gauge theory
(\ref{eq:Yi})  we conclude that such element
$\mathring{\psi}(\xi|\Xi) \in \Qaff_q( \gq)$ is given by 
\begin{equation}
\label{eq:element}
 \mathring{\psi}(\xi|\Xi) = \prod_{{\iv} \in \Ver} \prod_{\xi' \in \Xi_{\iv}}  \mathring \psi_{\iv}^+ (\xi/\xi')
\end{equation}
where $\mathring \psi_{\iv}^+ (\xi/\xi')$ is defined by  \eqref{eq:psi0}.

For example, in the case $\gq = \mathfrak{sl}_2$, the $q$-character of
the fundamental module $V_{1,q^{-\frac 12}}$ evaluated by the element
(\ref{eq:element}) gives us 
\begin{equation}
 \mathrm{ev}_{\mathring{\psi}(\xi|\Xi)} \chi_{q}(V_{1,q^{-\frac 1 2}}) =
 \y_{\iv}^{+}(q^{\frac 1 2} \xi) + \frac{1}{\y_{\iv}^{+} (q^{-\frac 1 2 }\xi)} 
\end{equation}

Notice by comparing (\ref{eq:yifived}) with (\ref{eq:element}) that
can represent the operator $\mathring{\psi}(\xi|\Xi) \in \Qaff_q( \gq)$
as follows 
 \begin{equation}
\label{eq:ginh}
 \mathring{\psi}(\xi | \Xi)    = \exp \left( -\sum_{{\iv} \in \Ver} \sum_{n=1}^{\infty}\
   \frac{\xi^{-n} q^{-\frac n 2} }{[n]_q} \la \ \tr_{\bv_{\iv}}      e^{\ii \lfive n \phi_{\iv}}  \ \ra
   h_{{\iv},n} \right)
 \end{equation}
For the gauge theory with fundamental matter we need to incorporate 
the extra factors of matter polynomials as in (\ref{eq:chi1A1}). We can
do this by multiplying (\ref{eq:element}) by  another element in $\Qaff_q( \gq)$, which depends
only on the masses and coupling constants. To find explicit expression
for $\gq = \mathfrak{sl}_2$,
it is useful to notice that the equation 
\begin{equation}
\label{eq:q-weights}
  s(q^{-\frac 1 2 } \xi) s(q ^{\frac 1 2} \xi) = 1 - \xi^{-1}
\end{equation}
is solved by the  function (defined as the expansion near $\xi = \infty$)
\begin{equation}
  s(\xi) = \exp \left( -\sum_{n=1}^{\infty} \frac 1 n \frac{ \xi^{-n}}{
      q^{-\frac n 2}     + q^{\frac n 2}} \right).
\end{equation}
Define an operator $\psi^{\vee}_1(\xi) \in \Qaff_q({\mathfrak{sl}_{2}})$ as follows
\begin{equation}
\label{eq:product-sigma}
\mathring{\psi}^{\vee}_1(\xi) := \exp \left( - \sum_{n=1}^{\infty} \frac{1}{[n]_q} \frac{
    \xi^{-n}}{q ^{\frac n 2} + q^{-\frac  n 2}} h_{1,n} \right)
\end{equation}

The eigenvalues of the operators $p_1(\xi),
\mathring{\psi}_1(x), \psi^{\vee}_1(x)$ on a fundamental eigenvector with 
generalized weight encoded by the Drinfeld polynomial \mbox{$P(\xi) = 1
  - \xi^{-1}$} are given by 
 \begin{equation}
\label{eq:cur-weights}
   \begin{aligned}
 &    p_1(\xi) | v  \rangle = (1 - \xi^{-1}) \\
&     \mathring{\psi}^{\vee}_1(\xi) | v \rangle  = s(\xi)  \\
 &  \mathring{\psi}_1(\xi) | v \rangle =  \frac{1 - q^{-\frac 1 2} \xi^{-1}}{ 1
 - q^{\frac 1 2}\xi^{-1}}
   \end{aligned}
 \end{equation}

Given a set
 \[M_1 = \{\mu_\fe\}\]
of the masses for fundamental multiplets
define the operator $  \mathring{\psi}_1^{\vee}(\xi | M) \in \Qaff_q( \mathfrak{sl}_2)$
\begin{equation}
  \mathring{\psi}^{\vee}_1(\xi | M_1) = \prod_{\mu_\fe \in M_1}  \mathring{\psi}^{\vee}_1(\xi/\mu_\fe)
\end{equation}
with eigenvalue given by the function
\begin{equation}
  s_{11}(\xi |M_1) = \prod_{\mu_\fe \in M_1} s( \xi/\mu_{\fe})
\end{equation}
that satisfies 
\begin{equation}
  s_{11}(\xi q^{\frac 1 2}|M) s_{11}(\xi q^{- \frac 1  2}|M) =
  \mathring{\sP}_{1}(\xi) \equiv \prod_{\fe} (1 - \mu_\fe /\xi) 
\end{equation}

Then  $\chi_{q}(V_{1,q^{-\frac 1 2 }})$ character evaluated on the element \mbox{$h(\xi) \in \Qaff_q( \mathfrak{sl}_2)$}
\begin{equation}
\label{eq:elementsl2}
h(\xi) = \qe^{-\frac 1 2 h_{1,0}} \mathring{\psi}_{1}(\xi | \Xi) / \mathring{\psi}_{1}^{\vee}(\xi | M)
\end{equation}
equals 
\begin{equation}
 \mathrm{ev}_{h(\xi)} \chi_{q}(V_{1,q^{-\frac 1 2}}) =
\qe^{-\frac 1 2} \frac{ \y_1(q^{\frac 1 2} \xi)}{ s_{11}( \xi q^{\frac 1 2} |
  M)} + \qe^{\frac 1 2 } \frac{s_{11}( \xi q^{-\frac 1 2} | M) }{\y_1 (q^{-\frac 1 2 }\xi)} 
\end{equation}

Finally, define the twisted $q$-character $\tilde \chi_{1,\zeta}$ multiplying
the $q$-character by a suitable scalar factor such the term
corresponding to the highest weight vector in the fundamental module
$V_{1,\zeta}$
equals to $\y_1( \xi /\zeta )$
\begin{equation}
\tilde \chi_{1, \zeta}    =c_{1}(\xi/\zeta)
 \chi_{q} (V_{1, \zeta}) 
\end{equation}
where
\begin{equation}
    c_1(\xi) = \qe^{\frac 1 2}  s_{11}( \xi | M)
\end{equation}

Hence, we obtain \emph{the first interpretation} of the  limit-shape equations (\ref{eq:SWq})

 For $\gq = \mathfrak{sl}_2$,  the twisted character $\tilde \chi_{1,\zeta}$ 
 evaluated on the element $h(\xi) \in \Qaff_q( \gq)$ is a
polynomial of degree $\bv_1$ that we denote $T_1(\xi)$
\begin{equation}
  \begin{aligned}
\mathrm{ev}_{h(\xi)}  \tilde \chi_{1, q^{-1/2} }= T_1 (\xi)     \qquad
  (\Rightarrow )\qquad  
\y_1^{+}(\xi q^{\frac 1 2}) + \frac{ \sP_1( \xi)}{ \y_1^{+}( \xi q^{-\frac 1 2
  })} = T_1(\xi)
  \end{aligned}
\end{equation}
This interpretation is completely in parallel with the construction of
$\Gq(\Cx)$-group element in \cite{NP2012a} and becomes the one in the
classical limit $q \to 1$.

The equation (\ref{eq:q-weights}) is the multiplicative $q$-version of the equation $a \cdot 
\Lambda^{\vee} = \Lambda^{\vee}$ relating the fundamental coweights
and the simple coroots. 
The higher rank $\gq$ requires a generalization of (\ref{eq:q-weights}). The
multiplicative $q$-version of the equation 
$$
\sum_{\bf j} a_{\bf ij} \Lambda_{\bf j}^{\vee} =
\alpha_{\iv}^{\vee}
$$ 
for the inverse Cartan matrix is 
\begin{equation}
\label{eq:defining-p}
  \frac{ s_{\bf ik} (q^{\frac 1 2} \xi) s_{\bf ik} (q^{-\frac 12} \xi)}
{\prod_{{\bf j}: \langle {\bf ij} \rangle = 1} s_{\bf jk} (\xi)} =
\begin{cases}
1 - \frac{1}{\xi}, \quad  {\iv} = {\bf k}\\
 1, \quad {\iv} \neq {\bf k}
\end{cases}
\end{equation}
Then $s_{\bf ik}(\xi)$ is the $q$-multiplicative decomposition of the
fundamental coweight $\Lambda_{\bf k}^{\vee}$ over the basis of simple
coroots $\alpha_{\iv}^{\vee}$.

Let $a(q)$ be the $q$-Cartan matrix defined by replacing each entry of
the Cartan matrix by its $q$-number, and ${\tilde a}(q)$ its inverse: 
\begin{equation}
a(q)_{\bf ij} = [ a_{\bf ij} ]_{q} , \qquad \sum_{\bf j} a(q)_{\bf ij} {\tilde a(q)}_{\bf jk} = \delta_{\bf ik}
\ . 
 \label{eq:qcar}
 \end{equation}
 Then, expanding the defining equation (\ref{eq:defining-p}) in power
series at $\xi = \infty$ we find 
\begin{equation}
  s_{\bf ij} (\xi)= \exp \, \left( -\sum_{n=1}^{\infty} \frac{ \xi^{-n}}{n} \tilde
    a_{\bf ij}(q^n) \right)
    \label{eq:sik}
\end{equation}
and for a set $M_{\bf j} = \{ \mu_{{\bf j}, \fe}\}$ define
\begin{equation}
  s_{\bf ij}(\xi | M_{\bf j}) = \prod_{\mu_{\fe} \in M_{\bf j}} s_{\bf ij}(\xi/\mu_{\fe})
\end{equation}
Consequently, the generalization of \eqref{eq:product-sigma} to the $q$-coweight operator   
$\mathring{\psi}^{\vee}(\xi)_{\iv}  \in \Qaff_{q} ( {\gq} )$ is
\begin{equation}
\mathring{\psi}_{\iv}^{\vee}({\xi})   = \exp \left( - \sum_{n=1}^{\infty} \frac{\xi^{-n}}{[n]_q} \tilde a_{\bf ij} (q^n) h_{{\bf j},n}   \right)
\end{equation}
The generalization of the formula \eqref{eq:elementsl2} to the element $h(\xi) \in \Qaff_q( \gq)$
 is 
\begin{equation}
\label{eq:helement}
 h ({\xi}) =    \prod_{{\iv} \in \Ver} \qe_{\iv}^{-\tilde a_{\bf ij} h_{{\bf j},0}}  
 \mathring{\psi}_{\iv} ({\xi} | \Xi_{\iv}) / \mathring{\psi}_{\iv}^{\vee} ({\xi} | M_{\iv})
\end{equation}
where ${\Xi}_{\iv}$  is the set of  zeroes of $Q_{\iv}^{+}(\xi)$ and $M_{\iv}$ is the set
of zeroes of ${\sP}_{\iv}^{+}({\xi})$.

The gauge theory $q$-character equations are written using the twisted
$q$-characters $\tilde \chi_{{\iv}, \zeta}$ for the fundamental $\Qaff_q( \gq)$ modules  
$V_{{\iv},\zeta}$
\begin{equation}
\label{eq:chi-char}
  \tilde \chi_{{\iv}, \zeta}    =c_{\iv}({\xi}/{\zeta})
 \chi_{q} (V_{{\iv}, {\zeta}}) 
\end{equation}
where the scalar factor is 
\begin{equation}
\label{eq:chi-charc}
  c_{\iv}(\xi) = \prod_{\bf j} \qe^{\tilde a_{\bf ij}} s_{\bf ij}({\xi} | M_{\bf j})
\end{equation}
We summarize the first interpretation by the proposition
\begin{proposition}
The element $h({\xi}) \in \Qaff_q( \gq)$ that encodes by (\ref{eq:helement}~(\ref{eq:ginh}))
the gauge theory chiral ring generating functions $\y_{\iv}^{+}(\xi)$ satisfies
the equations
\begin{equation}
 \mathrm{ev}_{h({\xi})} \tilde \chi_{{\iv}, \zeta} = T_{\iv}({\xi}), \quad {\iv} \in \Ver
\end{equation}
where $\tilde \chi_{{\iv},\zeta}$ is the twisted $\Qaff_q( \gq)$
$q$-character of the fundamental $\Qaff_q( \gq)$-module $V_{i,\zeta}$
defined by (\ref{eq:chi-char}~(\ref{eq:chi-charc})) and
(\ref{eq:qchardef}), and $T_{\iv}(\xi) \in \BC[\xi^{-1}]$ is a  polynomial
in $\xi^{-1}$ of degree $\bv_{\iv}$. 
\end{proposition}

\section{Examples and discussions}

\subsection{Finite $A_{r}$ quiver}
For $\gq = A_{r}$ the system of $r$ $q$-difference equations (\ref{eq:SWq}) 
can be reduced to a single $q$-difference equation of order $r+1$. This reduction is the $q$-analogue of the expansion of the characteristic polynomial of an ${\mathfrak sl}(r+1)$ matrix in  the fundamental
characters. Explicitly,  we obtain
\begin{equation}
\label{eq:q-determinant}
  \sum_{k=0}^{r+1}(-1)^k \y_{1,k}^{[r+1-k]} T_{k,(k-1)/2}
  \prod_{j=1}^{k-1} \sP_{j,j/2}^{[k-j]} = 0 
\end{equation}
where
\begin{equation}
  f_{i,j}^{[k]} = \prod_{j=1}^{k} f_{i,j+k-1}, \quad f = {\y,\sP}
\end{equation}
Now, in terms of Baxter functions (\ref{eq:Baxter})
\begin{equation}
  \y_{i,j} = \frac{Q_{i,j-\frac 1 2} }{Q_{i,j+\frac 1 2}}
\end{equation}
the $q$-determinant equation (\ref{eq:q-determinant}) reduces to the 
linear degree $r+1$ $q$-difference equation on $Q_{1}(\xi)$.
\begin{equation}
\label{eq:Ar-spectral}  \sum_{k=0}^{r+1}(-1)^k Q_{1,k-\frac 1 2 } T_{k,\tfrac{k-1}{2}}
  \prod_{j=1}^{k-1} \sP_{j,\tfrac{j}{2}}^{[k-j]} = 0 
\end{equation}
The above is known as the first equation in the hierarchy of QT-system
for the integrable $A_{r}$ spin chain with inhomogeneous parameters encoded in the roots of
$\sP_{\iv}(\xi)$. 

\subsection{ $A_{\infty}$ quiver}
For $\gq = A_{\infty}$, serving as the universal cover of $\hat A_{r-1}$, we construct the $q$-character of quantum 
affine algebra $\Qaff_q( A_{\infty})$ using the same
requirement (cancellation of poles) and from (\ref{eq:AinftyBethe})
 arrive to the formula
\newcommand{\jmax}{j_{\text{max}}}
\begin{equation}
\label{eq:chi-i-infty}
  \chi_{i} = 
\sum_{\lambda}
\prod_{j=1}^{{\ell}_{\lambda}} \left(\sP_{i-j+1,\tfrac j 2}^{[[\lambda_j]]}\, 
\frac { \y_{i + \lambda_j - j + 1, \tfrac{\lambda_j  + j - 1 }{2}}}
{ \y_{i + \lambda_j - j , \tfrac{\lambda_j+ j}{2}}} \right)
\y_{i - {\ell}_{\lambda}, \tfrac{{\ell}_{\lambda}}{2}}
\end{equation}
where the sum is over all partitions $\lambda$.
In \eqref{eq:chi-i-infty} we omitted symbol $\tilde{\tilde{}}$ compared to (\ref{eq:AinftyBethe}), and where 
\begin{equation}
  \sP^{[[\lambda_j]]}_{i,0} \equiv \prod_{k=i}^{k=i+\lambda_j -1}
  \sP_{k,\frac 1 2 (k-i)}
\end{equation}
By restricting the range of indices in (\ref{eq:chi-i-infty}) one
naturally recovers the characters of all fundamental modules for
$\Qaff_q( \gq)$ for $\gq = A_{r}$ with a finite $r$. 

It is useful to express the formula (\ref{eq:chi-i-infty}) in term of
the variables $t_i(\xi)$ defined by 
\begin{equation}
  t_{i,0} := \frac{ \y_{i,0}}{ \y_{i-1, \frac 1 2}}
\end{equation}
which are the $q$-versions of the $GL_{\infty}$ eigenvalues. We find
\begin{equation}
\label{eq:Ainf-q-char}
  \chi_{i} = \y_{i - {\ell}_{\lambda}, \frac 12 {\ell}_{\lambda}} 
\sum_{\lambda}
\prod_{j=1}^{{\ell}_{\lambda}} \left(\sP_{i-j+1,\frac 1 2
    j}^{[[\lambda_j]]}   t_{i+ \lambda_{j} - (j-1), \frac 1 2(
  \lambda_j + j - 1)}
 \right)
\end{equation}

Now we define a non-commutative equivalent of the determinant generating
function for all fundamental characters of $A$-type quivers.  For $j \in \tfrac 1 2 +  \BZ_{\geq 0}$ define 
\begin{equation}
  \begin{aligned}
&  \sP_{i,0}^{\stackrel{\rightarrow}{j}} \equiv \sP_{i,0}^{\stackrel{\rightarrow}{\frac 1 2}} \sP_{i+1,\frac
  1 2} \cdots
  \sP_{i+j-\frac 1 2, \frac  1 2 (j-\frac 1 2) } \\
&  \sP_{i,0}^{\stackrel{\leftarrow}{j}} \equiv   \sP_{i,0}^{\stackrel{\leftarrow}{\frac 1 2}}
  \sP_{i-1,\frac 1 2 } \dots   \sP_{i-(j-\frac 1 2),\frac 1 2 (j
    -\frac 1 2)}
  \end{aligned}
\end{equation}
where the polynomial $\sP_{i,0}$ has been factorized into a product of
two polynomials 
\begin{equation}
  \sP_{i,0}  =  \sP_{i,0}^{\stackrel{\leftarrow}{\frac 1 2}} \sP_{i,0}^{\stackrel{\rightarrow}{\frac 1 2}} 
\end{equation}
   To get the canonical (minimal degree equation) for $A_{r}$ theory we set $i$ to be the vertex with
the maximal $\bv_{i}$ and split arbitrarily  $\sP_{i,0}$ into the polynomials
$ \sP_{i,0}^{\stackrel{\leftarrow}{\frac 1 2}} $ of degree $\bv_{i} -
\bv_{i-1}$ and $\sP_{i,0}^{\stackrel{\leftarrow}{\frac 1 2}}$ of degree
$\bv_{i} - \bv_{i+1}$ (c.f. the degree profile in section 7.1.1 of
\cite{NP2012a}). 

Then we  consider the operator 
\begin{equation}
  \boldsymbol{\Theta_i}  = (\stackrel{\leftarrow}{\boldsymbol{\Theta}_{i}}  Y_{i,0} \stackrel{\rightarrow}{\boldsymbol{\Theta}}_{i})
\end{equation}
where 
\begin{equation}
  \begin{aligned}
&   \stackrel{\leftarrow}{\boldsymbol{\Theta}_{i}}  \  = \ \prod_{j \in \frac 1 2 + \BZ_{\geq 0}
    }^{\leftarrow} \left(1  - D^{-\frac 1 2}\, \sP_{i,0 }^{
      \stackrel{\leftarrow}{j}}  \, 
    \frac{  \y_{i -j - \frac 1 2, -\frac 14 + \frac j2} } { \y_{i - j
        + \frac 12,  \frac 14 + \frac j2}} \, D^{-\frac 1 2}
\right)\\
&   \stackrel{\rightarrow}{\boldsymbol{\Theta}}_{i}  \ = \ 
\prod_{j \in \frac 12 + \BZ_{\geq 0}
    }^{\rightarrow} \left(1  -  D^{\frac 12 } \, \sP_{i, 0}^{
      \stackrel{\rightarrow}{j}}  \, 
    \frac{  \y_{i + j + \frac 12, -\frac 14 + \frac j2} } { \y_{i + j
        - \frac 12, \frac 14 + \frac j2}}\,
 D^{\frac 12} \right)\\
  \end{aligned}
\end{equation}
and $D$ is the shift operator
\begin{equation}
  D = q^{\xi \partial_{\xi}}
\end{equation}
so that $D f_{i,j} D^{-1} =  f_{i, j - 1} $ where $f = \y, \sP$. 
The notation for the non-commutative products over an ordered index set $J$
is 
\begin{equation}
  \begin{aligned}
&  \prod_{ j \in J }^{\rightarrow}  f_{j} = \dots  f_{\bullet} f_{\bullet+1} \dots  \\
&  \prod_{j \in J}^{\leftarrow} f_{j}  =  \dots f_{\bullet + 1} f_{\bullet}  \dots 
  \end{aligned}
\end{equation}
The operator $\mathbf{\Theta}$, like the $A_{\infty}$ determinant or lattice
theta-function, is the generating function in the operator variable $D$ of all fundamental
$q$-characters: 
\begin{multline}
\label{eq:Theta_i-generic}
  \boldsymbol{\Theta}_{i} = \sum_{k < 0} (-1)^k \left( \prod_{j=1}^{-k} \sP_{i, \frac 12
    -k -j}^{\stackrel{\leftarrow}{j - \frac 12}} \right)   \chi_{i+k, -\frac k2} \, D^{k} \   \\
 + \    \chi_{i,0}\  + \\
    \sum_{k > 0}   (-1)^k \left( \prod_{j=1}^{k}
  \sP_{i, j \frac 1 2}^{\stackrel{\rightarrow}{j-\frac 12}} \right) \,  \chi_{i+k, -\frac k2} \, D^{k}
\end{multline}
Our gauge theory equations state that the characters $\chi_i(\xi)$ are the
(Laurent) polynomials $T_i(\xi)$ in $\xi$. We call $T_{i}({\xi})$ the \emph{Coulomb} polynomials. After substitution $\chi_{i}( {\xi} ) =
T_{i} (\xi)$ the operator version of the spectral curve is
\begin{equation}
\label{eq:q-spectral}
  \boldsymbol{\Theta}_i | \mathbf{Q}_i(\xi) \rangle= 0
\end{equation}

\begin{remark}
  If $A_{\infty}$ $q$-characters is specialized to $\hat A_{r}^{*}$ quiver using
  periodicity modulo $r+1$ than the polynomials 
$\sP_{i}(\xi)$ are of degree zero, i.e. $\sP_{i}(\xi) = \qe_{i}''$ where
$\qe_{i}$ is a certain complex number 
 made from coupling constants $\qe_{\iv}$, masses $\mu_e$ and $q$ using~(\ref{eq:P-source})(\ref{eq:mass-shift}). 
  
 The $q$-characters ${\chi}_{i}$ are defined for $i \in \BZ$, but there
 are only $r+1$ functionally independent ones. 

\end{remark}

\subsubsection{Details on the $A_1$ quiver}

The operator $\boldsymbol{\Theta}_1$ is 
\begin{equation}
\label{eq:theta_1-prod}
  \boldsymbol{\Theta}_1 = \left  ( 1 -D^{-\frac 1 2} \sP_{1,0}^{\stackrel{\leftarrow}{\frac 1 2}} \tfrac{1}{ \y_{1, \frac 1 2 }} D^{-\frac 1 2} \right)
  \y_{1, 0}
\left  (1 -  D^{\frac 1 2 } \sP_{1,0}^{\stackrel{\rightarrow}{\frac 1 2}} \tfrac{ 1}{ \y_{1, \frac 1 2}} D^{\frac 1 2}\right)
\end{equation}
and its expansion is
\begin{equation}
  \boldsymbol{\Theta}_1 = \sP_{1,\frac 1
    2}^{\stackrel{\leftarrow}{\frac 1 2}}   D^{-1}  +   \chi_{1,0} -  \sP_{1, -\frac 1 2}^{\stackrel{\rightarrow}{\frac 1 2}} D
\end{equation}
In the equation 
\begin{equation}
  \boldsymbol{\Theta}_1 |\boldsymbol{Q}_1 \rangle= 0
\end{equation}
we recognize the Baxter equation
\begin{equation}
\label{eq:baxter2}
-  \sP_{1,\frac 1
    2}^{\stackrel{\leftarrow}{\frac 1 2}}   \boldsymbol{Q}_{1,1}  +
 T_{1,0}
\boldsymbol{Q}_{1,0} -  \sP_{1, -\frac 1 2}^{\stackrel{\rightarrow}{\frac
    1 2}} \boldsymbol{Q}_{1,-1} = 0
\end{equation}
where the  polynomials $ \sP_{1,\frac 1
    2}^{\stackrel{\leftarrow}{\frac 1 2}} $, $ \sP_{1, -\frac 1 2}^{\stackrel{\rightarrow}{\frac
    1 2}}$ and $T_{1,0}$ would be conventionally
called $A, D$ and $T$ Baxter polynomials. 
The equation (\ref{eq:baxter2}) is the second order linear difference equation
with a basis  $\la \bQ_1^{(0)},\bQ_1^{(1)} \ra$ of  two independent solutions. We
choose one solution that it is annihilated  the last factor of
(\ref{eq:theta_1-prod}). Then we find 
\begin{equation}
\label{eq:bfQvsQ}
\frac {  \bQ_{1, -\frac 1 2}} { \bQ_{1, +\frac 1 2}} = \frac { \y_{i, + \frac 1
    2}}{  \sP_{1, 0}^{\stackrel{\rightarrow}{\frac    1 2}}}
\end{equation}
so that $\bQ_1$  is $Q_{1}$ of
(\ref{eq:Baxter}) up to a factor of  $\gamma_{q, \sP_{1,
    0}^{\stackrel{\rightarrow}{\frac    1 2}}}$.

Here the function $\gamma_{q; \sP}(\xi)$  is a generalization of
$q$-Gamma function, it is determined by the polynomial $\sP(\xi)$ as the solution of the
$q$-difference equation
\begin{equation}
\label{eq:qgamma}
\frac{\gamma_{q,\sP}(q^{\frac 1 2} \xi)} {\gamma_{q,\sP}(q^{-\frac 1 2} \xi)} = \frac{1}{\sP(\xi)}
\end{equation}

\subsubsection{Details on the $A_{3} = D_{3}$ and $A_r$ quivers }
The new feature of the $A_{r}$  quivers with $r \geq 3$ 
is the presence of the interior nodes. To be specific, consider the $A_{3}$ quiver 
with the maximal rank at the middle node $i=2$. 
The $q$-character $\chi_{i, 0}$ can be displayed by the diagram
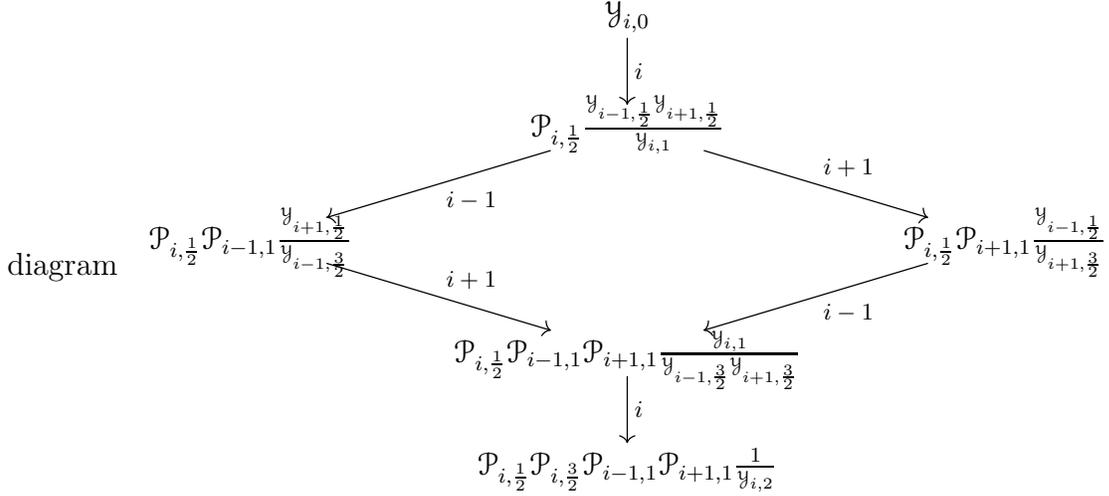
\begin{tikzcd}
{} &    \y_{i,0}\arrow{d}{i}  & \\
{}    &\sP_{i, \frac 1 2} \frac{ \y_{i-1, \frac 1 2} \y_{i+1, \frac 1 2}}{ \y_{i,1}} \arrow{dl}{i-1} \arrow{dr}{i+1}& {} \\
 \sP_{i,\frac 1 2} \sP_{i-1,1} \frac{\y_{i+1, \frac 1 2}}{\y_{i-1, \frac 3 2}} \arrow{dr}{i+1}         &    {}                         &         \sP_{i, \frac 1 2} \sP_{i+1, 1}  \frac{\y_{i-1, \frac 1 2}}{\y_{i+1,  \frac 3 2}}\arrow{dl}{i-1} \\
{} &   \sP_{i,\frac 1 2} \sP_{i-1, 1} \sP_{i+1, 1} \frac{\y_{i, 1}}{ \y_{i-1, \frac 3 2}  \y_{i+1, \frac 3 2}} \arrow{d}{i} &  {} \\
{} & \sP_{i, \frac 1 2} \sP_{i, \frac 3 2} \sP_{i-1, 1} \sP_{i+1, 1} \frac{1}{\y_{i,2}} & {} \\
\end{tikzcd}

where each node denotes a term in the character and each arrow with a label $j$ denotes the $q$-Weyl reflection in the quiver node $j$. The character $\chi_2$ for $A_3$ theory is also the
fundamental character of the vector representation for the $D_3$ theory. 

The $q$ characters $\chi_{i-1, \frac 1 2 }$ and $\chi_{i+1, -\frac 1 2 }$ are 

\begin{tikzcd}
   \y_{i-1, \frac 1 2}\arrow{d}{ i-1} \\
   \sP_{i-1, 1} \frac{ \y_{i,1}}{ \y_{i-1, \frac 3 2}}\arrow{d}{i} \\
   \sP_{i-1, 1} \sP_{i, \frac 3 2} \frac{ \y_{i+1, \frac 3 2}}{ \y_{i, 2}} \arrow{d}{i+1} \\
   \sP_{i-1, 1} \sP_{i, \frac 3 2} \sP_{i+1, 2} \frac{1}{ \y_{i+1, \frac 5 2}}
\end{tikzcd}
\begin{tikzcd}
   \y_{i+1, -\frac 1 2}\arrow{d}{ i+1} \\
   \sP_{i+1, 0} \frac{ \y_{i,0}}{ \y_{i+1, \frac 1 2}}\arrow{d}{i} \\
   \sP_{i+1, 0} \sP_{i, \frac 1 2} \frac{ \y_{i-1, \frac 1 2}}{ \y_{i, 1}} \arrow{d}{i-1} \\
   \sP_{i+1, 0} \sP_{i, \frac 1 2} \sP_{i-1, 1} \frac{1}{ \y_{i-1, \frac 3 2}}
\end{tikzcd}

Above we have defined the $q$-determinant operator $\boldsymbol{\Theta}_{i}$ 
\begin{multline}
\boldsymbol{\Theta}_{i} = \left (1 - D^{-\frac 1 2} \sP_{i,0}^{\stackrel{\leftarrow}{\frac 3 2}} \tfrac{1}{\y_{i - 1, 1}} D^{-\frac 1 2} \right)
\left  ( 1 -D^{-\frac 1 2} \sP_{i,0}^{\stackrel{\leftarrow}{\frac 1 2}} \tfrac{\y_{i-1, 0}}{ \y_{i, \frac 1 2 }} D^{-\frac 1 2} \right)
  \y_{i, 0}\\
\left  (1 -  D^{\frac 1 2 } \sP_{i,0}^{\stackrel{\rightarrow}{\frac 1 2}} \tfrac{ \y_{i+1, 0}}{ \y_{i, \frac 1 2}} D^{\frac 1 2}\right)
\left (1 - D^{\frac 1 2}  \sP_{i,0}^{\stackrel{\rightarrow}{\frac 3 2}}   \frac{1}{ \y_{i+1, 1}} D^{\frac 1 2}
\right)
\label{eq:A3-theta-product}
\end{multline}
and its expansion (\ref{eq:Theta_i-generic}) over the fundamental $q$-characters is 
\begin{equation}
  \boldsymbol{\Theta_i} =  \sP_{i,\frac 1 2}^{\stackrel{\leftarrow}{\frac 3 2}} \sP_{i,\frac 3 2}^{\stackrel{\leftarrow}{\frac 1 2}} D^{-2} - 
\sP_{i,\frac 1 2}^{\stackrel{\leftarrow}{\frac 1 2}} \chi_{i-1, \frac 1 2 }  D^{-1}  +   \chi_{i,0} -  \sP_{i, -\frac 1 2}^{\stackrel{\rightarrow}{\frac 1 2}} \chi_{i+1, -\frac 1 2 } D +  \sP_{i, -\frac 1 2}^{\stackrel{\rightarrow}{\frac 1 2}}   \sP_{i, -\frac 3 2}^{\stackrel{\rightarrow}{\frac 3 2}} D^2
\end{equation}

The operator $\boldsymbol{\Theta}_i$ annihilates the function $\mathbf{Q}_{i}(\xi)$ 
\begin{equation}
  \boldsymbol{\Theta}_i | \mathbf{Q}_i \rangle= 0
\end{equation}
Let $ \la \bQ_{i}^{(s)} \ra_{s \in (0,1,2,3)}$ be the basis of solutions for the $D$-spectral  equation (\ref{eq:q-spectral})
for $A_3$ theory,
 and let $\bQ_{i}^{(3)}$  be solution annihilated by the last factor in the (\ref{eq:A3-theta-product}). Then $\bQ_{i}^{(3)}$
 satisfies equation
\begin{equation}
\bQ_{i, \frac 1 2 }^{(3)} - \sP_{i,0}^{\stackrel{\rightarrow}{\frac 3 2}}  \frac{1}{ \y_{i+1, 1}} \bQ_{i, -\frac 1 2 }^{(3)} = 0
\end{equation}
as we can see from the last factor of (\ref{eq:A3-theta-product}). For a generic $A_{r}$ quiver the corresponding equation on $\bQ_r$ is
\begin{equation}
\bQ_{r,\frac 1 2} - \sP_{i,0}^{\stackrel{\rightarrow}{\frac 1 2 + r - i}}  \frac{1}{ \y_{r, \frac 1 2 (1+r-i)}} \bQ_{r, - \frac  1 2 } = 0
\end{equation}
or, equivalently
\begin{equation}
\label{eq:bfQr}
\frac{  \bQ_{r, -\frac 1 2}} 
       { \bQ_{r, +\frac 1 2}} = \frac { \y_{r, \frac 1 2 (1+r-i)}} {\sP_{i,0}^{\stackrel{\rightarrow}{\frac 1 2 + r - i}}}
\end{equation}

We conclude, that up to a product of $\gamma_q$ factors coming from the $\sP_{i,0}^{\stackrel{\rightarrow}{\frac 1 2 + r - i}}$ in the above equation and a simple shift, 
 the function $\bQ_r(\xi)$ is the function $Q_r(\xi)$ (see  (\ref{eq:Baxter})) associated to the right terminal node  $r$ of the linear quiver 
 \begin{equation}
  \bQ_{r}= \gamma_{q,\sP_{i,0}^{\stackrel{\rightarrow}{\frac 1 2 + r - i}}}  Q_{r, \frac 1 2 (1 + r  - i)}
 \end{equation}

\subsection{From Baxter equation to $Y$-functions from
  $D$-Wronskian}. 
Here  we will recover relation to the $\y$ and $Q$ functions of (\ref{eq:Baxter}) from the solutions 
to the $D$-spectral equation (\ref{eq:q-spectral}), see
\cite{Bazhanov:1998dq,Kuniba:2010ir} for details.

Define operator $L_k$ to be the product of $k$ right factors in the
$\boldsymbol{\Theta}_{i}$. 

For $ k \leq  r - i + 1 $ we find 
\begin{equation}
  L_{k} = \prod_{j = r - i + \frac 1 2 -(k-1)}^{r - i + \frac  12 } \left(1  -  D^{\frac 1 2 } \sP_{i, 0}^{
      \stackrel{\rightarrow}{j}}  
    \frac{  \y_{i + j + \frac 1 2, -\frac 1 4 + \frac 1 2 j} } { \y_{i + j
        - \frac 1 2, \frac 1 4 + \frac 1 2 j}}
 D^{\frac 1 2} \right)
\end{equation}
Let  $\la \bQ^{(r-j;k)}_{i} \ra_{j \in \{0,1,\dots,k-1\}} $ be the basis in
the kernel of $ L_{k}$
\begin{equation}
  \ker L_k = \oplus_{j=0}^{k-1} \BC \bQ_{i}^{(r - j;k)}
\end{equation}
Let $W_k$ be the $D$-Wronskian built on functions $\la \bQ^{(r-j;k)}_{i}
\ra_{j \in \{0,1,\dots, k-1\}} $
\begin{equation}
\label{eq:DWronskian}
  W_{k} = \det 
  \begin{pmatrix}
    \bQ^{(r-j;k)}_{i,j'} 
  \end{pmatrix}_{\substack{j \in  \{0,1,\dots, k-1\} \\ {j'\in
      \{0,1,\dots, k-1\}}}}
\end{equation}

Expand $L_k$ in the powers of $D$ 
\begin{multline}
L_k = 
1 + \sum_{j=1}^{k-1} c_j D^{j}  +  \\
    + \frac{(-1)^k}{ \y_{ r- k+1,   \frac 12 ( r - i  - k+1 )}} \left( \prod_{j = r - i + \frac 3 2 -k}^{r - i + \frac  12 }
  \sP_{i, -j + ( r - i -(k-1))}^{\stackrel{\rightarrow}{j}} \right) \, D^{k} 
\end{multline}
where $c_j$ are certain polynomials in $\y$ and $\sP$. The  equations
$(L_k \bQ^{(r-j;k)}_{i} = 0)_{j \in \{0,1,\dots, k-1\}}$  together with determinant 
definition (\ref{eq:DWronskian}) implies 
\begin{equation}
\left( 1  -  \frac{1}{ \y_{ r- k+1,   \frac 12 ( r - i  -k+1 )}} \left( \prod_{j = r - i + \frac 3 2 -k}^{r - i + \frac  12 }
  \sP_{i,- j + r - i - k+1}^{\stackrel{\rightarrow}{j}} \right)  D \right)   W_{r - k+1} = 0
\end{equation}
so that finally
\begin{equation}
\label{eq:kright}
\frac{   W_{r - (k-1), -1} } { W_{r - (k-1), 0 }} = \frac { \y_{ r- (k-1),  \frac 1 2 (r - i  -(k-1) )}}
{\prod_{j = r - i + \frac 3 2 -k}^{r - i + \frac  12 }
  \sP_{i, -j + ( r - i -( k - 1))}^{\stackrel{\rightarrow}{j}}}
\end{equation}
For $k > r - i + 1 $ we  find 
\begin{multline}
  L_k = (-1)^{r-i - k+1} \y_{r - k+1, -\frac 12 ( r - i - k+1)} \, \left( \prod_{j=1}^{k-1  - r+i} 
\sP^{\stackrel{\leftarrow}{ j - \frac 12 }}_{ i, k - r + i  - j-\frac 12} \right)
D^{r - i - k+1} + \\ + \sum_{j= r - i -k+2}^{ r - i} c_j D^{j} \ +  \\  (-1)^{r - i + 1} \left( \prod_{j= \frac  12 }^{r - i + \frac 12} \sP_{i,-j}^{\stackrel{\rightarrow}{j}} \right) D^{r - i + 1}
\end{multline}
The  equations
$(L_k \bQ^{(r-j;k)}_{i} = 0)_{j \in \{0,1,\dots, k-1\}}$ and 
 the determinant definition (\ref{eq:DWronskian}) imply 
\begin{equation}
  \frac{ W_{r - k +1 , k -  r + i -2}} { W_{r - k+1, k -r + i -1}} = 
\frac { \y_{r - k +1, -\frac 12 ( r - i - k+1)}  \left( \prod_{j=1}^{ k+i-r -1} 
\sP^{\stackrel{\leftarrow}{ j - \frac 12 }}_{ i,  k-\frac 12  - r+ i - j}\right)} 
{ \left( \prod_{j=\frac  1 2 }^{r - i + \frac 12} \sP_{i,-j}^{\stackrel{\rightarrow}{j}} \right) }
\end{equation}
or, equivalently, 
\begin{equation}
\label{eq:kleft}
    \frac{ W_{r - k + 1, -1}} { W_{r - k+1, 0}} = 
\frac { \y_{r - k+1, \frac 12 ( r - i - k+1)}  \left( \prod_{j=1}^{k-r-1+i} 
\sP^{\stackrel{\leftarrow}{ j - \frac 12 }}_{ i,- j + \frac 12 }\right)} 
{ \left(\prod_{j=\frac  12 }^{r - i + \frac 12} \sP_{i,-j + r - i  - k+1}^{\stackrel{\rightarrow}{j}}\right) }
\end{equation}
This formula \eqref{eq:kright} and \eqref{eq:kleft}  together with the $D$-Wronskian 
definition (\ref{eq:DWronskian}) allows us to compute all functions $\y_{j}$ 
 from the complete basis in the $\ker \boldsymbol{\Theta}_i$.

\subsection{ $\hat A_{0}^{*}$ quiver}

Consider  the type II* theory corresponding to the ${\hat A}_{0}$ quiver, 
i.e. the ${\CalN}=2^*$ theory with the gauge group $SU(N)$, where $N = {\bv}_{0}$.
There is one fundamental $q$-character in this case:
\begin{equation}
\begin{aligned}
& {\chi}({\xi}) = \y ({\xi} ) \sum_{\lambda} {\qe}^{|{\lambda}|}\, \prod_{{\square} \in {\lambda}} \frac{\y({\sigma}_{\square} {\mu}  {\xi}) \y({\sigma}_{\square} {\mu}^{-1}  {\xi})}{\y({\sigma}_{\square} q^{\frac 12} {\xi}) \y({\sigma}_{\square}  q^{-\frac 12} {\xi})} 
\\
& \qquad =    \sum_{\lambda} {\qe}^{|{\lambda}|}\,  \frac{\prod_{{\square} \in {\partial}_{+}{\lambda}} \y( {\sigma}_{\square}q^{\frac 12} {\xi})}{\prod_{{\blacksquare} \in {\partial}_{-}{\lambda}} \y( {\sigma}_{\blacksquare}q^{-\frac 12} {\xi})} \ , \\
& \\
\end{aligned}
\label{eq:chi1}
\end{equation}
 Where for ${\square} = (i,j) \in \lambda$,  
\begin{equation}
\begin{aligned}
 & \qquad {\sigma}_{\square} = {\mu}^{j-i} q^{\frac{1-i-j}{2}} \ , \ \textrm{and} \\
 & 
1 - \sum_{{\square} \in {\lambda}} {\sigma}_{\square} \left( q^{\frac 12} +  q^{-\frac 12} - {\mu} - {\mu}^{-1}  \right) = \\
& \qquad\qquad\qquad = \sum_{{\square} \in {\partial}_{+}{\lambda}} {\sigma}_{\square} q^{\frac 12} - \sum_{{\blacksquare} \in {\partial}_{-}{\lambda}} {\sigma}_{\blacksquare} q^{-\frac 12}
\end{aligned}
\label{eq:genrel}
\end{equation}
The \emph{outer} and \emph{inner} boundaries $\partial_{\pm}{\lambda}$ correspond to the generators and relations of the monomial ideal $I_{\lambda}$ corresponding to the partition $\lambda$. 

Let us now present the \emph{quantization} of the curve $R(t,x) =0$ found in \cite{NP2012a} for this theory. 

Introduce the notation (not to be confused with \eqref{eq:nota}):
\begin{equation}
y_{i,j} = \y ({\xi} q^{-\frac{i+j}{2}} {\mu}^{j-i} ) , \qquad i,j \in {\BZ}
\label{eq:yij}
\end{equation}
Then \eqref{eq:chi1} can be rewritten in the form:
\begin{equation}
{\chi} = \sum_{{\lambda}} {\qe}^{|{\lambda}|} y_{{\ell}_{\lambda}, 0} \prod_{j=1}^{{\ell}_{\lambda}}
\frac{y_{j-1, {\lambda}_{j}}}{y_{j, {\lambda}_{j}}} 
\label{eq:chi2}
\end{equation}
Now let us introduce the operator
\begin{equation}
U = q^{-{\xi}{\partial}_{\xi}}
\label{eq:top}
\end{equation}
such that
\begin{equation}
U y_{i,j} U^{-1} = y_{i+1, j+1}
\label{eq:tyij}
\end{equation}
Introduce also the notations
\begin{equation}
g_{j} = \frac{y_{-1,j}}{y_{0,j}}, \qquad {\tilde g}_{j} = \frac{y_{j,-1}}{y_{j,0}}, \qquad\qquad j \in {\BZ}
\label{eq:gj}
\end{equation}
Let us also introduce the shift operators
\begin{equation}
U_{\pm} = \left( {\mu} q^{{\pm} \half} \right)^{-{\xi}{\partial}_{\xi}}, \qquad U_{+} = U U_{-}
\end{equation}
which commute with $U$. 
The quantities $g_{j}, {\tilde g}_{j}$ can be written also in the
following form:
\begin{equation}
g_{j} = U_{-}^{-j} g_{0} U_{-}^{j}, \qquad {\tilde g}_{j} = U_{+}^{j} {\tilde g}_{0} U_{+}^{-j}
\end{equation}
Using $U$, $g$, ${\tilde g}$ we can present $\chi$ in a more suggestive form:
\begin{equation}\begin{aligned}
& {\chi} = \sum_{{\lambda}} {\qe}^{|{\lambda}|} y_{{\ell}_{\lambda}, 0} \prod_{i=1}^{{\ell}_{\lambda}} U^{i}\, g_{{\lambda}_{i}-i} \, U^{-i} \\
& \qquad = \sum_{\lambda} {\qe}^{|{\lambda}|} \, y_{{\ell}_{\lambda}, 0}\, 
U g_{{\lambda}_{1}-\scriptscriptstyle{1}} U g_{{\lambda}_{2} - \scriptscriptstyle{2}} \ldots U g_{{\lambda}_{{\ell}_{\lambda}}-\scriptscriptstyle{{\ell}_{\lambda}}} U^{-{\ell}_{\lambda}} \\
& \qquad\qquad = \sum_{\lambda} {\qe}^{\frac{{\ell}_{\lambda}^{2}}2} \, \, \left( \prod_{i=1}^{\longrightarrow\, {\ell}_{\lambda}}
{\qe}^{{\lambda}_{i}-i + \frac 12} U g_{{\lambda}_{i}-\scriptscriptstyle{i}} \right) y_{0, -{\ell}_{\lambda}} U^{-{\ell}_{\lambda}}\\
& \qquad\qquad\qquad\qquad = {\text{Coeff}_{U^{0}}} {\bf \Theta} ({\xi}, {\qe}; U) \end{aligned}
\label{eq:chi3}
\end{equation}
with the \emph{$q$-vertex operator} ${\bf \Theta}$, 
\begin{equation}
{\bf \Theta} ({\xi}, {\qe}; U) =  {\bf \Theta}_{-} \y({\xi}) {\bf \Theta}_{+} \end{equation}
whose factors are given by the ordered infinite products:
\begin{equation}\begin{aligned}
& {\bf \Theta}_{-} = \prod_{r \in {\BZ}_{\geq 0}+ \frac 12}^{\longleftarrow} \left( 1 + {\qe}^{r} U g_{r - \frac 12}  \right) \\
&  {\bf \Theta}_{+} =  \prod_{r \in {\BZ}_{\geq 0}+ \frac 12}^{\longrightarrow} \left( 1 + {\qe}^{r} {\tilde g}_{r-\frac 12} U^{-1} \right) \end{aligned}\label{eq:vvert}\end{equation}
The notations
$$
\prod^{\longleftarrow}\ , \qquad  \prod^{\longrightarrow}
$$ for the noncommutative products
read as follows:
\begin{equation}
\prod^{\longleftarrow}_{r \in {\BZ}_{\geq 0}+ \frac 12} a_{r} = \ldots a_{\frac 52}a_{\frac 32} a_{\frac 12}\ , \qquad \prod^{\longrightarrow}_{r \in {\BZ}_{\geq 0}+ \frac 12} {\tilde a}_{r} = {\tilde a}_{\frac 12} {\tilde a}_{\frac 32} {\tilde a}_{\frac 52} \ldots 
\end{equation} 
The vertex operator ${\bf \Theta}$ is the noncommutative analogue of the theta-product:
\begin{equation}
{\vt} (t; {\qe}) \equiv 
\sum_{n \in {\BZ}} {\qe}^{\frac{n^2}{2}} t^{n} = {\phi}({\qe}) \prod_{r \in {\BZ}_{\geq 0} + \frac 12} \left( 1 + t {\qe}^{r} \right) \left( 1 + t^{-1} {\qe}^{r} \right)
\label{eq:thetaprod}
\end{equation}  Just like the theta-function ${\vt}(t; {\qe})$, the operator ${\bf \Theta}$ can be expanded as a $U$-series:
\begin{equation}
{\bf \Theta}({\xi}, {\qe}; U) = \sum_{n \in {\BZ}} {\qe}^{\frac{n^2}2} {\chi}_{n}({\xi}) \, U^{n}
\label{eq:vth}
\end{equation}
where
\begin{equation}
{\chi}_{n}({\xi}) = {\chi} ({\xi} {\mu}_{-}^{n}) , \qquad {\mu}_{\pm} = {\mu} q^{\pm \frac 12}
\label{eq:chinx}
\end{equation}
Now let us define the function ${\tilde{\bf Q}}({\xi})$ by the property:
\begin{equation}
\left( 1 + {\qe}^{\frac 12} U g_{0} \right) y_{0,0} {\tilde{\bf Q}}= 0 \qquad
\Leftrightarrow\qquad {\qe}^{\frac 12} {\tilde{\bf Q}}({\xi}q^{-1}) \y ({\xi} {\mu}_{-}) = - \y ({\xi}){\tilde{\bf Q}}({\xi})
\label{eq:qqst}
\end{equation}
It is related to the $Q$-function
\begin{equation}
\y({\xi}) = \frac{Q({\xi})}{Q({\xi}q^{-1})}
\label{eq:qbax}
\end{equation}
in an obvious manner. 
Define
\begin{equation}
{\bf Q}({\xi}) = \left( {\bf \Theta}_{+}^{-1} {\tilde{\bf Q}} \right) ({\xi})
\end{equation}
Then ${\bf \Theta} ( {\xi}, {\qe}; U) {\bf Q}({\xi}) = 0$, which is an infinite order linear difference equation:
\begin{equation}
\sum_{n \in \BZ} {\qe}^{n^2 \over 2} {\chi} ( {\xi} {\mu}_{-}^{n} ) {\bf Q} ({\xi} q^{-n}) = 0
\label{eq:hir}
\end{equation}
which is an analogue of the Hirota difference equation \cite{Krichever:1996qd}. 
We discuss the four-dimensional limit of the equation \eqref{eq:hir} in
the section below.

\section{Four dimensional limit and the opers.}

\subsection{The $A_1$ case.} 

The $q$-character equation \eqref{eq:chi1A1}, up to non-essential $\ep$ shifts in
the variable $x$, assumes the following form in the four dimensional limit:
\begin{equation}
\y \left( x+ \frac {\ep} 2 \right) + {\qe}  \frac{\mathring{\sP}(x)}{\y \left(x- \frac \ep 2 \right)} = (1 + {\qe} ) T(x)
\label{eq:yqchar}
\end{equation}
with the degree $N_f = 2N$ monic polynomial $\mathring{\sP}(x)$ and
degree $N$ monic polynomial $T(x)$. The roots of polynomial
$\mathring{\sP}(x)$ are the  masses of the fundamental hypermultiplets, while the polynomial $T(x)$
\begin{equation}
T(x) = x^{N} + u_{1} x^{N-1} + u_{2}x^{N-2} + \ldots + u_{N} 
\label{eq:tofu}\end{equation} determines the vacuum of the theory. Now let us factorize the fundamental polynomial
\begin{equation}
\mathring{\sP}(x) = A(x-{\ep})D(x), \label{eq:twomass}
\end{equation}
where $A(x)$, $D(x)$ are degree $N$ monic polynomials. Obviously, there are many representations like \eqref{eq:twomass}. 
Define {\it Baxter function} $\bQ(x)$ to be an entire function of $x$,
such that
(c.f. with (\ref{eq:bfQvsQ}) where the partial matter polynomial $A(x)$ corresponds to $\sP_{1, 0}^{\stackrel{\rightarrow}{\frac    1 2}}(\xi)$)
\begin{equation}
\y(x) = A \left( x - \frac {\ep} 2 \right) \frac{\bQ \left( x+\frac \ep 2 \right)}{\bQ \left(x-\frac {\ep} 2 \right)}
\label{eq:qfroy}
\end{equation}
Comparing with (\ref{eq:Yi}) we find
\begin{equation}
  A(x) \frac{\bQ(x+\ep)}{\bQ(x)} = \frac{Q(x+\ep)}{Q(x )}
\end{equation}
therefore 
\begin{equation}
  \bQ(x) = \frac{Q(x)}{\Gamma_{A;\ep}(x)}
\end{equation}
where for a monic polynomial $A(x)$ the $\Gamma_{A;\ep}(x)$ function
satisfies 
\begin{equation}
  \Gamma_{A;\ep}(x + \ep) = A(x) \Gamma_{A;\ep}(x)
\end{equation}
Up to an $\ep$-periodic function the $\Gamma_{A;\ep}(x)$ function is a product of ordinary $\Gamma_{\ep}$-functions
\begin{equation}
   \prod_{f=1}^{N} \Gamma_{\ep}(x - m_f)
\end{equation}
its inverse is an entire function with zeroes at $m_{f} - \ep \BZ_{\geq
  0}$. 
The zeroes ${\xi}_{{\ba}, i}$ of $Q(x)$ behave as 
\begin{equation}
{\xi}_{{\ba}, i} = a_{\ba} + {\ep}(i-1) + x_{{\ba}, i}, \qquad x_{{\ba},i} \propto {\qe}^{i} \to 0, i \to \infty
\end{equation}
The factor $\Gamma_{A;\ep}(x)$ adds a string of zeroes going in the
opposite direction
\begin{equation}
  m_{f} - k \ep, \quad k \geq 0
\end{equation}
If $m_\ba - \ac_{\ba} \in \ep \BZ_{\geq 0}$, then an infinite string of zeroes
$m_{\ba} + \ep \BZ$ going in both directions can be removed from
$\bQ(x)$ by factoring out the periodic function $\sin \frac{\pi}{\ep} (x
- m_{\ba})$, this is the case studied in
\cite{Dorey:2011pa,Chen:2011sj}, in which $\bQ(x)$ becomes polynomial.

The equation \eqref{eq:yqchar} becomes the celebrated $T\bQ$-relation
(c.f. eq. (\ref{eq:baxter2}):
\begin{equation}
A(x)\bQ(x+{\ep}) + {\qe} D(x) \bQ(x-{\ep}) = (1+{\qe})T(x)\bQ(x)
\label{eq:tq}\end{equation}
Now define:
\begin{equation}
{\bf\Psi}(t) = \sum_{x \in \Gamma} \bQ(x) t^{-x/{\ep}}
\label{eq:psioft}
\end{equation}
where $\Gamma \subset {\BC}$ is some lattice, invariant under the shifts by $\ep$: ${\Gamma} + {\ep} = {\Gamma}$. Then \eqref{eq:tq} implies:
\begin{equation}
\left\{ A( - {\ep}t{\pa}_{t} ) t + {\qe} D( - {\ep} t{\pa}_{t} ) t^{-1} - ( 1 + {\qe}) T( - {\ep}t{\pa}_{t})  \right\} {\bf\Psi}(t) = 0
\label{eq:psioper}
\end{equation}
i.e. $\bf\Psi$ is a solution of the $N$-order differential equation with $4$ regular singularities $t = 0, {\qe}, 1, \infty$. The monodromy of \eqref{eq:psioper} around $t= 0$ and $t = \infty$
has $N$ distinct eigenvalues $e^{2{\pi}{\ii} m_{i}^{\pm}/{\ep}}$, $i=1, \ldots , N$, 
where $m^{\pm}$ are the roots of $D(x)$ and $A(x)$, respectively. 
The monodromy of \eqref{eq:psioper} around $t = {\qe}, 1$ has $N-1$ eigenvalues equal to $1$, and one non-trivial eigenvalue $e^{2{\pi}{\ii} {\mu}^{\pm}/{\ep}}$, respectively, where ${\mu}^{\pm}$ can be computed from the $U(1)$ Coulomb moduli $u_{1}$ \eqref{eq:tofu} and $m_{i}^{\pm}$.

\subsection{Other examples}

\subsubsection{$q$-characters for the $A_2$ theory}

We have two $\y$-functions and two matter polynomials $\sP_{1,2}(x)$, 
\begin{equation}
\sP_{i}(x) = (x - m_{i}^{+})(x - m_{i}^{-})\ . 
\label{eq:piofx}
\end{equation}
The equations determining the vacua are:
\begin{equation}
\begin{aligned}
& \y_{1} \left( x-\frac \ep 2 \right) + {\qe}_{1} \frac{\sP_{1}(x-\ep ) \y_{2}(x- 
  \ep  )}{\y_{1}(x -\frac 3 2 \ep )} + 
{\qe}_{1}{\qe}_{2} \frac{\sP_{1}(x -  \ep )\sP_{2}(x- \frac 3 2 \ep
  )}{\y_{2}(x-2 {\ep})}  = \\
   & \qquad\qquad\qquad\qquad =  ( 1 + {\qe}_{1} + {\qe}_{1}{\qe}_{2} )
T_{1} \left( x-\frac \ep 2 \right) \\
& \y_{2}\left( x + \frac \ep 2 \right) + {\qe}_{2} \frac{\sP_{2}(x) \y_{1}(x)}{\y_{2}\left( x-\frac \ep 2 \right)} + {\qe}_{1}{\qe}_{2} \frac{\sP_{1}\left( x-\frac \ep 2 \right)
  \sP_{2}(x)}{\y_{1}(x- {\ep})} =  \\
   & \qquad\qquad\qquad\qquad = ( 1 + {\qe}_{2} + {\qe}_{1}{\qe}_{2} ) T_{2}\left( x + \frac \ep 2 \right) \end{aligned}
\label{eq:qcha}
\end{equation}
where $T_{1,2}(x)$ are the monic degree $2$ polynomials:
\begin{equation}
T_{i}(x) = x^{2} - u_{i,1}x - u_{i,2}, \qquad i=1,2
\end{equation}
The parameters $u_{i,1}$ are the $U(1)$ Coulomb moduli, they are determined by the masses of the fundamental hypermultiplets and the mass of the bi-fundamental hypermultiplet. 

Shift the argument of the first equation in \eqref{eq:qcha} as 
$x \mapsto x + {\ep}/2$, multiply the result
by  $\qe_2 \sP_2(x)/\y_2\left( x-\frac \ep 2 \right)$ and subtract the result from
the second equation in \eqref{eq:qcha} to obtain
\begin{equation}
\begin{aligned}
 \label{eq:singl}
& \y_{2}\left( x + \frac \ep 2 \right) + ( 1 + {\qe}_{1} + {\qe}_{1}{\qe}_{2}
){\qe}_{2}\frac{\sP_{2}(x)T_{1}(x)}{\y_{2}\left( x-\frac \ep 2 \right)} - \\
& \qquad - 
{\qe}_{1}{\qe}_{2}^{2}
\frac{\sP_{1}\left( x-\frac \ep 2 \right)\sP_{2}(x - \ep )\sP_{2}(x)}{\y_{2} \left( x-\frac{3{\ep}}{2} \right)\y_{2}\left(x -
  \frac {\ep}2 \right)} = ( 1 + {\qe}_{2} + {\qe}_{1}{\qe}_{2} )  T_{2}\left( x + \frac {\ep}2 \right) 
\end{aligned}
\end{equation}
Now, write (cf. \eqref{eq:bfQr})
\begin{equation}
\y_{2}(x) = \sP_{2} \left( x+\frac {\ep} 2 \right) \frac{\bQ(x+\frac {\ep} 2 )}{\bQ(x - \frac {\ep} 2)} 
\label{eq:bax}
\end{equation}
to reduce \eqref{eq:singl} to the linear equation (cf. \eqref{eq:Theta_i-generic}, \eqref{eq:q-spectral}): 
\begin{multline}
\sP_{2}(x+{\ep}) \bQ(x+{\ep})
-( 1 + {\qe}_{2} + {\qe}_{1}{\qe}_{2} )  T_{2}(x + \tfrac 1 2 \ep) \bQ(x)  \\
 + {\qe}_{2} ( 1 + {\qe}_{1} + {\qe}_{1}{\qe}_{2} )T_{1}(x) \bQ(x - {\ep})
- {\qe}_{1}{\qe}_{2}^{2} \sP_{1}(x - \tfrac 1 2 {\ep}) \bQ(x - 2 \ep) = 0
\label{eq:baxd}
\end{multline}
Now let us make a Fourier transform:
\begin{equation}
{\boldsymbol{\Psi}}(t) = \sum_{x} \bQ(x) t^{-x/{\ep}}
\end{equation}
where the sum goes over some lattice $x \in a + {\BZ}{\ep}$.  Now the substitution: 
\begin{equation}
\label{eq:ftx}
x \mapsto  - {\ep} t D_{t}, \qquad e^{{\ep}D_{x}} \mapsto t 
\end{equation}
maps the Eq. \eqref{eq:baxd} to the second order differential equation, obeyed by the
function ${\bf \Psi}(t)$
\begin{equation}
\begin{aligned}
& \left[ t^3 \sP_{2}( - {\ep} tD_{t}  ) 
- ( 1 + {\qe}_{2} + {\qe}_{1}{\qe}_{2} )  t^2 T_{2}( - {\ep}tD_{t}  - \tfrac \ep 2) 
 + \right. \\
& \left.\qquad + t {\qe}_{2}(1+{\qe}_{1} + {\qe}_{1}{\qe}_{2}) T_{1} ( -
  {\ep}  tD_{t} - \ep ) 
 -  {\qe}_{1}{\qe}_{2}^{2} \sP_{1} ( - {\ep}tD_{t} - \tfrac 3 2 {\ep} ))   \right] {\bf \Psi}(t) = 0 \end{aligned}
 \label{eq:diffbax}
\end{equation}
The symbol of the left hand side vanishes at the zeroes of 
\begin{equation}
t^3  - ( 1+ {\qe}_{2} + {\qe}_{1}{\qe}_{2}) t^2  + {\qe}_{2}( 1+ {\qe}_{1} + {\qe}_{1}{\qe}_{2}) t -
{\qe}_{1}{\qe}_{2}^{2}  = (t-1)(t-{\qe}_{2})(t- {\qe}_{1}{\qe}_{2})
\end{equation}
There are, then, five regular singularities, at $t = 0, {\qe}_{1}{\qe}_{2}, 
{\qe}_{2}, 1, \infty$. Near $t = 0$ the dominant term is 
${\qe}_{1}{\qe}_{2}^{2} \sP_{1} ( - {\ep} tD_{t} - \frac 3 2 {\ep})$,
 the eigenfunctions of the monodromy of the Eq. \eqref{eq:diffbax} around $t = 0$ are
\begin{equation}
{\psi}^{\pm}_{(0)} \sim t^{-m_{1}^{\pm}/{\ep} - \frac 3 2 } \left( 1 + O(t) \right)
\label{eq:psipmz}
\end{equation}
and the corresponding eigenvalues are
\begin{equation}
e^{-2\pi {\ii} (m_{1}^{\pm}/{\ep} + \frac 3 2)}
\end{equation}
Analogously, near $t = \infty$ the dominant term is $t^3 \sP_{2}( - {\ep} tD_{t}  )$, and the eigenfunctions are
\begin{equation}
 {\psi}^{\pm}_{({\infty})} \sim t^{-m_{2}^{\pm}/{\ep}} \left( 1 + O(t^{-1}) \right)
 \label{eq:psipmi}
 \end{equation}
with the corresponding eigenvalues
\begin{equation}
e^{- 2\pi \ii m_{2}^{\pm}/{\ep}}
\end{equation}
Near the points $t = z_{i}$, $z_{1}= {\qe}_{1}{\qe}_{2}, z_{2} = {\qe}_{2}, z_{3}=1$ the equation
\eqref{eq:diffbax} has one singular solution
\begin{equation}
(t - z_{i})^{-{\mu_i}/{\ep}}
\end{equation}
and one non-singular solution. The monodromy around these points has, therefore, one trivial eigenvalue $1$ and one non-trivial eigenvalue
\begin{equation}
e^{-2\pi \ii {\mu_i}/{\ep}}, \qquad i = 1, 2,3
\end{equation}
The exponents ${\mu}_{i}$ are determined 
by the $U(1)$ Coulomb moduli and the masses, and in fact can be used as convenient parameters instead of $u_{1,1}, u_{2,1}$, 
subject to the obvious relation
\begin{equation}
\sum_{i=1}^{3} {\mu}_{i}  =   m_{2}^{+} + m_{2}^{-} - m_{1}^{+} -
m_{1}^{-} - 3  \ep 
\end{equation} 
The redefinition
\begin{equation}
{\Psi} (t) = t^{\frac{-m_{1}^{+}-m_{1}^{-} - 3 \ep}{2{\ep}}} \prod_{i=1}^{3} (t - z_{i})^{-\frac{{\mu}_{i}}{2\ep}} \, {\chi}(t)
\end{equation}
maps \eqref{eq:diffbax} to the $\mathfrak{sl}_2$ oper 
(c.f. \cite{Beilinson:2005,Beilinson,Beilinson:Hecke,Frenkel:1995zp,Frenkel:1996,Frenkel:2003qx,Frenkel:2003})
\begin{equation}
\left[ {\ep}^{2} {\partial}_{t}^{2}  + T(t) \right] \, {\chi} = 0
\label{eq:Oper}
\end{equation}
where 
\begin{equation}
\begin{aligned}
& T = \sum_{i=0}^{3} \frac{{\Delta}_{i}}{(t - z_{i})^2} + \frac{c_{i}}{t - z_{i}} \\
& z_{0} = 0, z_{1} = {\qe}_{1}{\qe}_{2}, z_{2} = {\qe}_{2}, z_{3}=1, z_{4} = \infty \\
& \end{aligned}
\label{eq:ooper}
\end{equation}

{\remark Observe that the equation obeyed by ${\bf\Psi}(t)$ is the ${\ep}_{2}\to 0$ limit of the equation obeyed by the degenerate conformal block of $A_{N-1}$ Toda conformal field theory, with $b^2 = {\ep}_{2}/{\ep}_{1}$, corresponding to the $5$-punctured sphere. According to the AGT correspondence this is a ${\CalZ}$-partition function of the linear quiver theory of type $A_2$. It should be interesting to understand the relation between ${\bf\Psi} (t)$ and ${\CalZ} ({\ba}, {\bm}; {\qe}, t, {\ep})$ in more general context. See \cite{Litvinov:2013} for some developments in this direction. }

\subsubsection{Type II* ${\hat A}_{0}$ case.}

We claim that in the four dimensional limit the equation \eqref{eq:hir}
can be transformed to an interesting differential equation. Recall that
our master equations state that ${\chi}({\xi}) = T(\xi)$ where $T(\xi)$ is a (Laurent)
polynomial  of degree $N$ for the ${\bv}_{0} = N$ theory. In the four dimensional limit, $T$ is a degree $N$ polynomial in $x$ (recall that ${\xi} = e^{\betir x}$):
\begin{equation}
T(x) = \frac{1}{{\phi}({\qe})} \left( x^{N} + u_{1}x^{N-1} + \ldots + u_{N} \right)
\label{eq:chipol}
\end{equation}
where we used the standard notation
$$
{\phi}({\qe}) = \prod_{n=1}^{\infty} ( 1 - {\qe}^{n} ), \qquad \frac{1}{{\phi}({\qe})} = \sum_{\lambda} {\qe}^{| {\lambda} |}  
$$
The equation \eqref{eq:hir} adapts to, again in the four dimensional limit, 
\begin{equation}
\sum_{n \in {\BZ}} T (x + n m_{-}) {\bf Q} ( x  - n {\ep} ) {\qe}^{\frac{n^2}{2}}  = 0\, ,
\label{eq:hir2}
\end{equation}
where
$$
m_{\pm} = m \pm \frac{\ep}{2}
$$
Now let us perform the Fourier transform:
\begin{equation}
{\bf\Psi}(t) = {\vt}(t ; {\qe})^{-\frac{m_{+}}{\ep}} t^{-\frac{u_{1}}{N{\ep}}}
\sum_{x \in a + {\BZ} {\ep}} t^{\frac{x}{\ep}} {\bf Q}(x)
\end{equation}
assuming this sum converges. Then \eqref{eq:hir2} becomes the differential equation:
$$
{\CalD}{\bf\Psi}(t) = 0
$$
where 
\begin{equation}
\begin{aligned}
& {\CalD} = \frac{1}{{\vt}(t ; {\qe})}\sum_{n \in {\BZ}}
{\qe}^{\frac{n^2}{2}} t^{n} T \left(  {\ep} t {\partial}_{t}  - \frac{u_{1}}{N} + (n - t
 {\partial}_{t} {\log}{\vt}(t ; {\qe})) m_{+} \right)  \\
 & \qquad = ( {\ep} t {\partial}_{t} )^N + v_{2}(t) ({\ep} t{\partial}_{t})^{N-2} + \ldots + v_{N} (t) \end{aligned}
\label{eq:hirdiff}
\end{equation}
is the $N$-order differential operator
on the torus $E_{\qe} = {\BC}^{\times} / {\qe}^{\BZ}$ with one puncture, which is at
$t = -{\qe}^{\frac 12}$ in our normalization. The coefficients $v_{i}(t)$ are $\qe$-periodic, $v_{i}({\qe}t) = v_{i}(t)$, and have the $i$'th order pole at $t = - {\qe}^{\frac 12}$.  In line with \cite{Nekrasov:2010ka,Nekrasov:2011bc}  we should view $\CalD$ as the $\mathfrak{pgl}_N$-oper. 

\begin{example}
Let us now consider the small $N$ cases. For $N=1$ the Eq. \eqref{eq:hirdiff}
reads:
\begin{equation}
{\ep} t{\partial}_{t}  {\bf\Psi} = 0 
 \Longrightarrow {\bf\Psi}(t) = \text{const}
 \end{equation}
For $N=2$ the Eq. \eqref{eq:hirdiff}
 is equivalent to:
\begin{equation}
 \left( {\ep} t{\partial}_{t} \right)^2 {\bf\Psi}(t) + \left( m_{+}m_{-} \left(  t{\partial}_{t} \right)^2  {\log}{\vt}(t; {\qe}) + u_{2} - u_{1}^2/4 \right) {\bf\Psi}(t) = 0
\label{eq:elloper1}
\end{equation}
\end{example}

\subsubsection{$D$ and $E$ quivers}
For the $\gq = D_{r}$, the quantum determinant equation, or the
linear QT-equation like (\ref{eq:q-determinant}) was
found in \cite{Kuniba:2002}, but for the exceptional series
$E_{6},E_{7},E_{8}$ the linearization problem seems to be open. The
$E_r$-series  $(q,t)$-characters  were computed by Nakajima
\cite{Nakajima:2010} on supercomputer using sophisticated
optimization.

\appendix

\section{Quantum affine algebras\label{se:quantum-affine}}

\subsection{Quantum affine algebra}
In this section we collect the conventions and the definitions related
to quantum affine algebras \cite{Drinfeld:1986, Jimbo:1985} and the $q$-characters 
\cite{Frenkel:1998}.

\subsubsection{Conventions}
Literature differs in the conventions for the  $q$-parameter of quantum
groups. 
The parameter $q$ of the present note relates to the definitions of
$q$-characters by Frenkel and
Reshetikhin \cite{Frenkel:1998} and Drinfeld \cite{Drinfeld:1986} and
Jimbo \cite{Jimbo:1985} and \cite{MR1062425}
\begin{equation}
\label{eq:rel1}
e^{\betir  \ep_1}   = q_{\mathrm{present}} = q_{\text{\cite{Drinfeld:1986}}} =
e^{h_{\text{\cite{Drinfeld:1986}}}}=  t_{\text{\cite{Jimbo:1985}}}^4 = 
q_{\text{\cite{Frenkel:1998}}}^{2}  
\end{equation}
 where $\lfive$ is the circumference of the circle
$\BS^{1}$ in the compactification of the five dimensional theory on the
twisted bundle $\widetilde{\BR^{4} {\times} \BS^{1}}_{\la \ep_1, \ep_2 ; \lfive
  \ra}$. 

Our multiplicative spectral parameter $\xi$ corresponds 
to the variable $z$ of Nakajima \cite{Nakajima:2004} which corresponds
to the variable $u^{-1}$ of Frenkel and Reshetikhin \cite{Frenkel:1998}, see more details in the
appendix
\begin{equation}
\label{eq:rel2}
  \xi_{\text{present}} =  u^{-1}_{\text{\cite{Frenkel:1998}}} = z_{\text{\cite{Nakajima:2001}}}
\end{equation}

 We mostly use the conventions of
\cite{Nakajima:2004, Frenkel:1998, Beck:1994}, with the dictionary of
conventions (\ref{eq:rel1})(\ref{eq:rel2}).  

Let $a_{ij}= 2 \delta_{ij} - I_{ij}$, where $I_{ij}$ is the incidence
matrix,  be the Cartan matrix $\gq$.  The indices $i,j \in \Ver$ label the nodes
of the Dynkin diagram $\Gamma$.   We restrict our definitions to simply laced ADE Lie
algebras $\gq$. 
Let
$(\alpha_i)_{i \in \Ver}$ denote the simple roots. Then $a_{ij}=
(\alpha_i,\alpha_j)$ where $(,)$ is the standard bilinear form on the
dual Cartan algebra in normalization where the squared length simple
roots is 2. 

By $[n]_q$ we denote a $q$-number, 
\begin{equation}
 [n]_q = \frac{ q^{\frac n2} - q^{-\frac n2}}{q^{\frac 12} - q^{-\frac 12}},
 \label{eq:qnum}
 \end{equation}
by $[n!]_q$ the $q$-factorial, 
\begin{equation}
[n!]_q = [n]_q [n-1]_q \dots [1]_q,
\label{eq:qfac}
\end{equation}
and by
${s \brack k }_q$ the $q$-binomial coefficient:
\begin{equation}
 {s \brack k}_q = \frac { [s!]_q }{ [(s - k)!]_q [k!]_q}
 \label{eq:qbin}
\end{equation}

\subsubsection{Chevalley-Drinfeld-Jimbo construction}. Let $\g$ be a generalized
 Kac-Moody algebra with symmetrizable Cartan matrix $\tilde a_{ij}$. In particular,
$\g$
could be a finite-dimensional simple Lie algebra of affine Kac-Moody Lie algebra.
 Then we can define the quantum algebra $\Qalg_q(\g)$ by a
 $q$-deformation of Serre relation for Chevalley generators of $\g$
 \cite{Drinfeld:1986,Jimbo:1985,MR1062425,Chari:1994}.   If  $\gq$ is
 finite-dimensional, then applying such construction for
$ \g = \hat \gq$ we obtain quantum affine algebra $\Qalg_q(\g) = \Qalg_q(\hat \gq) \cong \Qaff_q(\gq) $.

 To simplify presentation, we assume that $ \g$ is
 simply laced
 \begin{equation}
 a_{ij} =  a_{ji}   
 \end{equation}

The algebra $\Qalg_q ( \g)$ is an associative algebra generated by the Chevalley 
generators $(h_i, e_i^{+}, e_i^{-})_{i \in I_{\g}}$,  
called  the \emph{coroots}, the \emph{raising} and the \emph{lowering}
elements, respectively, where $i$ runs over the set of nodes 
$I_{\g}$ on Dynkin graph of $ \g$. The relations are the $q$-deformation of the
Chevalley relations\footnote{Here $q^{\frac 1 2 h_i} = k_i|_{\text{\cite{Chari:1994}}}=
  (k_{i}^2)_{\text{\cite{Jimbo:1985,MR1062425}}} = (e^{\frac 1 2 \hbar
    h_i})|_{\text{\cite{Drinfeld:1986}}} $ and $q^{\frac 1 2} = q_{\text{\cite{Chari:1994}}}$}
\begin{equation}
\label{eq:firstDrinfeld}
  \begin{aligned}
&  h_i h_j = h_j h_i,    \qquad q^{\frac {h_i} 2} e_{j}^{\pm} q^{-\frac {h_i}2}= q^{\frac 1 2 a_{ij}} e_{j}^{\pm}, \qquad
  [e_i^{+}, e_j^{-}] = \delta_{ij} [h_i]_q\\
&\sum_{r=0}^{1 - a_{ij}} (-1)^{r}  {{1 - a_{ij}}\brack {r}}_{q}
(e_i^{\pm})^{1 - a_{ij}-r } e_{j}^{\pm} (e_{i}^{\pm})^{r} =0, i \neq j\\
  \end{aligned}
\end{equation}
The \emph{co-multiplication} $\Delta: \Qalg_q(\g) \to
\Qalg_q( \g)$\footnote{Different sign choices are possible and different conventions
  exist in the literature}, the \emph{antipod}   $\gamma: \Qalg_q(
\g) \to \Qalg_q( \g)$ and  the \emph{counit}
$\varepsilon: \Qalg_q( \g) \to \BC$ are given by
\begin{equation}
\begin{aligned}
&  \Delta(q^{\frac {h_i} 2}) = q^{\frac {h_i} 2} \otimes q^{\frac {h_i}
    2},\qquad 
  \Delta(e_{i}^{+}) = e_i^{+} \otimes 1 + q^{\frac {h_i} 2} \otimes
  e_i^{+}, \qquad
  \Delta(e_{i}^{-}) = e_i^{-} \otimes q^{-\frac {h_i} 2}  + 1  \otimes e_i^{+}\\
&\gamma(e_i^{+}) = - e_i^{+} q^{-\frac{h_i}{2}}, \qquad \gamma(e_i^{-}) = - q^{\frac {h_i}{2}} e_i^{-}, 
\qquad q^{h_i} = 1\\
&  \varepsilon(e_{i}^{\pm}) = 0, \quad \varepsilon(q^{\frac{h_i}{2}}) = 1
\end{aligned}
\end{equation}

The Borel subalgebra  $\Qalg_q( \g^{\pm}) \subset \Qalg_q( \g)$ is a
subalgebra generated by the generators $(h_i,e_i^{\pm})$ in this
realization (\ref{eq:firstDrinfeld}).

\subsection{Drinfeld current realization}
In \emph{the  Drinfeld current realization}
\cite{Drinfeld:1987,Beck:1994},  applicable for arbitrary 
symmetrizable Kac-Moody $\gq$,  the quantum affine algebra
$\Qaff_q( \gq)$ is generated by the 
\emph{loop coroots} $(h_{i,n})_{i \in \Ver,  n \in \BZ}$,   \emph{loop
  raising and lowering generators} $(x^{\pm}_{i,n})_{i \in \Ver, n \in
  \BZ}$, the \emph{center} $c$ and the \emph{energy}  $d$, where $i \in
\Ver$ runs over the nodes of $\gq$, with commutation relations  (\cite{Chari:1994,Frenkel:1998, Herandez:2003})
\footnote{Here $q^{\frac 1 2 } =
q_{\text{\cite{Chari:1994,Frenkel:1998,Herandez:2003}}}$ and $q^{\frac 1
2 c} =c_{\text{\cite{Chari:1994}}}$ and $ h_{i,n} q^{- c|n|/4} = a_{i,n}|_{\text{\cite{Jimbo:1999}}}$}
\begin{align}
\label{eq:Uqdefhe}
   & [h_{i,0}, e^{\pm}_{j,n}] = \pm a_{ij} e^{\pm}_{j,m}, \qquad  [h_{i,n \neq 0},
   e^{\pm}_{j,m}] = \pm \frac 1 n [n a_{ij}]_q q^{\mp c |n|/4}
   e^{\pm}_{j, n+m}\\
\label{eq:Uqdefhh}   & [h_{i,0}, h_{j,0}] = 0, \qquad [h_{i,n \neq 0}, h_{j,m}] = \delta_{n+m,0} \frac 1 n [n a_{ij}]_q [nc]_q\\
\label{eq:Uqdefee1} & [e^+_{i,n}, e^{-}_{j,m}] = \delta_{ij} \frac { q^{(n-m) c/4}
  \psi^{+}_{i,n+m} - q^{-(n-m) c/4} \psi^{-}_{i,n+m}}{q^{\frac 1 2} -
  q^{-\frac 1 2}}\\
\label{eq:Uqdefee2}&  e^{\pm}_{i, n+1} e^{\pm}_{j,m} -  q^{\pm \frac 1 2 a_{ij}} e^{\pm}_{i,n} e^{\pm}_{j,m+1} =q^{\pm \frac 1 2 a_{ij}} e^{\pm}_{j,m}
 e^{\pm}_{i,n+1} -
 e^{\pm}_{j,m+1} e^{\pm}_{i,n}\\
&  \sum_{\sigma \in \mathcal{S}_{a}} \sum_{k=0}^{a } (-1)^k
{a  \brack k }_q e^{\pm}_{i, n_{\sigma(1)}} \dots e^{\pm}_{i,n_{\sigma(k)}}
e^{\pm}_{j,m} e^{\pm}_{i,n_{\sigma(k+1)}} \dots e^{\pm}_{i,
  n_{\sigma(a )}} = 0, \quad (i \neq j, a =1 - a_{ij})
\end{align}
where $(\psi_{i,n}^{\pm})_{i \in \Ver, n \in \BZ_{\geq 0}}$ are \emph{loop
  modes}\footnote{Here $\xi = z_{\text{\cite{Jimbo:1999}}} =
  u^{-1}_{\text{\cite{Chari:1994,Frenkel:1998}}} =
  z^{-1}_{\text{\cite{Herandez:2003}}}$}
of the \emph{current} $\psi_{i}^{\pm}(\xi) \in \Qaff_q( \gq)$
\begin{equation}
\label{eq:psi}  
\begin{aligned}
    \psi_{i}^{\pm}(\xi) := \sum_{n=0}^{\infty} \psi_{i,n}^{\pm} \xi^{\mp n}
    := q^{\pm \frac 1 2 h_i} \exp \left (  \pm (q^{\frac 1 2} -
      q^{-\frac 1 2})
      \sum_{n=1}^{\infty} h_{i,\pm n} \xi^{\mp n} \right )
  \end{aligned}
\end{equation}

The \emph{level zero specialization} $\Qaff_q(\gq)|_{c=0}$ is a
specialization to the center $c = 0$.  At $c=0$ the operators $(h_{i,n})_{i \in \Ver, n \in \BZ}$ generate the commutative
subalgebra of $\Qaff_q( \gq)|_{c=0}$, and the currents
$\psi_{i}^{\pm}(\xi)$ are commutative. In the limit $q \to 1$, the  $\Qaff_q(\gq)|_{c=0}$ becomes 
algebra $\mathbf{U}(L\gq)$ where $L\gq$ is the loop algebra of
$\gq$.

\subsubsection{Drinfeld currents\label{se:Drinfeld-currents}}

 We can write the
defining relations of $\Qaff_q(\gq)$ in terms of \emph{currents} $x_{i}^{\pm}(\xi)$, see
\cite{Herandez:2003} and \cite{Jimbo:1999},
\begin{equation}
  e_{i}^{\pm}(\xi) = \sum_{n \in \BZ} e^{\pm}_{i,n} \xi^{-n}
\end{equation}

In terms of the currents $(\psi^{\pm}_i(\xi), e^{\pm}_i(\xi))_{i \in
  \Ver}$, the relations (\ref{eq:Uqdefhe}),
 (\ref{eq:Uqdefhh}), (\ref{eq:Uqdefee1}) are respectively packed
into\footnote{It is convenient to compute $[\log \psi_i^{+}(\xi), e_j^{\pm}(\xi')]$
from (\ref{eq:Uqdefhe}) and then use the fact that for any operators
$h,x$, the commutator $[h,x] = \lambda x$,  where $\lambda$ is a number,
implies $e^{h} x e^{-h} =
e^{\lambda} x$. If operators $a,b$ commute as $[a,b] = \lambda$, where
$\lambda$ is a number, then $e^{a} e^{b}e^{-a} = e^{\lambda} e^{b} $}

  \begin{align}
&\label{eq:psie}
  \psi_{i}^{\sharp}(\xi) e^{\pm}_j(\xi')  = 
q^{\pm \frac 1 2 a_{ij}}  \frac{ 1   - q^{\mp \frac 1 2  a_{ij}} q^{\mp \sharp \frac c 4 }\xi'/\xi }
{1  - q^{\pm \frac 1 2 a_{ij}} q^{\mp \sharp \frac c 4 }\xi'/\xi} e^{\pm}_j(\xi') \psi_{i}^{\sharp}(\xi), \qquad
\sharp \in \{ +, -\} \\
&  \psi_{i}^{+}(\xi) \psi_{j}^{-}(\xi') = \frac{ (1 - q^{\frac 1 2
      a_{ij}} q^{\frac 1 2 c} \xi'/\xi) ( 1- q^{-\frac 1 2 a_{ij}}
    q^{-\frac 1 2 c} \xi'/\xi)}
{ (1 - q^{-\frac 1 2 a_{ij} } q^{ \frac 1 2 c} \xi'/\xi) (1 - q^{\frac
    1 2 a_{ij}} q^{-\frac 1 2 c} \xi'/\xi)} \psi_{j}^{-}(\xi')
\psi_{i}^{+}(\xi); \\
& \qquad\qquad\qquad\qquad\qquad\qquad   \psi_{i}^{\pm}(\xi) \psi_{j}^{\pm}(\xi') =
\psi_{j}^{\pm}(\xi')  \psi_{i}^{\pm}(\xi) \\
&{} [e_i^{+}(\xi), e_j^{-}(\xi')] = \frac{\delta_{i,j}}{q^{\frac 1 2} -
  q^{-\frac 1 2}}\left ( \psi^{+}_i (\xi' q^{\frac c 4}) \delta( q^{\frac c 2 }
\xi'/\xi)  -  \psi^{-}_i (\xi' q^{-\frac c 4}) \delta( q^{-\frac c 2 }
\xi'/\xi) \right) \\
&\label{eq:epem} e_i^{\pm}(\xi) e_{j}^{\pm}(\xi') =
q^{\pm \frac 1 2 a_{ij}} \frac{1  - q^{\mp \frac 1 2 a_{ij}} \xi'/\xi}{1 - q^{\pm \frac 1 2 a_{ij}} \xi'/\xi} e^{\pm}_j(\xi') e_i^{\pm}(\xi)
  \end{align}
where
\begin{equation}
  \delta(z):= \sum_{n \in \BZ} z^{n} 
\end{equation}
and
\begin{equation}
\begin{aligned}
\sum_{k=0}^{a } (-1)^k
{a  \brack k }_q \mathrm{Sym}_{\xi_1, \dots, \xi_{a}} e^{\pm}_{i}(\xi_1) \dots e^{\pm}_{i}(\xi_k)
e^{\pm}_{j}(\xi') e^{\pm}_{i}(\xi_{k+1}) \dots e^{\pm}_{i}(\xi_{a}) = 0, \\
\qquad\qquad\qquad\qquad\qquad\qquad\quad (i \neq j, a =1 - a_{ij})  \end{aligned}
\end{equation}

We will also use the current
\begin{equation}
\label{eq:psi0} 
 \begin{aligned}
    \mathring \psi_{i}^{\pm}(\xi) := \sum_{n=0}^{\infty} \mathring \psi_{i,n}^{\pm} \xi^{\mp n}
    :=\exp \left (  \pm (q^{\frac 1 2} -
      q^{-\frac 1 2})
      \sum_{n=1}^{\infty} h_{i,\pm n} \xi^{\mp n} \right )
  \end{aligned}  
\end{equation}
and
\begin{equation}
\label{ra:pdef}
  p_i^{\pm}(\xi) := \exp \left( - \sum_{n=1}^{\infty} \frac { h_{i,\pm n}} {
      [n]_q } \xi^{\mp n}
\right)
\end{equation}
The currents  $\psi_{i}^{\pm} \in U_q(\gq)$ and $p_i^{\pm} \in
U_q(\gq)$ are related 
\begin{equation}
  \psi_{i}^{\pm}(\xi) =  q^{\pm \frac 1 2 h_i} \frac{ p_i^{\pm}(q^{\frac
      12}\xi) }{
    p_i^{\pm}(q^{-\frac 1 2 } \xi)}.
\end{equation}

\subsubsection{Yangian and elliptic version}
Here we define the algebras $\Yang_{\ep}(\g)$,  $\Qaff_q(\g)$ and
$\Qell_{q,p}(\g)$ uniformly called $\mathbf{U}_{\ep}\gq(\Cx)$ for a Kac-Moody Lie algebra $\gq$ with symmetric
Cartan matrix $a_{ij}$ using Drinfeld currents on $\Cx$, where 
 $\Cx=\BC$ for Yangian $\Yang_{\ep}(\g)$, $\Cx = \BC/\frac{2\pi}{\lfive}
 \BZ$ for quantum affine   $\Qaff_q(\g)$, and
 $\Cx = \BC/(\frac{2 \pi}{\lfive})(\BZ + \tau_p \BZ)$ for quantum
 elliptic $\Qell_{q,p}(\g)$. Instead of multiplicative variable 
 \begin{equation}
   \xi = e^{i \lfive x}
 \end{equation}
we use additive variable $x \in \Cx$ in the domain $\Cx$ which might
have 0, 1, or 2 periods.

The basic (quasi)-periodic function $s(x)$, with $0$, $1$ or $2$ periods
would be given by 
\begin{equation}
  s(x) =
  \begin{cases}
    x, \quad \Cx = \BC\\
   \frac{2}{\lfive}  \sin \frac {\lfive x} 2 , \quad \Cx =
    \BC/\frac{2\pi}{\lfive} \BZ \\
   \frac{ \theta_1( e^{i \lfive x}; p)}{\partial_x \theta_1( e^{i \lfive x}; p)|_{x=0}},    \quad \Cx = \BC/\frac{2 \pi}{\lfive}(\BZ + \tau_p \BZ)
  \end{cases}
\end{equation}
where $\theta_1(\xi;p)$ for $p =e^{2 \pi i \tau_p}$ is defined in
(\ref{eq:theta}). We have the hierarchy of degenerations:
 elliptic $\to$ trigonometric $\to$ linear
\begin{equation}
  s_{\BC/\frac{2 \pi}{\lfive}(\BZ + \tau_p \BZ)}(x) \stackrel{p \to
    0}{\longrightarrow} s_{  \BC/\frac{2\pi}{\lfive} \BZ}(x)
\stackrel{\lfive \to 0}{ \longrightarrow } s_{\BC}(x) 
\end{equation}

Then we would simply replace the defining relations 
(\ref{eq:Uqdefhe})(\ref{eq:Uqdefhh})(\ref{eq:Uqdefee1})(\ref{eq:Uqdefee2})
by 
  \begin{align}
&\label{eq:psiegen}
  \psi_{i}^{\sharp}(\xi) e^{\pm}_j(\xi')  = \frac{s( x - x' \pm \frac 1 2
  a_{ij} \ep \pm \sharp \frac c 4 \ep)}{ s( x - x' \mp \frac 1 2 a_{ij} \ep \pm
  \sharp \frac c 4 \ep)} e^{\pm}_j(\xi') \psi_{i}^{\sharp}(\xi), \qquad
\sharp \in \{ +, -\} \\
&  \psi_{i}^{+}(\xi) \psi_{j}^{-}(\xi') = \frac{ s(x - x' -\frac 1 2
  a_{ij} \ep - \frac 1 2 c \ep) s( x - x' + \frac 1 2 a_{ij} \ep + \frac
  1 2 c \ep)}
{ s(x - x' + \frac 1 2 a_{ij} \ep - \frac 1 2 c \ep)s(x - x' -\frac 1 2 a_{ij} \ep +
  \frac 1 2 c \ep)} \psi_{j}^{-}(\xi')
\psi_{i}^{+}(\xi); \\
& \qquad\qquad\qquad\qquad\qquad\qquad   \psi_{i}^{\pm}(\xi) \psi_{j}^{\pm}(\xi') =
\psi_{j}^{\pm}(\xi')  \psi_{i}^{\pm}(\xi) \\
&{} [e_i^{+}(\xi), e_j^{-}(\xi')] = \frac{\delta_{ij} \ep}{ s( \ep)}\left(
\psi_i^{+}(x' + \tfrac c 4 \ep) \delta(x - x' - \tfrac c 2 \ep) -
\psi_i^{-}(x' -\tfrac c 4 \ep) \delta( x - x' + \tfrac  c 2 \ep)\right)\\
&\label{eq:epemgen} e_i^{\pm}(\xi) e_{j}^{\pm}(\xi') =\frac{ s( x - x' \pm
  \frac 1 2 a_{ij} \ep)}{ s( x - x' \mp \frac 1 2 a_{ij} \ep)} e^{\pm}_j(\xi') e_i^{\pm}(\xi)
  \end{align}
See more on elliptic currents in \cite{Enriquez:1997}.

\subsection{Level zero representations}

\subsubsection{Level zero representations and generalized weights} 
For illustration, we first consider a fundamental evaluation
 module  $L_{1,q^{-\frac 1 2} \mu}$  for $\Qaff_q(\mathfrak{sl}_2)$. Let $\mu \in \BC^{\times}$
be an evaluation parameter, and $|\omega_0 \rangle, |\omega_1 \rangle $
be basis vectors in $L_{1,q^{-\frac 1 2} \mu}$ so that $|\omega_0
\rangle $ is the highest vector.  
In this basis the  currents
$\psi(\xi), e^{\pm}(\xi)$ are represented by matrices
\begin{equation}
\label{eq:diagonal}
  \psi^{\pm}(\xi) =
  \begin{pmatrix}
    q^{\frac 1 2} \frac{ 1 - \mu q^{-1}/\xi}{1 - \mu/\xi} &  0 \\
    0 & q^{-\frac 1 2} \frac{1 - \mu q /\xi}{1 - \mu/\xi}
  \end{pmatrix}_{\pm}
\end{equation}
and
\begin{equation}
  e^{+}(\xi) =
  \begin{pmatrix}
    0 & \delta(\mu/\xi) \\
    0 & 0 
  \end{pmatrix}, \qquad 
e^{-}(\xi) =
\begin{pmatrix}
  0 & 0 \\
 \delta(\mu/\xi) & 0 
\end{pmatrix}
\end{equation}
where symbols $f(\xi)_{\pm}$ for a function $f(\xi)$ denotes expansions near
$\xi = \infty$ and $\xi = 0$ respectively 
\begin{equation}
  f(\xi)_{+} = \sum_{n \geq n_{+}} f_{n} \xi^{-n}, \qquad   f(\xi)_{-} = \sum_{n \geq n_{-}} f_{n} \xi^{n}
\end{equation}
The key property of these expansions is localization to the poles, such
as 
\begin{equation}
 \left( \frac{1}{1 - 1/\xi} \right)_{+} -  \left( \frac{1}{1 - 1/\xi}
 \right)_{ -} = \sum_{n \geq 0} \xi^{-n} + \sum_{n > 0} \xi^{n}  =
 \delta(\xi) 
\end{equation}

For example, checking the commutation relation (\ref{eq:epem}), in the
right hand side we find the matrix elements such as 
\begin{equation}
    \left(  q^{\frac 1 2} \frac{ 1 - \mu q^{-1}/\xi}{1 - \mu/\xi}
\right)_{+} -  
     \left(  q^{\frac 1 2} \frac{ 1 - \mu q^{-1}/\xi}{1 - \mu/\xi}
 \right)_{-} = ( q^{\frac 1 2} - q^{-\frac 1 2} )\sum_{n \in \BZ}
 (\mu/\xi)^n =( q^{\frac 1 2} - q^{-\frac 1 2} ) \delta(\mu/\xi) 
\end{equation}
so that indeed 
\begin{multline}
  [e^{+}(\xi), e^{-}(\xi')] =
 \delta(\mu/\xi) \delta(\mu/\xi')  \begin{pmatrix}
   1 & 0 \\
    0 & -1 
  \end{pmatrix} =  \frac{ \delta(\xi/\xi') \delta(\mu/\xi)}{q^{\frac 1
      2} - q^{-\frac 1 2}}   \begin{pmatrix}
q^{\frac 1 2} - q^{-\frac 1 2} & 0 \\
0 & q^{-\frac 1 2} - q^{\frac 1 2}  
\end{pmatrix}= \\
=\frac{1}{q^{\frac 1 2} - q^{-\frac 1 2}} \delta(\xi/\xi')(
\psi^{+}(\xi)  - \psi^{-}(\xi))
\end{multline} 

The diagonal elements in (\ref{eq:diagonal}) are called the
\emph{current weights} or \emph{loop weights} of an operator
$\psi(\xi)$.

The highest weight of the evaluation representation $L_{1,\mu q^{-\frac
    1 2}}$ is encoded by the Drinfeld polynomial
  \[ 
P_1 =  ( 1 -  \mu q^{-\frac 1 2}/\xi)
\]
The  operator $\psi(\xi) \in \Qaff_q({\mathfrak{sl}_2})$ acts by 
  \begin{equation}
    \begin{aligned}
    \psi(\xi)|\omega_0 \rangle = q^{\frac 1 2} \frac {1 - \mu q^{-1}/\xi}{
      1  - \mu/\xi} |\omega_0 \rangle \\
    \psi(\xi)|\omega_1 \rangle = q^{-\frac 1 2} \frac {1 - \mu q^{+1}/\xi}{
      1  - \mu/\xi} |\omega_1 \rangle \\
    \end{aligned}
  \end{equation}
Hence, by definition, the $q$-character of the $\Qaff_q({\mathfrak{sl}_2})$ fundamental module $V_{1,aq^{-\frac 1 2}}$ is 
\begin{equation}
  \chi_{q}[ V_{1, \mu q^{-\frac 1 2}}] = Y_{1, \mu q^{-\frac 1 2}} +
  \frac{1}{Y_{1,\mu q^{\frac 1 2}}}
\end{equation}
Notice that the sum of eigenvalues of $\psi(\xi)$ does not encode as
much information about the $\Qaff_q({\mathfrak{sl_2}})$ module
$V_{1,\mu q^{-\frac 1 2}}$ as it turns out to be simply 
\begin{equation}
  \tr_{V} \psi_1(z) = q^{\frac 12 } + q^{-\frac  12}
\end{equation}

In terms of the current modes $\psi_{1,n}^{\pm}, e^{\pm}_{1,n}$ for the
same example of the evaluation representation $L_{1, q^{-\frac 1 2} \mu}$ of
  $\Qaff_q({\mathfrak{sl}_2})$ with
  evaluation parameter $\mu q^{-\frac 1 2}$ we find the explicit action by the  $\Qaff_q({\mathfrak{sl}_2})$
generators on $V_{1, \mu q^{-\frac 1 2}}$ is given by
\begin{equation}
  \begin{aligned}
&  e^{-}_{1,n} | \omega_0 \rangle = \mu^n |\omega_1 \rangle \quad
&&  e^{+}_{1,n} | \omega_1 \rangle = \mu^n |\omega_0 \rangle  \\
&  h_{1,n} | \omega_0 \rangle = \mu^{n} q^{-\frac{n}{2}} \frac{
  [n]_{q}}{n} |\omega_0 \rangle \quad (n > 0), \qquad   &&h_{1,0} | \omega_0
\rangle = | \omega_0 \rangle \\
& h_{1,n} | \omega_1 \rangle = \mu^{n} (q^{-\frac{n}{2}} \frac{ [n]_{q}}{n}
 - \frac {[2n]_q}{n}) |\omega_1 \rangle \quad (n > 0), \qquad && h_{1,0} | \omega_1
\rangle =  - | \omega_1 \rangle
   \end{aligned}
\end{equation}
or equivalently by the matrices
\begin{equation}
  \begin{aligned}
&  e^{-}_{1,n} =
\mu^n  \begin{pmatrix}
    0  & 0 \\
    1 & 0 
  \end{pmatrix}\qquad 
&&e_{1,n}^{+} =
\mu^n \begin{pmatrix}
  0 & 1 \\
  0 & 0  
\end{pmatrix}\\
&h_{1,n \neq 0 } = \frac{\mu^n}{n} 
\begin{pmatrix}
  \frac{ 1 - q^{-n}}{ q^{1/2 } - q^{-1/2}} & 0 \\
 0 & \frac{ 1 - q^{n}}{ q^{1/2}  - q^{-1/2}}
\end{pmatrix} \qquad 
&&h_{1,0} =
\begin{pmatrix}
  1 & 0\\
  0 & - 1
\end{pmatrix}\\
&\psi^{\pm}(\xi) =
\begin{pmatrix}
\frac {   q^{\frac 1 2 } (1 - \mu q^{-1}/\xi)}{1 - \mu /\xi} &  0 \\
0 &    \frac{ q^{-\frac 1 2 }(1 - \mu q /\xi)} {1 - \mu/\xi} 
\end{pmatrix}^{\pm}\quad \Rightarrow
&&\psi_{1,n  \in \pm \BZ_{>0}}^{\pm} =
\pm \begin{pmatrix}
 \mu^{n} (q^{\frac 1 2} - q^{-\frac 1 2}) & 0 \\
0 & \mu^{n} (q^{\frac 1 2} - q^{-\frac 1 2})
\end{pmatrix}
  \end{aligned}
\end{equation}
which obviously satisfy the commutation relations (\ref{eq:Uqdefhe})(\ref{eq:Uqdefhh})(\ref{eq:Uqdefee1})(\ref{eq:Uqdefee2}) of
$\Qaff_q({\mathfrak{sl}_2})$ at $c =1$
\begin{equation}
  [h_{1,n \neq 0}, e^{\pm}_{1,m}] = \frac{ [2n]_q}{n} e_{m + n}\qquad
  [e^{+}_{1,m}, e^{-}_{1,n}] = \frac{1}{q^{\frac 1 2} - q^{-\frac 1 2}}
  (\psi_{1,m+n}^{+} - \psi_{1,m+n}^{-})
\end{equation}

\subsubsection{Level zero fundamental highest weight modules}

The basic level zero fundamental representations of quantum affine
algebra $\Qaff_q(\g)$ are highest weight modules $L_{i, \mu}$, labeled
by a node $i \in I_{\g}$ of Dynkin diagram of $\g$ and an evaluation
parameter $\mu \in \BC^{\times}$. A module $L_{i, \mu}$ is a highest 
weight $\Qaff_q(\g)$-module with the current weights of $(\psi_i^{\pm})_{i \in I_{\g}}$ on the highest
vector $| \omega_{i,\mu} \rangle $ given by 
\begin{equation}
  \psi^{\pm}_{j} (\xi) | \omega_{i,\mu} \rangle =
  \begin{cases}
    1, \quad j \neq i \\
q^{\frac 1 2 }    \frac{ 1 - \mu q^{-\frac 1 2}/\xi}{ 1  - \mu q^{\frac 1 2}/\xi}, \quad j
    = i 
  \end{cases}
\end{equation}
The fundamental module $L_{i, \mu}$ is a quantum affine algebra analogue
of the $i$-th fundamental highest weight evaluation module at $\xi = \mu$ for the loop
algebra $\g((\xi))$.

\subsection{Loop weights}
Following \cite{Chari:1994},\cite{Frenkel:1998}  we call a vector $| w
\rangle \in W$, for a $\Qaff_q( \gq)$-module
$W$, a generalized eigenvector of the operators $\psi_i^{\pm}(\xi) \in \Qaff_q(
\gq)$   with the  generalized eigenvalues $\Psi_i^{\pm}(\xi) \in \BC[[\xi^{\mp}]]$ if there exists a positive integer $n$, such that 
\begin{equation}
 (\psi_i^{\pm}(\xi) -
 \Psi_i^{\pm}(\xi) )^n | w\rangle = 0 \quad \text{for all $i \in \Ver$}.
\end{equation}
in other words it belongs to the Jordan block of uniformly bounded size $n$ for all $\psi_i^{\pm}(\xi)$'s. 
The values  $\Psi_i^+(\infty)$ and $\Psi_i^-(0)$ are the eigenvalues
of $q^{\frac 1 2h_i} $  and $q^{-\frac 1 2 h_i}$ since
\begin{equation}
  \begin{aligned}
  & q^{\frac 1 2 h_i} |w\rangle  =  \Psi_i^{+}(\infty) |w\rangle  \quad \\
  &q^{-\frac 12 h_i} |w \rangle =  \Psi_i^{-}(0) |w\rangle.
  \end{aligned}
\end{equation}

All integrable finite-dimensional modules of $\Qaff_q( \gq)$ with highest
weight have been classified \cite{Chari:1994} (theorem 3.3). 
  The generalized eigenvalue $\Psi^{\pm}(\xi)$ of the operator
 $\psi^{\pm}(\xi)$  on the highest vector $| v \rangle$ of an integrable
  finite-dimensional module  $V$ of $\Qaff_q( \gq)$ always has the form 
  \begin{equation}
    \Psi_i^{\pm}(\xi) = q^{\frac 1 2 \deg P_i} \left( \frac{ P_i( q^{\frac 1
          2} \xi)}{ P_i(
      q^{-\frac 1 2} \xi)}\right)^{\pm}
  \end{equation}
where $P_i^{+}(\xi): = \prod_{k=1}^{\deg P_i} (1 -  \xi_k/\xi)$ is a
polynomial in $\xi^{-1}$ with $P_i(\infty) = 1$ and 
$()^{\pm}$ denotes the expansion at $\xi = \infty$ and $\xi = 0$
respectively. The collection of polynomials $P_i(\xi)_{i \in \Ver}$,
defining an integrable highest weight module of $\Qaff_q( \gq)$ is called Drinfeld polynomial.

The equation (\ref{ra:pdef}) implies that the highest vector $|v_0 \rangle $ in the $i$-th fundamental representation with Drinfeld polynomial
  $P_j(\xi) = 1$ for $j \neq i$ and $P_i(\xi) = 1 - 1/\xi$, 
such that the eigenvalue of  $p_j^{\pm}(\xi)$ on $|v_0 \rangle$ is
equal to  $P_j(\xi)$,  is a common eigenvector of the generators $h_{j,n}$ with the corresponding eigenvalues
  \begin{equation}
\label{eq:hj-eigens}
    h_{j,n} =
    \begin{cases}
      0, \quad j \neq i \\
      1, \quad j = i, n = 0 \\
      \frac{ [n]_q}{n}, \quad j = i, n \neq  0\\
    \end{cases}
  \end{equation}
E.~Frenkel and N.~Reshetikhin \cite{Frenkel:1998} have shown that for 
any finite-dimensional $\Qaff_q( \gq)$ module $V$ the generalized 
eigenvalues of the operators  $\psi^{\pm}_i(\xi) \in U_q (\hat \gq)$ 
always have the form 
\begin{equation}
\label{eq:monom}
\mathrm{ev}_{\psi(\xi)}  \prod_{\xi_k} (Y_{i, \xi_k})^{\pm} 
\end{equation}
where \emph{evaluation on $\psi(\xi)$} is defined multiplicatively by
\begin{equation}
\mathrm{ev}_{\psi(\xi)} (Y_{i,\zeta}) :=   q^{\frac 1 2} \frac {1 - q^{-\frac 1 2} \zeta/\xi}{ 1 -
 q^{\frac 1 2}  \zeta/\xi }
\end{equation}
Similarly, for stripped element $\mathring{\psi}(\xi)$
  \begin{equation}
\label{eq:evaluation}
\mathrm{ev}_{\mathring{\psi}(\xi)} (Y_{i,\zeta}) :=   \frac {1 - q^{-\frac 1 2} \zeta/\xi}{ 1 -
 q^{\frac 1 2}  \zeta/\xi }
\end{equation}

The $q$-character $\chi_{q}(V)$ of $U_{q}(\hat \gq)$ finite-dimensional module $V$ is
\emph{defined} as a sum of $\dim V$ monomials of the form
\begin{equation}
\label{eq:qchardef}
\chi_q(V) = \sum_{|v \rangle }   \prod_{i, \xi_{i} \in \Xi_{|v\rangle}}
(Y_{i, \xi_{i}})^{\pm 1} 
\end{equation}
one  monomial  for each generalized eigenvector $|v \rangle$ in
$\Qaff_q( \gq)$ module $V$ for the commuting set of operators $\psi_{i}(\xi)
\in \Qaff_q( \gq)$, such that each such monomial encodes the generalized
eigenvalue of the operator 
\begin{equation}
 \psi(\xi) = \prod_{i} \psi_{i \in \Ver}(\xi) \in U_{q}(\hat \gq)
\end{equation}
by the evaluation (\ref{eq:evaluation}).

\subsection{Universal $R$-matrix}
Abstractly, quantum affine algebra $A = \Qaff_q( \gq)$ is a Hopf algebra, and hence is
equipped with the \emph{comultiplication} operation $\Delta: A \to A
\otimes A$. The comultiplication operation is what allows to define tensor products of
representations for associative algebras. If $\rho_i: A \to \End(V_i)$ are two representations,
then the tensor product $\rho: A \to \End(V_1) \otimes \End(V_2)$ is
defined by $x \mapsto \rho_1 \otimes \rho_2 ( \Delta(x))$. If an
abstract Hopf algebra $A$ is co-commutative, then $A$-modules $V_1
\otimes V_2$ and $V_2 \otimes V_1$ are naturally isomorphic, but there
is no natural reason for such isomorphism if $A$ is not co-commutative. 
Quantum affine algebras have extra structure in addition to the structure of generic
Hopf algebras, called \emph{quasi-triangular structure} that ensures
that tensor products $V_1 \otimes V_2$ and $V_2 \otimes V_1$ are isomorphic. Such
isomorphism is not a  simple permutation of the factors, but is given
by a certain linear map that depends on representations $V_1, V_2$. Namely, by
definition, a quasi-triangular structure on a Hopf algebra $A$ is an
element $\CalR \in
A \otimes A$, called \emph{universal R-matrix} such that 
    \begin{equation}
\label{eq:Rmatrix}
    \begin{aligned}
&      \CalR \Delta(x) = \Delta^{op}(x) \CalR \\
  &    (\Delta \otimes \mathrm{Id}_A )(\CalR) = \CalR_{13} \CalR_{23} = a_i \otimes
 a_j \otimes  b_i b_j\\
 &     (\mathrm{Id}_A \otimes \Delta) (\CalR) = \CalR_{13} \CalR_{12} = a_i a_j
 \otimes b_j \otimes b_i
    \end{aligned}
  \end{equation}
where $x \in A$ and if $\CalR = \sum_i a_i \otimes b_i$ then $\CalR_{12} = \sum_i a_i \otimes b_i \otimes 1, 
    \CalR_{13} = \sum_i a_i \otimes 1 \otimes b_i,
    \CalR_{23} = \sum_i 1 \otimes a_i \otimes b_i$, 
and $\Delta^{op}$ denotes the co-multiplication in the opposite
order. The map $R_{V_1, V_2}' =P R_{V_1, V_2}$, where $P: V_1 \otimes V_2
\to V_2 \otimes V_1$ is a permutation of factors, gives isomorphism
$R_{V_1, V_2}': V_1 \otimes V_2 \to V_2 \otimes V_1$.
The relations (\ref{eq:Rmatrix}) imply that the universal R-matrix $\CalR \in A
\otimes A$ satisfies universal Yang-Baxter equation
\begin{equation}
  \CalR_{12} \CalR_{13} \CalR_{23} = \CalR_{23} \CalR_{13} \CalR_{12} 
\end{equation}
The axioms on $R$-matrix ensure that the category of representations of
$A$  is \emph{a braided tensor category} (which means this tensor
category is equipped with 
a commutativity isomorphism $\sigma_{V_1, V_2} : V_1 \otimes V_2 \to V_2
\otimes V_1$, an associativity isomorphism $\alpha_{V_1, V_2, V_3}: (V_1
\otimes V_2) \otimes V_3 \to V_1 \otimes (V_2 \otimes V_3)$ and unit
morphisms $\lambda_{V}: 1 \otimes V \to V, \rho_V: V \otimes 1 \to V$,
which satisfy certain compatibility axioms. For more details see
\cite{Etingof:1998}).

For any finite-dimensional Hopf algebra $A$,  Drinfeld constructed \cite{Drinfeld:1986}
structure of quasi-triangular Hopf algebra on \emph{Drinfeld double}
\[D(A) = A \otimes A^{* op}\]
 If $a_i$ is a basis in $A$ and $a^i$ is a
dual basis in $A^{*}$ then $D(A)$ is a quasi-triangular Hopf algebra
with the R-matrix 
\begin{equation}
  \mathcal{R}  = \sum_{i} (a_i \otimes 1_{A^*}) \otimes (1_A \otimes a^i)
\end{equation}

The quasi-triangular structure on $\Qaff_q( \gq)$ (for $\gq$ of finite
type) can be found using the fact that $\Qaff_q( \gq)$ is almost
Drinfeld double of it Borel subalgebra $\Qaff_q( \gq^{+})$
\cite{Etingof:1998,Tanisaki_1991}. The structure of Hopf algebra on
$\Qaff_q( \gq^{\pm})$ naturally induces the structure of Hopf algebra on
its dual $\Qaff_q( \gq^{\pm})^{*}$. 
Moreover, V.~Drinfeld has shown \cite{Drinfeld:1986} that there exists an isomorphism of Hopf
algebras $\theta: \Qaff_q( \gq^{-}) \to \Qaff_q(
\gq^{+})^{*\mathrm{op}}$ such that for $h$ in the extended Cartan subalgebra of
$\hat \gq$ we have $\theta(q^{h}) = q^{h}$ if we identify the extended Cartan
subalgebra\footnote{If $\gq$ of finite type has rank $r$, the extended Cartan
  subalgebra of $\hat \gq$ has rank $r+2$ and is generated by the
  coroots of $\gq$, the center and the energy element}
with its dual using the standard Cartan-Killing quadratic
form. The isomorphism $\theta$ provides a non-degenerate bilinear form
(called Drinfeld pairing) $\Qaff_q( \gq^{+}) \otimes \Qaff_q( \gq^{-})
\to \BC$.  Using Drinfeld pairing we present the Drinfeld double of the Borel
$\Qaff_q( \gq^{+})$ as 
\begin{equation}
  D(\Qaff_q( \gq^{+})) = \Qaff_q( \gq^{+}) \otimes \Qaff_q( \gq^{-})
\end{equation}
The Drinfeld double $D(\Qaff_q( \gq^{+}))$ contains two-sided ideal
$H$ generated by $q^{h} \otimes 1 - 1 \otimes q^{h}$ for $h$ in the
extended Cartan of $\hat \gq$. One can check that the map $f: D(\Qaff_q( \gq^{+}))/H \to
\Qaff_q( \gq)$ is isomorphism. The pushforward by $f$ of the universal
$R$-matrix on $D(\Qaff_q( \gq^{+}))/H$ will give a universal $R$-matrix
on $\Qaff_q( \gq)$. An explicit formula can be found in
\cite{Khoroshkin_1992b}.

\subsection{Dual interpretation} 
 The $q$-character $\chi_q(V)$ for a representation $V$ of $\Qaff_q( \gq)$ can be interpreted not only as a formal expression
 encoding generalized eigenvalues of diagonal generators of $\Qaff_q(
 \gq)$ but also as a certain element of $\Qaff_q( \gq)$, in fact, that
 is how the  $q$-character was originally introduced in
 \cite{Frenkel:1998}. The tensor product $U_q (\hat \gq) \otimes U_q
 (\hat \gq)$ contains a special element $\CalR$ known as the universal
 R-matrix \cite{Frenkel:1998}.

Let $(V,\pi_V)$ denote a finite-dimensional representation of $U_q (\hat
\gq)$. Let $V(\xi)$ denote the twist of representation $V$ by spectral
parameter $\xi$. 
 Define the transfer matrix $t_V(\xi)$ associated to $V$ 
 as an element of $\Qaff_q( \gq)$
given by the trace over $V$:
\begin{equation}
\label{eq:transfer}
  t_V(\xi) = \tr_V q^{\rho} (\pi_{V(\xi)} \otimes \mathrm{id}) \mathcal{R} 
\end{equation}
where $\rho = \sum \tilde h_i$ with $\tilde h_i = \tilde a_{ij}h_j$.

The explicit formula for universal R-matrix is given in
\cite{Khoroshkin_1993} (eq 42),   \cite{Khoroshkin:1994uj}
\begin{equation}
  \CalR = \CalR^{+} \CalR^{0} \CalR^{-} T
\end{equation}
where 
\begin{equation}
  \begin{aligned}
&  \CalR^{0} = \exp \left( -(q^{1/2} - q^{-1/2}) \sum_{n >0}
    \frac{n}{[n]_q} \tilde a_{ij}(q^n) h_{i,n} \otimes h_{i,-n} \, , 
 \right)\\
& T = q^{-\frac 1 2 \tilde a_{ij}h_i \otimes h_j}
  \end{aligned}
\end{equation}
and the factors $\CalR^{\pm} \in \Qaff_q({\gq^\pm}) \otimes U_q
(\hat{\gq^\mp})$.

The $q$-character $\chi_q(V(\xi))$ is essentially the diagonal
projection of $t_{V}(\xi)$, i.e. 
\begin{equation}
\begin{aligned}
&   \chi_q(V(\xi)) = \tr_{V} \left(
q^{\rho} \Gamma_{V} (\pi_{V} \otimes 1) (T)  \right) \\
& \qquad  \Gamma_{V} =  \exp \left( - (q^{1/2} - q^{-1/2})
  \sum_{n>0} \frac{n}{[n]_q} \tilde a_{ij}(q^{n}) \xi^{-n}
  \pi_{V}(h_{i,n}) \otimes h_{j,-n} \right) \end{aligned}
\end{equation}
Now we introduce special elements $\hat Y_{i,a}$  of $\Qaff_q( \gq)$ by
definition 
\begin{equation}
  \hat Y_{i,\zeta} = q^{ (\rho, \omega_i)} q^{-\frac 1 2 \tilde h_i} \exp
  \left ( -(q^{1/2} - q^{-1/2}) \sum_{n > 0} \tilde h_{i,-n} \zeta^{n} \xi^{-n} 
\right)
\end{equation}
where $\omega_i$ are fundamental weights. 

The $q$-character is a sum of $\dim V$ terms. Each term in
$\chi_q(V(\xi))$ corresponds to an eigenvector of $V$. 
The eigenvalue of a given eigenvector of $V$ is encoded by  Drinfeld
polynomial loop-weights which can be factorized into a product of basic
monomials $(P) = (1,\dots, 1,  {1 - \zeta/\xi}, 1, \dots, 1)$ with $1  -
\zeta/\xi$, say, in $i$-th position. Each such factor produces a factor $\hat Y_{i,\zeta}$ in
the given term in the $q$-character associated with this eigenvalue.

\subsubsection{Twisted $\qe$-characters}

For our purposes it is  natural to consider the twisted transfer
matrix defined by the formula similar to the  (\ref{eq:transfer})  in which  $q^{\sum \tilde h_i }$ is replaced by $\prod \qe_{\iv}^{-\tilde h_i}$:
\begin{equation}
  t_{\qe,V(\xi)} = \tr_{V(\xi)} \prod_{i \in i} \qe_{\iv}^{-\tilde h_i}
  (\pi_{V(\xi)} \otimes \mathrm{id} )\mathcal{R}
\end{equation}
and then set 
\begin{equation}
  \hat Y_{i}(\xi) =  q^{-\frac 12 \tilde h_i} \exp
  \left ( -(q^{\frac 12} - q^{-\frac 12}) \sum_{n > 0} \tilde h_{i,-n}  \xi^{-n} 
\right)
\end{equation}
The $\qe$-twisted $q$-character is a sum of monomials associated to
eigenvectors. Each monomial is a product of factors $\hat Y_{i,a}^{\pm}$
encoding the generalized eigenvalues with an additional 
$\qe_{\iv}$-dependent factor precisely like in
(\ref{eq:chi-character}) for the quiver theory without
fundamental matter, with  $\sP_i = \qe_{\iv}$. For example, 
the $\qe$-twisted $q$-character associated to the $A_1$ quiver and the
fundamental evaluation module is
\begin{equation}
t_{\qe,L_1} =    \qe^{-\frac 12} \hat Y_1(\xi) + \qe^{\frac 12}  \frac{1} {\hat Y_1(q^{-1} \xi)}
\end{equation}

\begin{proposition}
The equations
\begin{equation}
 \text{  $\qe$-twisted $q$-characters $\chi_i$ = $c$-number polynomials $T_i(\xi)$}
\end{equation} can be interpreted as follows:  there exists a special eigenvector\footnote{It is not clear whether
  this eigenvector belongs to a natural $\Qaff_q( \gq)$-module though} of the $\Qaff_q( \gq)$ elements $\hat Y_i(\xi)$  with the generalized eigenvalues equal to our
$c$-number functions $Y_i(\xi)$ and the eigenvalue of the $\qe$-twisted $q$-character
viewed as an element of $\Qaff_q( \gq)$ equal to the polynomial $T_i(\xi)$.
\end{proposition}
It looks like our formulation is $T$-$Q$ dual (in a sense of \cite{Krichever:1996qd}) to the results of the recent paper \cite{Frenkel:2013dh} on the category $\CalO$ of representations
of the quantum affine algebras, at least for the finite dimensional $\gq$.

\bibliography{lib-sp1}

\def\cprime{$'$}
\providecommand{\href}[2]{#2}\begingroup\raggedright\begin{thebibliography}{100}

\bibitem{Moore:1997dj}
G.~W. Moore, N.~Nekrasov, and S.~Shatashvili, ``{Integrating over Higgs
  branches},'' \href{http://dx.doi.org/10.1007/PL00005525}{{\em Commun. Math.
  Phys.} {\bfseries 209} (2000) 97--121},
\href{http://arxiv.org/abs/hep-th/9712241}{{\ttfamily arXiv:hep-th/9712241}}.

\bibitem{Gerasimov:2006zt}
A.~A. Gerasimov and S.~L. Shatashvili, ``{Higgs Bundles, Gauge Theories and
  Quantum Groups},'' \href{http://dx.doi.org/10.1007/s00220-007-0369-1}{{\em
  Commun.Math.Phys.} {\bfseries 277} (2008) 323--367},
\href{http://arxiv.org/abs/hep-th/0609024}{{\ttfamily arXiv:hep-th/0609024
  [hep-th]}}.

\bibitem{Gerasimov:2007ap}
A.~A. Gerasimov and S.~L. Shatashvili, ``{Two-dimensional Gauge Theories and
  Quantum Integrable Systems},''
\href{http://arxiv.org/abs/0711.1472}{{\ttfamily arXiv:0711.1472 [hep-th]}}.

\bibitem{Nekrasov:2009zz}
N.~Nekrasov and S.~Shatashvili, ``{Bethe Ansatz and supersymmetric vacua},''
\href{http://dx.doi.org/10.1063/1.3149487}{{\em AIP Conf. Proc.} {\bfseries
  1134} (2009) 154--169}.

\bibitem{Nekrasov:2009rc}
N.~A. Nekrasov and S.~L. Shatashvili, ``{Quantization of Integrable Systems and
  Four Dimensional Gauge Theories},''
\href{http://arxiv.org/abs/0908.4052}{{\ttfamily arXiv:0908.4052 [hep-th]}}.

\bibitem{Nekrasov:2009ui}
N.~A. Nekrasov and S.~L. Shatashvili, ``{Quantum integrability and
  supersymmetric vacua},'' \href{http://dx.doi.org/10.1143/PTPS.177.105}{{\em
  Prog. Theor. Phys. Suppl.} {\bfseries 177} (2009) 105--119},
\href{http://arxiv.org/abs/0901.4748}{{\ttfamily arXiv:0901.4748 [hep-th]}}.

\bibitem{Nekrasov:2009uh}
N.~A. Nekrasov and S.~L. Shatashvili, ``{Supersymmetric vacua and Bethe
  ansatz},'' \href{http://dx.doi.org/10.1016/j.nuclphysbps.2009.07.047}{{\em
  Nucl. Phys. Proc. Suppl.} {\bfseries 192-193} (2009) 91--112},
\href{http://arxiv.org/abs/0901.4744}{{\ttfamily arXiv:0901.4744 [hep-th]}}.

\bibitem{Nekrasov:2010ka}
N.~Nekrasov and E.~Witten, ``{The Omega Deformation, Branes, Integrability, and
  Liouville Theory},'' \href{http://dx.doi.org/10.1007/JHEP09(2010)092}{{\em
  JHEP} {\bfseries 09} (2010) 092},
\href{http://arxiv.org/abs/1002.0888}{{\ttfamily arXiv:1002.0888 [hep-th]}}.

\bibitem{Nekrasov:2011bc}
N.~Nekrasov, A.~Rosly, and S.~Shatashvili, ``{Darboux coordinates, Yang-Yang
  functional, and gauge theory},''
  \href{http://dx.doi.org/10.1016/j.nuclphysbps.2011.04.150}{{\em
  Nucl.Phys.Proc.Suppl.} {\bfseries 216} (2011) 69--93},
\href{http://arxiv.org/abs/1103.3919}{{\ttfamily arXiv:1103.3919 [hep-th]}}.

\bibitem{Witten:1989wf}
E.~Witten, ``{Gauge theories and integrable lattice models},''
\href{http://dx.doi.org/10.1016/0550-3213(89)90232-0}{{\em Nucl.Phys.}
  {\bfseries B322} (1989) 629}.

\bibitem{Gorsky:1993dq}
A.~Gorsky and N.~Nekrasov, ``{Relativistic Calogero-Moser model as gauged WZW
  theory},'' \href{http://dx.doi.org/10.1016/0550-3213(94)00499-5}{{\em Nucl.
  Phys.} {\bfseries B436} (1995) 582--608},
\href{http://arxiv.org/abs/hep-th/9401017}{{\ttfamily arXiv:hep-th/9401017}}.

\bibitem{Gorsky:1993pe}
A.~Gorsky and N.~Nekrasov, ``{Hamiltonian systems of Calogero type and
  two-dimensional Yang-Mills theory},''
  \href{http://dx.doi.org/10.1016/0550-3213(94)90429-4}{{\em Nucl. Phys.}
  {\bfseries B414} (1994) 213--238},
\href{http://arxiv.org/abs/hep-th/9304047}{{\ttfamily arXiv:hep-th/9304047}}.

\bibitem{Gorsky:1994dj}
A.~Gorsky and N.~Nekrasov, ``{Elliptic Calogero-Moser system from
  two-dimensional current algebra},''
\href{http://arxiv.org/abs/hep-th/9401021}{{\ttfamily arXiv:hep-th/9401021}}.

\bibitem{Nekrasov:2004sem}
N.~Nekrasov, ``{On the BPS/CFT correspondence},'' Feb. 3, 2004.
\newblock
  \url{{http://www.science.uva.nl/research/itf/strings/stringseminar2003-4.html}}.
  {lecture at the string theory group seminar, University of Amsterdam}.

\bibitem{Nakajima:1994r}
H.~Nakajima, ``Gauge theory on resolutions of simple singularities and simple
  {L}ie algebras,'' \href{http://dx.doi.org/10.1155/S1073792894000085}{{\em
  Internat. Math. Res. Notices} no.~2, (1994) 61--74}.
  \url{http://dx.doi.org/10.1155/S1073792894000085}.

\bibitem{Nakajima:1998}
H.~Nakajima, ``Quiver varieties and {K}ac-{M}oody algebras,''
  \href{http://dx.doi.org/10.1215/S0012-7094-98-09120-7}{{\em Duke Math. J.}
  {\bfseries 91} no.~3, (1998) 515--560}.
  \url{http://dx.doi.org/10.1215/S0012-7094-98-09120-7}.

\bibitem{Vafa:1994tf}
C.Vafa and E.Witten, ``{A Strong coupling test of S duality},''
  \href{http://dx.doi.org/10.1016/0550-3213(94)90097-3}{{\em Nucl.Phys.}
  {\bfseries B431} (1994) 3--77},
  \href{http://arxiv.org/abs/hep-th/9408074}{{\ttfamily arXiv:hep-th/9408074}}.

\bibitem{Losev:1995cr}
A.~Losev, G.~W. Moore, N.~Nekrasov, and S.~Shatashvili, ``{Four-dimensional
  avatars of two-dimensional RCFT},''
  \href{http://dx.doi.org/10.1016/0920-5632(96)00015-1}{{\em Nucl. Phys. Proc.
  Suppl.} {\bfseries 46} (1996) 130--145},
\href{http://arxiv.org/abs/hep-th/9509151}{{\ttfamily arXiv:hep-th/9509151}}.

\bibitem{Nekrasov:2002qd}
N.~A. Nekrasov, ``{Seiberg-Witten prepotential from instanton counting},'' {\em
  Adv.Theor.Math.Phys.} {\bfseries 7} (2004) 831--864,
  \href{http://arxiv.org/abs/hep-th/0206161}{{\ttfamily arXiv:hep-th/0206161
  [hep-th]}}.
To Arkady Vainshtein on his 60th anniversary.

\bibitem{Losev:2003py}
A.~S. Losev, A.~Marshakov, and N.~A. Nekrasov, ``{Small instantons, little
  strings and free fermions},''
\href{http://arxiv.org/abs/hep-th/0302191}{{\ttfamily arXiv:hep-th/0302191
  [hep-th]}}.

\bibitem{Nekrasov:2003rj}
N.~Nekrasov and A.~Okounkov, ``{Seiberg-Witten theory and random partitions},''
\href{http://arxiv.org/abs/hep-th/0306238}{{\ttfamily arXiv:hep-th/0306238
  [hep-th]}}.

\bibitem{Alday:2009aq}
L.~F. Alday, D.~Gaiotto, and Y.~Tachikawa, ``{Liouville Correlation Functions
  from Four-dimensional Gauge Theories},''
  \href{http://dx.doi.org/10.1007/s11005-010-0369-5}{{\em Lett. Math. Phys.}
  {\bfseries 91} (2010) 167--197},
\href{http://arxiv.org/abs/0906.3219}{{\ttfamily arXiv:0906.3219 [hep-th]}}.

\bibitem{NP2012a}
N.~Nekrasov and V.~Pestun, ``{Seiberg-Witten} geometry of $\mathcal{N}=2$
  quiver gauge theories,'' \href{http://arxiv.org/abs/1211.2240}{{\ttfamily
  arXiv:1211.2240 [hep-th]}}.

\bibitem{Seiberg:1994aj}
N.~Seiberg and E.~Witten, ``Monopoles, duality and chiral symmetry breaking in
  {N=2} supersymmetric {QCD},'' {\em Nucl. Phys.} {\bfseries B431} (1994)
  484--550,
\href{http://arxiv.org/abs/hep-th/9408099}{{\ttfamily hep-th/9408099}}.

\bibitem{Seiberg:1994rs}
N.~Seiberg and E.~Witten, ``Electric - magnetic duality, monopole condensation,
  and confinement in {N=2} supersymmetric {Y}ang-{Mi}lls theory,'' {\em Nucl.
  Phys.} {\bfseries B426} (1994) 19--52,
\href{http://arxiv.org/abs/hep-th/9407087}{{\ttfamily hep-th/9407087}}.

\bibitem{Losev:1997tp}
A.~Losev, N.~Nekrasov, and S.~L. Shatashvili, ``{Issues in topological gauge
  theory},'' \href{http://dx.doi.org/10.1016/S0550-3213(98)00628-2}{{\em Nucl.
  Phys.} {\bfseries B534} (1998) 549--611},
\href{http://arxiv.org/abs/hep-th/9711108}{{\ttfamily arXiv:hep-th/9711108}}.

\bibitem{Lossev:1997bz}
A.~Losev, N.~Nekrasov, and S.~L. Shatashvili, ``{Testing Seiberg-Witten
  solution},''
\href{http://arxiv.org/abs/hep-th/9801061}{{\ttfamily arXiv:hep-th/9801061}}.

\bibitem{Nekrasov:1996cz}
N.~Nekrasov, ``{Five dimensional gauge theories and relativistic integrable
  systems},'' \href{http://dx.doi.org/10.1016/S0550-3213(98)00436-2}{{\em Nucl.
  Phys.} {\bfseries B531} (1998) 323--344},
\href{http://arxiv.org/abs/hep-th/9609219}{{\ttfamily arXiv:hep-th/9609219}}.

\bibitem{Poghossian:2010pn}
R.~Poghossian, ``{Deforming SW curve},''
  \href{http://dx.doi.org/10.1007/JHEP04(2011)033}{{\em JHEP} {\bfseries 1104}
  (2011) 033},
\href{http://arxiv.org/abs/1006.4822}{{\ttfamily arXiv:1006.4822 [hep-th]}}.

\bibitem{Dorey:2011pa}
N.~Dorey, S.~Lee, and T.~J. Hollowood, ``{Quantization of Integrable Systems
  and a 2d/4d Duality},'' \href{http://dx.doi.org/10.1007/JHEP10(2011)077}{{\em
  JHEP} {\bfseries 1110} (2011) 077},
\href{http://arxiv.org/abs/1103.5726}{{\ttfamily arXiv:1103.5726 [hep-th]}}.

\bibitem{Chen:2011sj}
H.-Y. Chen, N.~Dorey, T.~J. Hollowood, and S.~Lee, ``{A New 2d/4d Duality via
  Integrability},'' \href{http://dx.doi.org/10.1007/JHEP09(2011)040}{{\em JHEP}
  {\bfseries 1109} (2011) 040},
\href{http://arxiv.org/abs/1104.3021}{{\ttfamily arXiv:1104.3021 [hep-th]}}.

\bibitem{Bao:2011rc}
L.~Bao, E.~Pomoni, M.~Taki, and F.~Yagi, ``{M5-Branes, Toric Diagrams and Gauge
  Theory Duality},'' \href{http://dx.doi.org/10.1007/JHEP04(2012)105}{{\em
  JHEP} {\bfseries 1204} (2012) 105},
\href{http://arxiv.org/abs/1112.5228}{{\ttfamily arXiv:1112.5228 [hep-th]}}.

\bibitem{Mironov:2012ba}
A.~Mironov, A.~Morozov, B.~Runov, Y.~Zenkevich, and A.~Zotov, ``{Spectral
  Duality Between Heisenberg Chain and Gaudin Model},''
\href{http://arxiv.org/abs/1206.6349}{{\ttfamily arXiv:1206.6349 [hep-th]}}.

\bibitem{Gorsky:1995zq}
A.~Gorsky, I.~Krichever, A.~Marshakov, A.~Mironov, and A.~Morozov,
  ``{Integrability and Seiberg-Witten exact solution},''
  \href{http://dx.doi.org/10.1016/0370-2693(95)00723-X}{{\em Phys.Lett.}
  {\bfseries B355} (1995) 466--474},
\href{http://arxiv.org/abs/hep-th/9505035}{{\ttfamily arXiv:hep-th/9505035
  [hep-th]}}.

\bibitem{Gorsky:1996hs}
A.~Gorsky, A.~Marshakov, A.~Mironov, and A.~Morozov, ``{N=2 supersymmetric QCD
  and integrable spin chains: Rational case $N_f \leq 2N_c$},''
  \href{http://dx.doi.org/10.1016/0370-2693(96)00480-7}{{\em Phys.Lett.}
  {\bfseries B380} (1996) 75--80},
\href{http://arxiv.org/abs/hep-th/9603140}{{\ttfamily arXiv:hep-th/9603140
  [hep-th]}}.

\bibitem{Gorsky:1997jq}
A.~Gorsky, S.~Gukov, and A.~Mironov, ``{Multiscale N=2 SUSY field theories,
  integrable systems and their stringy / brane origin. 1.},''
  \href{http://dx.doi.org/10.1016/S0550-3213(98)00055-8}{{\em Nucl.Phys.}
  {\bfseries B517} (1998) 409--461},
\href{http://arxiv.org/abs/hep-th/9707120}{{\ttfamily arXiv:hep-th/9707120
  [hep-th]}}.

\bibitem{Gorsky:1997mw}
A.~Gorsky, S.~Gukov, and A.~Mironov, ``{SUSY field theories, integrable systems
  and their stringy / brane origin. 2.},''
  \href{http://dx.doi.org/10.1016/S0550-3213(98)00106-0}{{\em Nucl.Phys.}
  {\bfseries B518} (1998) 689--713},
\href{http://arxiv.org/abs/hep-th/9710239}{{\ttfamily arXiv:hep-th/9710239
  [hep-th]}}.

\bibitem{Bazhanov:1998dq}
V.~V. Bazhanov, S.~L. Lukyanov, and A.~B. Zamolodchikov, ``{Integrable
  structure of conformal field theory. 3. The Yang-Baxter relation},''
  \href{http://dx.doi.org/10.1007/s002200050531}{{\em Commun.Math.Phys.}
  {\bfseries 200} (1999) 297--324},
\href{http://arxiv.org/abs/hep-th/9805008}{{\ttfamily arXiv:hep-th/9805008
  [hep-th]}}.

\bibitem{Dorey:2007zx}
P.~Dorey, C.~Dunning, and R.~Tateo, ``{The ODE/IM Correspondence},''
  \href{http://dx.doi.org/10.1088/1751-8113/40/32/R01}{{\em J.Phys.} {\bfseries
  A40} (2007) R205},
\href{http://arxiv.org/abs/hep-th/0703066}{{\ttfamily arXiv:hep-th/0703066
  [hep-th]}}.

\bibitem{Gerasimov:2005qz}
A.~Gerasimov, S.~Kharchev, D.~Lebedev, and S.~Oblezin, ``{On a class of
  representations of the Yangian and moduli space of monopoles},''
\href{http://dx.doi.org/10.1007/s00220-005-1417-3}{{\em Commun.Math.Phys.}
  {\bfseries 260} (2005) 511--525}.

\bibitem{Gerasimov:2005}
A.~{Gerasimov}, S.~{Kharchev}, D.~{Lebedev}, and S.~{Oblezin}, ``{On a Class of
  Representations of Quantum Groups},'' {\em ArXiv Mathematics e-prints} (Jan.,
  2005) , \href{http://arxiv.org/abs/arXiv:math/0501473}{{\ttfamily
  arXiv:math/0501473}}.

\bibitem{Galakhov:2012hy}
D.~Galakhov, A.~Mironov, A.~Morozov, A.~Smirnov, A.~Mironov, {\em et~al.},
  ``{Three-dimensional extensions of the Alday-Gaiotto-Tachikawa relation},''
  \href{http://dx.doi.org/10.1007/s11232-012-0088-4}{{\em Theor.Math.Phys.}
  {\bfseries 172} (2012) 939--962},
\href{http://arxiv.org/abs/1104.2589}{{\ttfamily arXiv:1104.2589 [hep-th]}}.

\bibitem{Mironov:2009uv}
A.~Mironov and A.~Morozov, ``{Nekrasov Functions and Exact Bohr-Zommerfeld
  Integrals},'' \href{http://dx.doi.org/10.1007/JHEP04(2010)040}{{\em JHEP}
  {\bfseries 1004} (2010) 040},
\href{http://arxiv.org/abs/0910.5670}{{\ttfamily arXiv:0910.5670 [hep-th]}}.

\bibitem{Mironov:2009dv}
A.~Mironov and A.~Morozov, ``{Nekrasov Functions from Exact BS Periods: The
  Case of SU(N)},''
  \href{http://dx.doi.org/10.1088/1751-8113/43/19/195401}{{\em J.Phys.}
  {\bfseries A43} (2010) 195401},
\href{http://arxiv.org/abs/0911.2396}{{\ttfamily arXiv:0911.2396 [hep-th]}}.

\bibitem{Teschner:2010je}
J.~Teschner, ``{Quantization of the Hitchin moduli spaces, Liouville theory,
  and the geometric Langlands correspondence I},'' {\em Adv.Theor.Math.Phys.}
  {\bfseries 15} (2011) 471--564,
\href{http://arxiv.org/abs/1005.2846}{{\ttfamily arXiv:1005.2846 [hep-th]}}.

\bibitem{Muneyuki:2011qu}
K.~Muneyuki, T.-S. Tai, N.~Yonezawa, and R.~Yoshioka, ``{Baxter's T-Q equation,
  $SU(N)/SU(2)^{N-3}$ correspondence and $\Omega$-deformed Seiberg-Witten
  prepotential},'' \href{http://dx.doi.org/10.1007/JHEP09(2011)125}{{\em JHEP}
  {\bfseries 1109} (2011) 125},
\href{http://arxiv.org/abs/1107.3756}{{\ttfamily arXiv:1107.3756 [hep-th]}}.

\bibitem{Drinfeld:1987}
V.~G. Drinfeld, ``A new realization of {Y}angians and of quantum affine
  algebras,'' {\em Dokl. Akad. Nauk SSSR} {\bfseries 296} no.~1, (1987) 13--17.

\bibitem{Knight:1995}
H.~Knight, ``Spectra of tensor products of finite-dimensional representations
  of {Y}angians,'' \href{http://dx.doi.org/10.1006/jabr.1995.1123}{{\em J.
  Algebra} {\bfseries 174} no.~1, (1995) 187--196}.
  \url{http://dx.doi.org/10.1006/jabr.1995.1123}.

\bibitem{Frenkel:1998}
E.~Frenkel and N.~Reshetikhin, ``The {$q$}-characters of representations of
  quantum affine algebras and deformations of {$\mathscr W$}-algebras,''
  \href{http://dx.doi.org/10.1090/conm/248/03823}{{\em Recent developments in
  quantum affine algebras and related topics ({R}aleigh, {NC}, 1998)}
  {\bfseries 248} (1999) 163--205}.
  \url{http://dx.doi.org/10.1090/conm/248/03823}.

\bibitem{Chari:1994}
V.~Chari and A.~Pressley, ``Quantum affine algebras and their
  representations,'' \href{http://arxiv.org/abs/hep-th/9411145}{{\ttfamily
  hep-th/9411145}}.

\bibitem{Witten:1988hf}
E.~Witten, ``{Quantum field theory and the Jones polynomial},''
\href{http://dx.doi.org/10.1007/BF01217730}{{\em Commun. Math. Phys.}
  {\bfseries 121} (1989) 351}.

\bibitem{Kirillov:1991ec}
A.~Kirillov and N.~Y. Reshetikhin, ``Representations of the algebra
  $u_q(sl_2)$, orthogonal polynomials and invariants of links,''.
  \url{http://www.worldscientific.com/worldscibooks/10.1142/1056}.

\bibitem{Reshetikhin:1991tc}
N.~Reshetikhin and V.~Turaev, ``{Invariants of three manifolds via link
  polynomials and quantum groups},''
\href{http://dx.doi.org/10.1007/BF01239527}{{\em Invent.Math.} {\bfseries 103}
  (1991) 547--597}.

\bibitem{Reshetikhin:1990pr}
N.~Y. Reshetikhin and V.~Turaev, ``{Ribbon graphs and their invariants derived
  from quantum groups},''
\href{http://dx.doi.org/10.1007/BF02096491}{{\em Commun.Math.Phys.} {\bfseries
  127} (1990) 1--26}.

\bibitem{Witten:2011zz}
E.~Witten, ``{Fivebranes and Knots},''
\href{http://arxiv.org/abs/1101.3216}{{\ttfamily arXiv:1101.3216 [hep-th]}}.

\bibitem{Nekrasov:2004js}
N.~Nekrasov, H.~Ooguri, and C.~Vafa, ``{S-duality and topological strings},''
  \href{http://dx.doi.org/10.1088/1126-6708/2004/10/009}{{\em JHEP} {\bfseries
  10} (2004) 009},
\href{http://arxiv.org/abs/hep-th/0403167}{{\ttfamily arXiv:hep-th/0403167}}.

\bibitem{Nekrasov:2005bb}
N.~Nekrasov, ``{Z-theory: Chasing M-F-Theory},''
\href{http://dx.doi.org/10.1016/j.crhy.2004.12.011}{{\em Comptes Rendus
  Physique} {\bfseries 6} (2005) 261--269}.

\bibitem{Hernandez:2008}
D.~Hernandez, ``Quantum toroidal algebras and their representations,''
  \href{http://dx.doi.org/10.1007/s00029-009-0502-4}{{\em Selecta Math. (N.S.)}
  {\bfseries 14} no.~3-4, (2009) 701--725},
  \href{http://arxiv.org/abs/abs/0801.2397}{{\ttfamily abs/0801.2397}}.
  \url{http://dx.doi.org/10.1007/s00029-009-0502-4}.

\bibitem{Cherednik:2004}
I.~{Cherednik}, ``{Introduction to double Hecke algebras},'' {\em ArXiv
  Mathematics e-prints} (Apr., 2004) ,
  \href{http://arxiv.org/abs/arXiv:math/0404307}{{\ttfamily
  arXiv:math/0404307}}.

\bibitem{Pestun:berlin}
talk~by V.~Pestun, ``Integrable systems for {4d N=2 ADE} quiver theories from
  instanton counting,'' Aug 2012.
\newblock \url{http://people.physik.hu-berlin.de/~ahoop/pestun.pdf}.

\bibitem{Pestun:SCGP}
talk~by V.~Pestun, ``{Supersymmetric Four-dimensional Quiver Gauge Theories and
  Quantum {ADE} Spin Chains},'' Sep 2012.
\newblock \url{http://scgp.stonybrook.edu/archives/4709}.

\bibitem{Nikita:talk2012i}
talk~by N.~Nekrasov, ``{Seiberg-Witten geometry of N=2 quiver theories, and
  quantization},'' May 2012.
\newblock
  \url{http://brahms.mth.kcl.ac.uk/cgi-bin/main.pl?action=seminars&id=1108}.

\bibitem{Nikita:talk2012ii}
talk~by N.~Nekrasov, ``{Seiberg-Witten Geometry of N=2 Superconformal Theories,
  and ADE Bundles on Curves},'' March 2012.
\newblock
  \url{http://media.scgp.stonybrook.edu/video/video.php?f=20120510_1_qtp.mp4}.

\bibitem{Samson:talk2012}
talk~by S.Shatashvili, ``{Gauge theory angle at integrability},'' May 2012.
\newblock
  \url{http://www.kcl.ac.uk/nms/depts/mathematics/research/theorphysics/pastevents/stringgauge.aspx}.

\bibitem{Samson:talk2013i}
talk~by S.Shatashvili, ``{Integrability and quantization},'' February 2013.
\newblock
  \url{https://indico.desy.de/contributionDisplay.py?contribId=16&confId=6969}.

\bibitem{Samson:talk2013ii}
talk~by S.Shatashvili, ``{Integrability and supersymmetric vacua (IV)},'' July
  2013.
\newblock \url{http://cdsagenda5.ictp.trieste.it/full_display.php?ida=a13168}.

\bibitem{Fucito:2012xc}
F.~Fucito, J.~F. Morales, and D.~R. Pacifici, ``{Deformed Seiberg-Witten Curves
  for ADE Quivers},''
\href{http://arxiv.org/abs/1210.3580}{{\ttfamily arXiv:1210.3580 [hep-th]}}.

\bibitem{Nekrasov:thesis}
N.~Nekrasov, ``{Four dimensional holomorphic theories},''
{\em Princeton University} {\bfseries PhD. thesis} (1996) .

\bibitem{Baulieu:1997nj}
L.~Baulieu, A.~Losev, and N.~Nekrasov, ``{Chern-Simons and twisted
  supersymmetry in various dimensions},''
  \href{http://dx.doi.org/10.1016/S0550-3213(98)00096-0}{{\em Nucl. Phys.}
  {\bfseries B522} (1998) 82--104},
\href{http://arxiv.org/abs/hep-th/9707174}{{\ttfamily arXiv:hep-th/9707174}}.

\bibitem{Schwarz:1995zw}
J.~H. Schwarz, ``{Anomaly - free supersymmetric models in six-dimensions},''
  \href{http://dx.doi.org/10.1016/0370-2693(95)01610-4}{{\em Phys.Lett.}
  {\bfseries B371} (1996) 223--230},
\href{http://arxiv.org/abs/hep-th/9512053}{{\ttfamily arXiv:hep-th/9512053
  [hep-th]}}.

\bibitem{Blum:1997}
J.~Blum and K.~Intriligator, ``{New phases of string theory and 6-D RG fixed
  points via branes at orbifold singularities},''
  \href{http://dx.doi.org/10.1016/S0550-3213(97)00449-5}{{\em Nucl.Phys.}
  {\bfseries B506} (1997) 199--222},
\href{http://arxiv.org/abs/hep-th/9705044}{{\ttfamily arXiv:hep-th/9705044}}.

\bibitem{Blum:1997mm}
J.~D. Blum and K.~A. Intriligator, ``{New phases of string theory and 6-D RG
  fixed points via branes at orbifold singularities},''
  \href{http://dx.doi.org/10.1016/S0550-3213(97)00449-5}{{\em Nucl.Phys.}
  {\bfseries B506} (1997) 199--222},
\href{http://arxiv.org/abs/hep-th/9705044}{{\ttfamily arXiv:hep-th/9705044
  [hep-th]}}.

\bibitem{Seiberg:1996qx}
N.~Seiberg, ``{Nontrivial fixed points of the renormalization group in
  six-dimensions},''
  \href{http://dx.doi.org/10.1016/S0370-2693(96)01424-4}{{\em Phys.Lett.}
  {\bfseries B390} (1997) 169--171},
\href{http://arxiv.org/abs/hep-th/9609161}{{\ttfamily arXiv:hep-th/9609161
  [hep-th]}}.

\bibitem{Kac_1990}
V.~G. Kac, \href{http://dx.doi.org/10.1017/CBO9780511626234}{{\em
  Infinite-dimensional {L}ie algebras}}.
\newblock Cambridge University Press, Cambridge, third~ed., 1990.
\newblock \url{http://dx.doi.org/10.1017/CBO9780511626234}.

\bibitem{Nakajima:1999}
H.~Nakajima, {\em Lectures on Hilbert Schemes of Points on Surfaces}.
\newblock AMS, 1999.
\newblock \href{http://arxiv.org/abs/AMS University Lecture Series, ISBN
  0-8218-1956-9}{{\ttfamily AMS University Lecture Series, ISBN
  0-8218-1956-9}}.

\bibitem{Nakajima:2003Lectures}
H.~{Nakajima} and K.~{Yoshioka}, ``{Lectures on Instanton Counting},'' {\em
  ArXiv Mathematics e-prints} (Nov., 2003) ,
  \href{http://arxiv.org/abs/math/0311058}{{\ttfamily math/0311058}}.

\bibitem{Nakajima:2003}
H.~Nakajima, ``{$t$}-analogs of {$q$}-characters of quantum affine algebras of
  type {$A_n,D_n$},''. \url{http://dx.doi.org/10.1090/conm/325/05669}.

\bibitem{Shadchin:2005mx}
S.~Shadchin, ``{On certain aspects of string theory/gauge theory
  correspondence},'' \href{http://arxiv.org/abs/hep-th/0502180}{{\ttfamily
  arXiv:hep-th/0502180 [hep-th]}}.
Ph.D. Thesis.

\bibitem{Pestun:2007rz}
V.~Pestun, ``{Localization of gauge theory on a four-sphere and supersymmetric
  Wilson loops},'' \href{http://dx.doi.org/10.1007/s00220-012-1485-0}{{\em
  Commun.Math.Phys.} {\bfseries 313} (2012) 71--129},
\href{http://arxiv.org/abs/0712.2824}{{\ttfamily arXiv:0712.2824 [hep-th]}}.

\bibitem{Nekrasov:1998ss}
N.~Nekrasov and A.~S. Schwarz, ``{Instantons on noncommutative ${\BR}^{4}$ and
  $(2,0)$-superconformal six dimensional theory},''
  \href{http://dx.doi.org/10.1007/s002200050490}{{\em Commun. Math. Phys.}
  {\bfseries 198} (1998) 689--703},
\href{http://arxiv.org/abs/hep-th/9802068}{{\ttfamily arXiv:hep-th/9802068}}.

\bibitem{Atiyah:1978ri}
M.~Atiyah, N.~J. Hitchin, V.~Drinfeld, and Y.~Manin, ``{Construction of
  Instantons},''
\href{http://dx.doi.org/10.1016/0375-9601(78)90141-X}{{\em Phys.Lett.}
  {\bfseries A65} (1978) 185--187}.

\bibitem{Kozlowski:2010tv}
K.~Kozlowski and J.~Teschner, ``{TBA for the Toda chain},''
\href{http://arxiv.org/abs/1006.2906}{{\ttfamily arXiv:1006.2906 [math-ph]}}.

\bibitem{Polyakov:1987ez}
A.~M. Polyakov, {\em Gauge fields and strings}.
\newblock Harwood, Chur, Switzerland, 1987.

\bibitem{Shadchin:2005hp}
S.~Shadchin, ``{Status report on the instanton counting},'' {\em SIGMA}
  {\bfseries 2} (2006) 008,
\href{http://arxiv.org/abs/hep-th/0601167}{{\ttfamily arXiv:hep-th/0601167}}.

\bibitem{Givental:1997}
A.~{Givental}, ``{A mirror theorem for toric complete intersections},'' {\em
  arXiv} (Jan., 1997) 1016,
  \href{http://arxiv.org/abs/alg-geom/9701016}{{\ttfamily alg-geom/9701016}}.

\bibitem{Bazhanov:1989yk}
V.~Bazhanov and N.~Reshetikhin, ``{Restricted Solid on Solid Models Connected
  With Simply Based Algebras and Conformal Field Theory},''
\href{http://dx.doi.org/10.1088/0305-4470/23/9/012}{{\em J.Phys.} {\bfseries
  A23} (1990) 1477}.

\bibitem{Kirillov:1987zz}
A.~Kirillov and N.~Y. Reshetikhin, ``{Exact solution of the integrable XXZ
  Heisenberg model with arbitrary spin. I. The ground state and the excitation
  spectrum},''
\href{http://dx.doi.org/10.1088/0305-4470/20/6/038}{{\em J.Phys.} {\bfseries
  A20} (1987) 1565--1585}.

\bibitem{Kuniba:1994na}
A.~Kuniba and J.~Suzuki, ``{Analytic Bethe Ansatz for fundamental
  representations of Yangians},''
  \href{http://dx.doi.org/10.1007/BF02101234}{{\em Commun.Math.Phys.}
  {\bfseries 173} (1995) 225--264},
\href{http://arxiv.org/abs/hep-th/9406180}{{\ttfamily arXiv:hep-th/9406180
  [hep-th]}}.

\bibitem{Kuniba:2010ir}
A.~Kuniba, T.~Nakanishi, and J.~Suzuki, ``{T-systems and Y-systems in
  integrable systems},''
  \href{http://dx.doi.org/10.1088/1751-8113/44/10/103001}{{\em J.Phys.}
  {\bfseries A44} (2011) 103001},
\href{http://arxiv.org/abs/1010.1344}{{\ttfamily arXiv:1010.1344 [hep-th]}}.

\bibitem{Fucito:2011pn}
F.~Fucito, J.~Morales, D.~R. Pacifici, and R.~Poghossian, ``{Gauge theories on
  $\Omega$-backgrounds from non commutative Seiberg-Witten curves},''
  \href{http://dx.doi.org/10.1007/JHEP05(2011)098}{{\em JHEP} {\bfseries 1105}
  (2011) 098},
\href{http://arxiv.org/abs/1103.4495}{{\ttfamily arXiv:1103.4495 [hep-th]}}.

\bibitem{Hernandez:2011}
D.~Hernandez, ``The algebra {$U_q(\hat{sl}_\infty)$} and applications,''
  \href{http://dx.doi.org/10.1016/j.jalgebra.2010.04.002}{{\em J. Algebra}
  {\bfseries 329} (2011) 147--162}.
  \url{http://dx.doi.org/10.1016/j.jalgebra.2010.04.002}.

\bibitem{Frenkel:2002}
E.~Frenkel and E.~Mukhin, ``The {H}opf algebra {${\rm
  Rep}\,U_q\widehat{\mathfrak{g}\mathfrak{l}_\infty}$},''
  \href{http://dx.doi.org/10.1007/PL00012603}{{\em Selecta Math. (N.S.)}
  {\bfseries 8} no.~4, (2002) 537--635}.
  \url{http://dx.doi.org/10.1007/PL00012603}.

\bibitem{Faddeev:1996iy}
L.~Faddeev, ``{How algebraic Bethe ansatz works for integrable model},''
\href{http://arxiv.org/abs/hep-th/9605187}{{\ttfamily arXiv:hep-th/9605187
  [hep-th]}}.

\bibitem{Frenkel:2013dh}
E.~{Frenkel} and D.~{Hernandez}, ``{Baxter's Relations and Spectra of Quantum
  Integrable Models},'' {\em ArXiv e-prints} (Aug., 2013) ,
  \href{http://arxiv.org/abs/1308.3444}{{\ttfamily arXiv:1308.3444 [math.QA]}}.

\bibitem{Baxter:1985}
R.~J. Baxter, {\em Exactly solved models in statistical mechanics}, vol.~1 of
  {\em Ser. Adv. Statist. Mech.}
\newblock World Sci. Publishing, Singapore, 1985.

\bibitem{Frenkel:2001}
E.~Frenkel and E.~Mukhin, ``Combinatorics of {$q$}-characters of
  finite-dimensional representations of quantum affine algebras,''
  \href{http://dx.doi.org/10.1007/s002200000323}{{\em Comm. Math. Phys.}
  {\bfseries 216} no.~1, (2001) 23--57}.
  \url{http://dx.doi.org/10.1007/s002200000323}.

\bibitem{Hernandez:2004}
D.~Hernandez, ``The {$t$}-analogs of {$q$}-characters at roots of unity for
  quantum affine algebras and beyond,''
  \href{http://dx.doi.org/10.1016/j.jalgebra.2004.02.022}{{\em J. Algebra}
  {\bfseries 279} no.~2, (2004) 514--557}.
  \url{http://dx.doi.org/10.1016/j.jalgebra.2004.02.022}.

\bibitem{Etingof:2000}
P.~{Etingof} and A.~{Varchenko}, ``{Dynamical Weyl groups and applications},''
  {\em ArXiv Mathematics e-prints} (Oct., 2000) ,
  \href{http://arxiv.org/abs/arXiv:math/0011001}{{\ttfamily
  arXiv:math/0011001}}.

\bibitem{Smirnov:2003}
F.~A. {Smirnov}, ``{Baxter Equations and Deformation of Abelian
  Differentials},'' \href{http://dx.doi.org/10.1142/S0217751X04020543}{{\em
  International Journal of Modern Physics A} {\bfseries 19} (2004) 396--417},
  \href{http://arxiv.org/abs/math-ph/0302014}{{\ttfamily math-ph/0302014}}.

\bibitem{Drinfeld:1985}
V.~G. Drinfeld, ``Hopf algebras and the quantum {Y}ang-{B}axter equation,''
  {\em Dokl. Akad. Nauk SSSR} {\bfseries 283} no.~5, (1985) 1060--1064.

\bibitem{Jimbo:1985}
M.~Jimbo, ``A {$q$}-difference analogue of {$U({\mathfrak g})$} and the
  {Y}ang-{B}axter equation,'' \href{http://dx.doi.org/10.1007/BF00704588}{{\em
  Lett. Math. Phys.} {\bfseries 10} no.~1, (1985) 63--69}.
  \url{http://dx.doi.org/10.1007/BF00704588}.

\bibitem{Sklyanin:1978pj}
E.~Sklyanin and L.~Faddeev, ``{Quantum Mechanical Approach to Completely
  Integrable Field Theory Models},''
{\em Sov.Phys.Dokl.} {\bfseries 23} (1978) 902--904.

\bibitem{Sklyanin:1979}
E.~K. Sklyanin, L.~A. Takhtajan, and L.~D. Faddeev, ``Quantum inverse problem
  method. {I},'' {\em Teoret. Mat. Fiz.} {\bfseries 40} no.~2, (1979) 194--220.

\bibitem{Takhtajan:1979iv}
L.~Takhtajan and L.~Faddeev, ``{The Quantum method of the inverse problem and
  the Heisenberg XYZ model},''
{\em Russ.Math.Surveys} {\bfseries 34} (1979) 11--68.

\bibitem{Sklyanin:1980ij}
E.~Sklyanin, ``{Quantum version of the method of inverse scattering problem},''
\href{http://dx.doi.org/10.1007/BF01091462}{{\em J.Sov.Math.} {\bfseries 19}
  (1982) 1546--1596}.

\bibitem{Sklyanin:1982tf}
E.~Sklyanin, ``{Some algebraic structures connected with the Yang-Baxter
  equation},''
{\em Funct.Anal.Appl.} {\bfseries 16} (1982) 263--270.

\bibitem{Faddeev:1987ih}
L.~Faddeev, N.~Y. Reshetikhin, and L.~Takhtajan, ``{Quantization of Lie Groups
  and Lie Algebras},''
{\em Leningrad Math.J.} {\bfseries 1} (1990) 193--225.

\bibitem{Kulish:1983md}
P.~Kulish and N.~Y. Reshetikhin, ``{Quantum linear problem for the Sine-Gordon
  equation and higher representation},''
\href{http://dx.doi.org/10.1007/BF01084171}{{\em J.Sov.Math.} {\bfseries 23}
  (1983) 2435--2441}.

\bibitem{Beck:1994}
J.~Beck, ``Braid group action and quantum affine algebras,'' {\em Comm. Math.
  Phys.} {\bfseries 165} no.~3, (1994) 555--568.
  \url{http://projecteuclid.org/getRecord?id=euclid.cmp/1104271413}.

\bibitem{Chari:1991}
V.~Chari and A.~Pressley, ``Quantum affine algebras,'' {\em Comm. Math. Phys.}
  {\bfseries 142} no.~2, (1991) 261--283.
  \url{http://projecteuclid.org/getRecord?id=euclid.cmp/1104248585}.

\bibitem{Jing:1998}
N.~Jing, ``Quantum {K}ac-{M}oody algebras and vertex representations,''
  \href{http://dx.doi.org/10.1023/A:1007493921464}{{\em Lett. Math. Phys.}
  {\bfseries 44} no.~4, (1998) 261--271}.
  \url{http://dx.doi.org/10.1023/A:1007493921464}.

\bibitem{Nakajima:2001q}
H.~Nakajima, ``Quiver varieties and finite-dimensional representations of
  quantum affine algebras,''
  \href{http://dx.doi.org/10.1090/S0894-0347-00-00353-2}{{\em J. Amer. Math.
  Soc.} {\bfseries 14} no.~1, (2001) 145--238}.
  \url{http://dx.doi.org/10.1090/S0894-0347-00-00353-2}.

\bibitem{Ginzburg:1995b}
V.~Ginzburg, M.~Kapranov, and {\'E}.~Vasserot, ``Langlands reciprocity for
  algebraic surfaces,'' \href{http://dx.doi.org/10.4310/MRL.1995.v2.n2.a4}{{\em
  Math. Res. Lett.} {\bfseries 2} no.~2, (1995) 147--160},
  \href{http://arxiv.org/abs/q-alg/9502013}{{\ttfamily q-alg/9502013}}.
  \url{http://dx.doi.org/10.4310/MRL.1995.v2.n2.a4}.

\bibitem{Nakajima:2000}
H.~Nakajima, ``Quiver varieties and quantum affine algebras [translation of
  {S}\=ugaku {\bf 52} (2000), no. 4, 337--359; mr1802956],'' {\em Sugaku
  Expositions} {\bfseries 19} no.~1, (2006) 53--78. Sugaku Expositions.

\bibitem{Nakajima:2002b}
H.~Nakajima, ``Geometric construction of representations of affine algebras,''
  \href{http://arxiv.org/abs/math/0212401}{{\ttfamily math/0212401}}.

\bibitem{Drinfeld:1986}
V.~G. Drinfeld, ``Quantum groups,'' {\em Proceedings of the {I}nternational
  {C}ongress of {M}athematicians, {V}ol. 1, 2 ({B}erkeley, {C}alif., 1986)}
  (1987) 798--820.

\bibitem{Gautam:2013}
S.~Gautam and V.~Toledano~Laredo, ``Yangians and quantum loop algebras,''
  \href{http://dx.doi.org/10.1007/s00029-012-0114-2}{{\em Selecta Math. (N.S.)}
  {\bfseries 19} no.~2, (2013) 271--336},
  \href{http://arxiv.org/abs/1012.3687}{{\ttfamily 1012.3687}}.
  \url{http://dx.doi.org/10.1007/s00029-012-0114-2}.

\bibitem{Guay:2012}
N.~Guay and X.~Ma, ``From quantum loop algebras to {Y}angians,''
  \href{http://dx.doi.org/10.1112/jlms/jds021}{{\em J. Lond. Math. Soc. (2)}
  {\bfseries 86} no.~3, (2012) 683--700}.
  \url{http://dx.doi.org/10.1112/jlms/jds021}.

\bibitem{Felder:1996}
G.~Felder and A.~Varchenko, ``On representations of the elliptic quantum group
  {$E_{\tau,\eta}({\rm sl}_2)$},'' {\em Comm. Math. Phys.} {\bfseries 181}
  no.~3, (1996) 741--761.
  \url{http://projecteuclid.org/getRecord?id=euclid.cmp/1104287911}.

\bibitem{Felder:1995}
G.~Felder, ``Elliptic quantum groups,'' {\em X{I}th {I}nternational {C}ongress
  of {M}athematical {P}hysics ({P}aris, 1994)} (1995) 211--218.

\bibitem{Jimbo:1999}
M.~Jimbo, H.~Konno, S.~Odake, and J.~Shiraishi, ``Elliptic algebra
  {$U_{q,p}(\widehat{\mathfrak{sl}_2})$}: {D}rinfeld currents and vertex
  operators,'' \href{http://dx.doi.org/10.1007/s002200050514}{{\em Comm. Math.
  Phys.} {\bfseries 199} no.~3, (1999) 605--647},
  \href{http://arxiv.org/abs/math/9802002}{{\ttfamily math/9802002}}.
  \url{http://dx.doi.org/10.1007/s002200050514}.

\bibitem{Marshakov:1997cj}
A.~Marshakov and A.~Mironov, ``{5-d and 6-d supersymmetric gauge theories:
  Prepotentials from integrable systems},''
  \href{http://dx.doi.org/10.1016/S0550-3213(98)00149-7}{{\em Nucl.Phys.}
  {\bfseries B518} (1998) 59--91},
\href{http://arxiv.org/abs/hep-th/9711156}{{\ttfamily arXiv:hep-th/9711156
  [hep-th]}}.

\bibitem{Hollowood:2003cv}
T.~J. Hollowood, A.~Iqbal, and C.~Vafa, ``{Matrix models, geometric engineering
  and elliptic genera},''
  \href{http://dx.doi.org/10.1088/1126-6708/2008/03/069}{{\em JHEP} {\bfseries
  0803} (2008) 069},
\href{http://arxiv.org/abs/hep-th/0310272}{{\ttfamily arXiv:hep-th/0310272
  [hep-th]}}.

\bibitem{Konno:2009}
H.~Konno, ``The elliptic quantum group
  {$U_{q,p}(\widehat{\mathfrak{sl}_2})$},''.
  \url{http://www.mis.hiroshima-u.ac.jp/~konno/RIMS09.pdf}.

\bibitem{Felder:1995iv}
G.~Felder and A.~Varchenko, ``{Integral representation of solutions of the
  elliptic Knizhnik-Zamolodchikov-Bernard equations},''
\href{http://arxiv.org/abs/hep-th/9502165}{{\ttfamily arXiv:hep-th/9502165
  [hep-th]}}.

\bibitem{Felder:1997}
G.~Felder, V.~Tarasov, and A.~Varchenko, ``Solutions of the elliptic q{KZB}
  equations and {B}ethe ansatz. {I},''
  \href{http://arxiv.org/abs/q-alg/9606005}{{\ttfamily q-alg/9606005}}.

\bibitem{Felder:1999}
G.~Felder and A.~Varchenko, ``The elliptic gamma function and {${\rm SL}(3,{\bf
  Z})\ltimes{\bf Z}^3$},'' \href{http://dx.doi.org/10.1006/aima.2000.1951}{{\em
  Adv. Math.} {\bfseries 156} no.~1, (2000) 44--76},
  \href{http://arxiv.org/abs/math/9907061}{{\ttfamily math/9907061}}.
  \url{http://dx.doi.org/10.1006/aima.2000.1951}.

\bibitem{Feigin:1996}
B.~Feigin and E.~Frenkel, ``Quantum {$W$}-algebras and elliptic algebras,''
  {\em Comm. Math. Phys.} {\bfseries 178} no.~3, (1996) 653--678,
  \href{http://arxiv.org/abs/q-alg/9508009}{{\ttfamily arXiv:q-alg/9508009
  [q-alg]}}. \url{http://projecteuclid.org/getRecord?id=euclid.cmp/1104286770}.

\bibitem{Chari:1996}
V.~Chari and A.~Pressley, ``Yangians: their representations and characters,''
  \href{http://dx.doi.org/10.1007/BF00116515}{{\em Acta Appl. Math.} {\bfseries
  44} no.~1-2, (1996) 39--58}. \url{http://dx.doi.org/10.1007/BF00116515}.
  Representations of Lie groups, Lie algebras and their quantum analogues.

\bibitem{Nakajima:2001}
H.~Nakajima, ``{$T$}-analogue of the {$q$}-characters of finite dimensional
  representations of quantum affine algebras,''.
  \url{http://dx.doi.org/10.1142/9789812810007_0009}.

\bibitem{Nakajima:2004}
H.~Nakajima, ``Quiver varieties and {$t$}-analogs of {$q$}-characters of
  quantum affine algebras,''
  \href{http://dx.doi.org/10.4007/annals.2004.160.1057}{{\em Ann. of Math. (2)}
  {\bfseries 160} no.~3, (2004) 1057--1097}.
  \url{http://dx.doi.org/10.4007/annals.2004.160.1057}.

\bibitem{Nakajima:1994}
H.~Nakajima, ``Instantons on {ALE} spaces, quiver varieties, and {K}ac-{M}oody
  algebras,'' \href{http://dx.doi.org/10.1215/S0012-7094-94-07613-8}{{\em Duke
  Math. J.} {\bfseries 76} no.~2, (1994) 365--416}.
  \url{http://dx.doi.org/10.1215/S0012-7094-94-07613-8}.

\bibitem{Nakajima:2010}
H.~Nakajima, ``{$t$}-analogs of {$q$}-characters of quantum affine algebras of
  type {$E_6,E_7,E_8$},''.
  \url{http://dx.doi.org/10.1007/978-0-8176-4697-4_10}.

\bibitem{Herandez:2003}
D.~Hernandez, ``Representations of quantum affinizations and fusion product,''
  \href{http://dx.doi.org/10.1007/s00031-005-1005-9}{{\em Transform. Groups}
  {\bfseries 10} no.~2, (2005) 163--200}.
  \url{http://dx.doi.org/10.1007/s00031-005-1005-9}.

\bibitem{Hernandez:2005a}
D.~Hernandez, ``Drinfeld coproduct, quantum fusion tensor category and
  applications,'' \href{http://dx.doi.org/10.1112/plms/pdm017}{{\em Proc. Lond.
  Math. Soc. (3)} {\bfseries 95} no.~3, (2007) 567--608}.
  \url{http://dx.doi.org/10.1112/plms/pdm017}.

\bibitem{Chari:2001}
V.~Chari, ``Braid group actions and tensor products,''
  \href{http://dx.doi.org/10.1155/S107379280210612X}{{\em Int. Math. Res. Not.}
  no.~7, (2002) 357--382}. \url{http://dx.doi.org/10.1155/S107379280210612X}.

\bibitem{Chari:2004}
V.~Chari and A.~A. Moura, ``Characters and blocks for finite-dimensional
  representations of quantum affine algebras,''
  \href{http://dx.doi.org/10.1155/IMRN.2005.257}{{\em Int. Math. Res. Not.}
  no.~5, (2005) 257--298}. \url{http://dx.doi.org/10.1155/IMRN.2005.257}.

\bibitem{Hernandez:2011b}
D.~{Hernandez} and M.~{Jimbo}, ``{Asymptotic representations and Drinfeld
  rational fractions},'' {\em ArXiv e-prints} (Apr., 2011) ,
  \href{http://arxiv.org/abs/1104.1891}{{\ttfamily arXiv:1104.1891 [math.QA]}}.

\bibitem{Bazhanov:1996dr}
V.~V. Bazhanov, S.~L. Lukyanov, and A.~B. Zamolodchikov, ``{Integrable
  structure of conformal field theory. 2. Q operator and DDV equation},''
  \href{http://dx.doi.org/10.1007/s002200050240}{{\em Commun.Math.Phys.}
  {\bfseries 190} (1997) 247--278},
\href{http://arxiv.org/abs/hep-th/9604044}{{\ttfamily arXiv:hep-th/9604044
  [hep-th]}}.

\bibitem{Varagnolo:1999}
M.~Varagnolo and E.~Vasserot, ``On the {$K$}-theory of the cyclic quiver
  variety,'' \href{http://dx.doi.org/10.1155/S1073792899000525}{{\em Internat.
  Math. Res. Notices} no.~18, (1999) 1005--1028},
  \href{http://arxiv.org/abs/math/9902091}{{\ttfamily math/9902091}}.
  \url{http://dx.doi.org/10.1155/S1073792899000525}.

\bibitem{Varagnolo:2005}
M.~Varagnolo, ``Quiver varieties and {Y}angians,''
  \href{http://dx.doi.org/10.1023/A:1007674020905}{{\em Lett. Math. Phys.}
  {\bfseries 53} no.~4, (2000) 273--283},
  \href{http://arxiv.org/abs/math/0005277}{{\ttfamily math/0005277}}.
  \url{http://dx.doi.org/10.1023/A:1007674020905}.

\bibitem{Guay:2005}
N.~Guay, ``Cherednik algebras and {Y}angians,''
  \href{http://dx.doi.org/10.1155/IMRN.2005.3551}{{\em Int. Math. Res. Not.}
  no.~57, (2005) 3551--3593}. \url{http://dx.doi.org/10.1155/IMRN.2005.3551}.

\bibitem{Guay:2007}
N.~Guay, ``Affine {Y}angians and deformed double current algebras in type
  {A},'' \href{http://dx.doi.org/10.1016/j.aim.2006.08.007}{{\em Adv. Math.}
  {\bfseries 211} no.~2, (2007) 436--484}.
  \url{http://dx.doi.org/10.1016/j.aim.2006.08.007}.

\bibitem{Guay:2009a}
N.~Guay, ``Quantum algebras and quivers,''
  \href{http://dx.doi.org/10.1007/s00029-009-0496-y}{{\em Selecta Math. (N.S.)}
  {\bfseries 14} no.~3-4, (2009) 667--700}.
  \url{http://dx.doi.org/10.1007/s00029-009-0496-y}.

\bibitem{Braden:2001yc}
H.~Braden, A.~Gorsky, A.~Odessky, and V.~Rubtsov, ``{Double elliptic dynamical
  systems from generalized Mukai-Sklyanin algebras},''
  \href{http://dx.doi.org/10.1016/S0550-3213(02)00248-1}{{\em Nucl.Phys.}
  {\bfseries B633} (2002) 414--442},
\href{http://arxiv.org/abs/hep-th/0111066}{{\ttfamily arXiv:hep-th/0111066
  [hep-th]}}.

\bibitem{Braden:2003gv}
H.~W. Braden and T.~J. Hollowood, ``{The Curve of compactified 6-D gauge
  theories and integrable systems},'' {\em JHEP} {\bfseries 0312} (2003) 023,
\href{http://arxiv.org/abs/hep-th/0311024}{{\ttfamily arXiv:hep-th/0311024
  [hep-th]}}.

\bibitem{Varagnolo:1996b}
M.~Varagnolo and E.~Vasserot, ``Schur duality in the toroidal setting,'' {\em
  Comm. Math. Phys.} {\bfseries 182} no.~2, (1996) 469--483,
  \href{http://arxiv.org/abs/q-alg/9506026}{{\ttfamily q-alg/9506026}}.
  \url{http://projecteuclid.org/getRecord?id=euclid.cmp/1104288156}.

\bibitem{Cherednik:1992}
I.~Cherednik, ``Double affine {H}ecke algebras, {K}nizhnik-{Z}amolodchikov
  equations, and {M}acdonald's operators,''
  \href{http://dx.doi.org/10.1155/S1073792892000199}{{\em Internat. Math. Res.
  Notices} no.~9, (1992) 171--180}.
  \url{http://dx.doi.org/10.1155/S1073792892000199}.

\bibitem{Saito:1996b}
Y.~Saito, ``Quantum toroidal algebras and their vertex representations,''
  \href{http://dx.doi.org/10.2977/prims/1195144759}{{\em Publ. Res. Inst. Math.
  Sci.} {\bfseries 34} no.~2, (1998) 155--177},
  \href{http://arxiv.org/abs/q-alg/9611030}{{\ttfamily q-alg/9611030}}.
  \url{http://dx.doi.org/10.2977/prims/1195144759}.

\bibitem{Cherednik:1995m}
I.~Cherednik, ``Nonsymmetric {M}acdonald polynomials,''
  \href{http://dx.doi.org/10.1155/S1073792895000341}{{\em Internat. Math. Res.
  Notices} no.~10, (1995) 483--515}.
  \url{http://dx.doi.org/10.1155/S1073792895000341}.

\bibitem{Cherednik:1995}
I.~Cherednik, ``Double affine {H}ecke algebras and {M}acdonald's conjectures,''
  \href{http://dx.doi.org/10.2307/2118632}{{\em Ann. of Math. (2)} {\bfseries
  141} no.~1, (1995) 191--216}. \url{http://dx.doi.org/10.2307/2118632}.

\bibitem{Varagnolo:1996}
M.~Varagnolo and E.~Vasserot, ``Double-loop algebras and the {F}ock space,''
  \href{http://dx.doi.org/10.1007/s002220050242}{{\em Invent. Math.} {\bfseries
  133} no.~1, (1998) 133--159},
  \href{http://arxiv.org/abs/q-alg/9612035}{{\ttfamily q-alg/9612035}}.
  \url{http://dx.doi.org/10.1007/s002220050242}.

\bibitem{Saito:1998}
Y.~Saito, K.~Takemura, and D.~Uglov, ``Toroidal actions on level {$1$} modules
  of {$U_q(\widehat{\rm sl}_n)$},''
  \href{http://dx.doi.org/10.1007/BF01237841}{{\em Transform. Groups}
  {\bfseries 3} no.~1, (1998) 75--102},
  \href{http://arxiv.org/abs/q-alg/9702024}{{\ttfamily q-alg/9702024}}.
  \url{http://dx.doi.org/10.1007/BF01237841}.

\bibitem{Kashiwara:1995}
M.~Kashiwara, T.~Miwa, and E.~Stern, ``Decomposition of {$q$}-deformed {F}ock
  spaces,'' \href{http://dx.doi.org/10.1007/BF01587910}{{\em Selecta Math.
  (N.S.)} {\bfseries 1} no.~4, (1995) 787--805},
  \href{http://arxiv.org/abs/q-alg/9508006}{{\ttfamily q-alg/9508006}}.
  \url{http://dx.doi.org/10.1007/BF01587910}.

\bibitem{Nagao:2007}
K.~Nagao, ``{$K$}-theory of quiver varieties, {$q$}-{F}ock space and
  nonsymmetric {M}acdonald polynomials,'' {\em Osaka J. Math.} {\bfseries 46}
  no.~3, (2009) 877--907, \href{http://arxiv.org/abs/0709.1767}{{\ttfamily
  0709.1767}}.
  \url{http://projecteuclid.org/getRecord?id=euclid.ojm/1256564211}.

\bibitem{Miki:1999}
K.~Miki, ``Toroidal braid group action and an automorphism of toroidal algebra
  {$U_q({\rm sl}_{n+1,\rm tor})\ (n\geq 2)$},''
  \href{http://dx.doi.org/10.1023/A:1007556926350}{{\em Lett. Math. Phys.}
  {\bfseries 47} no.~4, (1999) 365--378}.
  \url{http://dx.doi.org/10.1023/A:1007556926350}.

\bibitem{Berman:1996}
S.~Berman, Y.~Gao, and Y.~S. Krylyuk, ``Quantum tori and the structure of
  elliptic quasi-simple {L}ie algebras,''
  \href{http://dx.doi.org/10.1006/jfan.1996.0013}{{\em J. Funct. Anal.}
  {\bfseries 135} no.~2, (1996) 339--389}.
  \url{http://dx.doi.org/10.1006/jfan.1996.0013}.

\bibitem{Miki:2007}
K.~Miki, ``A {$(q,\gamma)$} analog of the {$W_{1+\infty}$} algebra,''
  \href{http://dx.doi.org/10.1063/1.2823979}{{\em J. Math. Phys.} {\bfseries
  48} no.~12, (2007) 123520, 35}. \url{http://dx.doi.org/10.1063/1.2823979}.

\bibitem{Awata:1996}
H.~Awata, H.~Kubo, S.~Odake, and J.~Shiraishi, ``Quantum {$\mathcal{W}_N$}
  algebras and {M}acdonald polynomials,'' {\em Comm. Math. Phys.} {\bfseries
  179} no.~2, (1996) 401--416.
  \url{http://projecteuclid.org/getRecord?id=euclid.cmp/1104286998}.

\bibitem{Shiraishi:1996}
J.~Shiraishi, H.~Kubo, H.~Awata, and S.~Odake, ``A quantum deformation of the
  {V}irasoro algebra and the {M}acdonald symmetric functions,''
  \href{http://dx.doi.org/10.1007/BF00398297}{{\em Lett. Math. Phys.}
  {\bfseries 38} no.~1, (1996) 33--51}.
  \url{http://dx.doi.org/10.1007/BF00398297}.

\bibitem{Nakajima:1995}
H.~Nakajima, ``Heisenberg algebra and {H}ilbert schemes of points on projective
  surfaces,'' \href{http://dx.doi.org/10.2307/2951818}{{\em Ann. of Math. (2)}
  {\bfseries 145} no.~2, (1997) 379--388},
  \href{http://arxiv.org/abs/alg-geom/9507012}{{\ttfamily alg-geom/9507012}}.
  \url{http://dx.doi.org/10.2307/2951818}.

\bibitem{Ginzburg:1993}
V.~Ginzburg and {\'E}.~Vasserot, ``Langlands reciprocity for affine quantum
  groups of type {$A_n$},''
  \href{http://dx.doi.org/10.1155/S1073792893000078}{{\em Internat. Math. Res.
  Notices} no.~3, (1993) 67--85}.
  \url{http://dx.doi.org/10.1155/S1073792893000078}.

\bibitem{Grojnowski:1995}
I.~Grojnowski, ``Instantons and affine algebras. {I}. {T}he {H}ilbert scheme
  and vertex operators,''
  \href{http://dx.doi.org/10.4310/MRL.1996.v3.n2.a12}{{\em Math. Res. Lett.}
  {\bfseries 3} no.~2, (1996) 275--291},
  \href{http://arxiv.org/abs/alg-geom/9506020}{{\ttfamily alg-geom/9506020}}.
  \url{http://dx.doi.org/10.4310/MRL.1996.v3.n2.a12}.

\bibitem{Baranovsky:2000}
V.~Baranovsky, ``Moduli of sheaves on surfaces and action of the oscillator
  algebra,'' {\em J. Differential Geom.} {\bfseries 55} no.~2, (2000) 193--227.
  \url{http://projecteuclid.org/getRecord?id=euclid.jdg/1090340878}.

\bibitem{Schiffmann:2009}
O.~{Schiffmann} and E.~{Vasserot}, ``{The elliptic Hall algebra and the
  equivariant K-theory of the Hilbert scheme of $A^2$},'' {\em ArXiv e-prints}
  (May, 2009) , \href{http://arxiv.org/abs/0905.2555}{{\ttfamily
  arXiv:0905.2555 [math.QA]}}.

\bibitem{Carlsson:2008}
E.~Carlsson and A.~Okounkov, ``Exts and vertex operators,''
  \href{http://dx.doi.org/10.1215/00127094-1593380}{{\em Duke Math. J.}
  {\bfseries 161} no.~9, (2012) 1797--1815},
  \href{http://arxiv.org/abs/0801.2565}{{\ttfamily 0801.2565}}.
  \url{http://dx.doi.org/10.1215/00127094-1593380}.

\bibitem{Carlsson:2013}
E.~{Carlsson}, N.~{Nekrasov}, and A.~{Okounkov}, ``{Five dimensional gauge
  theories and vertex operators},'' {\em ArXiv e-prints} (Aug., 2013) ,
  \href{http://arxiv.org/abs/1308.2465}{{\ttfamily arXiv:1308.2465 [math.RT]}}.

\bibitem{Feigin:2009com}
B.~Feigin, K.~Hashizume, A.~Hoshino, J.~Shiraishi, and S.~Yanagida, ``A
  commutative algebra on degenerate {$\Bbb{CP}^1$} and {M}acdonald
  polynomials,'' \href{http://dx.doi.org/10.1063/1.3192773}{{\em J. Math.
  Phys.} {\bfseries 50} no.~9, (2009) 095215, 42},
  \href{http://arxiv.org/abs/0904.2291}{{\ttfamily 0904.2291}}.
  \url{http://dx.doi.org/10.1063/1.3192773}.

\bibitem{Wyllard:2009hg}
N.~Wyllard, ``{A(N-1) conformal Toda field theory correlation functions from
  conformal N = 2 SU(N) quiver gauge theories},''
  \href{http://dx.doi.org/10.1088/1126-6708/2009/11/002}{{\em JHEP} {\bfseries
  0911} (2009) 002},
\href{http://arxiv.org/abs/0907.2189}{{\ttfamily arXiv:0907.2189 [hep-th]}}.

\bibitem{Awata:2009ur}
H.~Awata and Y.~Yamada, ``{Five-dimensional AGT Conjecture and the Deformed
  Virasoro 'lgebra},'' \href{http://dx.doi.org/10.1007/JHEP01(2010)125}{{\em
  JHEP} {\bfseries 1001} (2010) 125},
\href{http://arxiv.org/abs/0910.4431}{{\ttfamily arXiv:0910.4431 [hep-th]}}.

\bibitem{Fateev:2011hq}
V.~Fateev and A.~Litvinov, ``{Integrable structure, W-symmetry and AGT
  relation},'' \href{http://dx.doi.org/10.1007/JHEP01(2012)051}{{\em JHEP}
  {\bfseries 1201} (2012) 051},
\href{http://arxiv.org/abs/1109.4042}{{\ttfamily arXiv:1109.4042 [hep-th]}}.

\bibitem{Smirnov:2013}
A.~{Smirnov}, ``{On the Instanton R-matrix},'' {\em ArXiv e-prints} (Feb.,
  2013) , \href{http://arxiv.org/abs/1302.0799}{{\ttfamily arXiv:1302.0799
  [math.AG]}}.

\bibitem{Awata:2011dc}
H.~Awata, B.~Feigin, A.~Hoshino, M.~Kanai, J.~Shiraishi, {\em et~al.}, ``{Notes
  on Ding-Iohara algebra and AGT conjecture},''
\href{http://arxiv.org/abs/1106.4088}{{\ttfamily arXiv:1106.4088 [math-ph]}}.

\bibitem{Maulik:2012}
D.~{Maulik} and A.~{Okounkov}, ``{Quantum Groups and Quantum Cohomology},''
  {\em ArXiv e-prints} (Nov., 2012) ,
  \href{http://arxiv.org/abs/1211.1287}{{\ttfamily arXiv:1211.1287 [math.AG]}}.

\bibitem{Schiffmann:2012b}
O.~{Schiffmann} and E.~{Vasserot}, ``{Cherednik algebras, W algebras and the
  equivariant cohomology of the moduli space of instantons on A\^{}2},'' {\em
  ArXiv e-prints} (Feb., 2012) ,
  \href{http://arxiv.org/abs/1202.2756}{{\ttfamily arXiv:1202.2756 [math.QA]}}.

\bibitem{Alba:2010qc}
V.~A. Alba, V.~A. Fateev, A.~V. Litvinov, and G.~M. Tarnopolskiy, ``{On
  combinatorial expansion of the conformal blocks arising from AGT
  conjecture},'' \href{http://dx.doi.org/10.1007/s11005-011-0503-z}{{\em
  Lett.Math.Phys.} {\bfseries 98} (2011) 33--64},
\href{http://arxiv.org/abs/1012.1312}{{\ttfamily arXiv:1012.1312 [hep-th]}}.

\bibitem{Feigin:2010a}
B.~Feigin, E.~Feigin, M.~Jimbo, T.~Miwa, and E.~Mukhin, ``Quantum continuous
  {$\mathfrak{gl}_\infty$}: semiinfinite construction of representations,''
  \href{http://dx.doi.org/10.1215/21562261-1214375}{{\em Kyoto J. Math.}
  {\bfseries 51} no.~2, (2011) 337--364},
  \href{http://arxiv.org/abs/1002.3100}{{\ttfamily 1002.3100}}.
  \url{http://dx.doi.org/10.1215/21562261-1214375}.

\bibitem{Feigin:2010b}
B.~Feigin, E.~Feigin, M.~Jimbo, T.~Miwa, and E.~Mukhin, ``Quantum continuous
  {$\mathfrak{gl}_\infty$}: tensor products of {F}ock modules and
  {$\mathcal{W}_n$}-characters,''
  \href{http://dx.doi.org/10.1215/21562261-1214384}{{\em Kyoto J. Math.}
  {\bfseries 51} no.~2, (2011) 365--392},
  \href{http://arxiv.org/abs/1002.3113}{{\ttfamily 1002.3113}}.
  \url{http://dx.doi.org/10.1007/BF01237841}.

\bibitem{Burban:2005}
I.~Burban and O.~Schiffmann, ``On the {H}all algebra of an elliptic curve,
  {I},'' \href{http://dx.doi.org/10.1215/00127094-1593263}{{\em Duke Math. J.}
  {\bfseries 161} no.~7, (2012) 1171--1231},
  \href{http://arxiv.org/abs/math/0505148}{{\ttfamily math/0505148}}.
  \url{http://dx.doi.org/10.1215/00127094-1593263}.

\bibitem{Schiffmann:2010}
O.~Schiffmann, ``Drinfeld realization of the elliptic {H}all algebra,''
  \href{http://dx.doi.org/10.1007/s10801-011-0302-8}{{\em J. Algebraic Combin.}
  {\bfseries 35} no.~2, (2012) 237--262},
  \href{http://arxiv.org/abs/1004.2575}{{\ttfamily 1004.2575}}.
  \url{http://dx.doi.org/10.1007/s10801-011-0302-8}.

\bibitem{Feigin:2009b}
B.~L. Feigin and A.~I. Tsymbaliuk, ``Equivariant {$K$}-theory of {H}ilbert
  schemes via shuffle algebra,''
  \href{http://dx.doi.org/10.1215/21562261-1424875}{{\em Kyoto J. Math.}
  {\bfseries 51} no.~4, (2011) 831--854},
  \href{http://arxiv.org/abs/0904.1679}{{\ttfamily 0904.1679}}.
  \url{http://dx.doi.org/10.1215/21562261-1424875}.

\bibitem{Ding:1996mq}
J.-t. Ding and K.~Iohara, ``{Generalization and deformation of Drinfeld quantum
  affine algebras},''
\href{http://dx.doi.org/10.1023/A:1007341410987}{{\em Lett.Math.Phys.}
  {\bfseries 41} (1997) 181--193}.

\bibitem{Feigin:1997}
B.~Feigin and A.~Odesskii, ``A family of elliptic algebras,''
  \href{http://dx.doi.org/10.1155/S1073792897000354}{{\em Internat. Math. Res.
  Notices} no.~11, (1997) 531--539}.
  \url{http://dx.doi.org/10.1155/S1073792897000354}.

\bibitem{Enriquez:1998}
B.~Enriquez, ``On correlation functions of {D}rinfeld currents and shuffle
  algebras,'' \href{http://dx.doi.org/10.1007/BF01236465}{{\em Transform.
  Groups} {\bfseries 5} no.~2, (2000) 111--120},
  \href{http://arxiv.org/abs/math/9809036}{{\ttfamily math/9809036}}.
  \url{http://dx.doi.org/10.1007/BF01236465}.

\bibitem{Negut:2013}
A.~{Negut}, ``{An Isomorphism between the Quantum Toroidal and Shuffle
  Algebras, and a Conjecture of Kuznetsov},'' {\em ArXiv e-prints} (Feb., 2013)
  , \href{http://arxiv.org/abs/1302.6202}{{\ttfamily arXiv:1302.6202
  [math.RT]}}.

\bibitem{Okounkov:2004}
A.~Okounkov and R.~Pandharipande, ``Quantum cohomology of the {H}ilbert scheme
  of points in the plane,''
  \href{http://dx.doi.org/10.1007/s00222-009-0223-5}{{\em Invent. Math.}
  {\bfseries 179} no.~3, (2010) 523--557},
  \href{http://arxiv.org/abs/math/0411210}{{\ttfamily math/0411210}}.
  \url{http://dx.doi.org/10.1007/s00222-009-0223-5}.

\bibitem{Saito:2013}
Y.~{Saito}, ``{Elliptic Ding-Iohara Algebra and the Free Field Realization of
  the Elliptic Macdonald Operator},'' {\em ArXiv e-prints} (Jan., 2013) ,
  \href{http://arxiv.org/abs/1301.4912}{{\ttfamily arXiv:1301.4912 [math.QA]}}.

\bibitem{Saito:2013b}
Y.~{Saito}, ``{Elliptic Ding-Iohara Algebra and Commutative Families of the
  Elliptic Macdonald Operator},'' {\em ArXiv e-prints} (Sept., 2013) ,
  \href{http://arxiv.org/abs/1309.7094}{{\ttfamily arXiv:1309.7094 [math.QA]}}.

\bibitem{Feigin:2011}
B.~Feigin, M.~Jimbo, T.~Miwa, and E.~Mukhin, ``Quantum toroidal
  {$\mathfrak{gl}_1$}-algebra: plane partitions,'' {\em Kyoto J. Math.}
  {\bfseries 52} no.~3, (2012) 621--659,
  \href{http://arxiv.org/abs/1110.5310}{{\ttfamily 1110.5310}}.

\bibitem{Feigin:2013m}
G.~S. Mutafyan and B.~L. Fe{\u\i}gin, ``The quantum toroidal algebra
  {$\widehat{\widehat{{\mathfrak {gl}}_1}}$}: calculation of the characters of
  some representations as generating functions of plane partitions,''
  \href{http://dx.doi.org/10.1007/s10688-013-0006-z}{{\em Funktsional. Anal. i
  Prilozhen.} {\bfseries 47} no.~1, (2013) 62--76}.
  \url{http://dx.doi.org/10.1007/s10688-013-0006-z}.

\bibitem{Feigin:2012}
B.~Feigin, M.~Jimbo, T.~Miwa, and E.~Mukhin, ``Representations of quantum
  toroidal {${\mathfrak{gl}}_n$},''
  \href{http://dx.doi.org/10.1016/j.jalgebra.2012.12.029}{{\em J. Algebra}
  {\bfseries 380} (2013) 78--108},
  \href{http://arxiv.org/abs/1204.5378}{{\ttfamily 1204.5378}}.
  \url{http://dx.doi.org/10.1016/j.jalgebra.2012.12.029}.

\bibitem{Feigin:2013fga}
B.~Feigin, M.~Jimbo, T.~Miwa, and E.~Mukhin, ``{Branching rules for quantum
  toroidal gl(n)},''
\href{http://arxiv.org/abs/1309.2147}{{\ttfamily arXiv:1309.2147 [math.QA]}}.

\bibitem{Fuji:2012nx}
H.~Fuji, S.~Gukov, and P.~Sulkowski, ``{Super-A-polynomial for knots and BPS
  states},'' \href{http://dx.doi.org/10.1016/j.nuclphysb.2012.10.005}{{\em
  Nucl.Phys.} {\bfseries B867} (2013) 506--546},
\href{http://arxiv.org/abs/1205.1515}{{\ttfamily arXiv:1205.1515 [hep-th]}}.

\bibitem{Smirnov:1991me}
F.~A. Smirnov, ``{Dynamical symmetries of massive integrable models, 1.
  Form-factor bootstrap equations as a special case of deformed Knizhnik-
  Zamolodchikov equations},''
{\em Int.J.Mod.Phys.} {\bfseries A71B} (1992) 813--837.

\bibitem{Smirnov:1992vz}
F.~Smirnov, ``{Form-factors in completely integrable models of quantum field
  theory},''
{\em Adv.Ser.Math.Phys.} {\bfseries 14} (1992) 1--208.

\bibitem{Frenkel:1992}
I.~B. Frenkel and N.~Y. Reshetikhin, ``Quantum affine algebras and holonomic
  difference equations,'' {\em Comm. Math. Phys.} {\bfseries 146} no.~1, (1992)
  1--60. \url{http://projecteuclid.org/getRecord?id=euclid.cmp/1104249974}.

\bibitem{Feigin:1994in}
B.~Feigin, E.~Frenkel, and N.~Reshetikhin, ``{Gaudin model, Bethe ansatz and
  correlation functions at the critical level},''
  \href{http://dx.doi.org/10.1007/BF02099300}{{\em Commun.Math.Phys.}
  {\bfseries 166} (1994) 27--62},
\href{http://arxiv.org/abs/hep-th/9402022}{{\ttfamily arXiv:hep-th/9402022
  [hep-th]}}.

\bibitem{Bazhanov:1994ft}
V.~V. Bazhanov, S.~L. Lukyanov, and A.~B. Zamolodchikov, ``{Integrable
  structure of conformal field theory, quantum KdV theory and thermodynamic
  Bethe ansatz},'' \href{http://dx.doi.org/10.1007/BF02101898}{{\em
  Commun.Math.Phys.} {\bfseries 177} (1996) 381--398},
\href{http://arxiv.org/abs/hep-th/9412229}{{\ttfamily arXiv:hep-th/9412229
  [hep-th]}}.

\bibitem{Huang:2012kn}
M.-x. Huang, ``{On Gauge Theory and Topological String in Nekrasov-Shatashvili
  Limit},'' \href{http://dx.doi.org/10.1007/JHEP06(2012)152}{{\em JHEP}
  {\bfseries 1206} (2012) 152},
\href{http://arxiv.org/abs/1205.3652}{{\ttfamily arXiv:1205.3652 [hep-th]}}.

\bibitem{Chervov:2006xk}
A.~Chervov and D.~Talalaev, ``{Quantum spectral curves, quantum integrable
  systems and the geometric Langlands correspondence},''
\href{http://arxiv.org/abs/hep-th/0604128}{{\ttfamily arXiv:hep-th/0604128
  [hep-th]}}.

\bibitem{Krichever:1996qd}
I.~Krichever, O.~Lipan, P.~Wiegmann, and A.~Zabrodin, ``{Quantum integrable
  systems and elliptic solutions of classical discrete nonlinear equations},''
  \href{http://dx.doi.org/10.1007/s002200050165}{{\em Commun.Math.Phys.}
  {\bfseries 188} (1997) 267--304},
\href{http://arxiv.org/abs/hep-th/9604080}{{\ttfamily arXiv:hep-th/9604080
  [hep-th]}}.

\bibitem{Beilinson:2005}
A.~{Beilinson} and V.~{Drinfeld}, ``{Opers},'' {\em ArXiv Mathematics e-prints}
  (Jan., 2005) , \href{http://arxiv.org/abs/math/0501398}{{\ttfamily
  math/0501398}}.

\bibitem{Beilinson}
A.~A. Beilinson and V.~G. Drinfeld, ``Quantization of {H}itchin's fibration and
  {L}anglands' program,''. \url{http://dx.doi.org/10.1007/978-94-017-0693-3_1}.

\bibitem{Beilinson:Hecke}
A.~A. Beilinson and V.~G. Drinfeld, ``Quantization of hitchin's integrable
  system and hecke eigensheaves,''.
  \url{http://www.math.uchicago.edu/~mitya/langlands/QuantizationHitchin.pdf}.

\bibitem{Frenkel:1995zp}
E.~Frenkel, ``{Affine algebras, Langlands duality and Bethe ansatz},''
\href{http://arxiv.org/abs/q-alg/9506003}{{\ttfamily arXiv:q-alg/9506003
  [q-alg]}}.

\bibitem{Frenkel:1996}
E.~Frenkel and N.~Reshetikhin, ``Quantum affine algebras and deformations of
  the {V}irasoro and {$\mathcal{W}$}-algebras,'' {\em Comm. Math. Phys.}
  {\bfseries 178} no.~1, (1996) 237--264,
  \href{http://arxiv.org/abs/q-alg/9505025}{{\ttfamily q-alg/9505025}}.
  \url{http://projecteuclid.org/getRecord?id=euclid.cmp/1104286563}.

\bibitem{Frenkel:2003qx}
E.~Frenkel, ``{Opers on the projective line, flag manifolds and Bethe
  Ansatz},''
\href{http://arxiv.org/abs/math/0308269}{{\ttfamily arXiv:math/0308269
  [math-qa]}}.

\bibitem{Frenkel:2003}
E.~{Frenkel}, ``{Affine Kac-Moody algebras, integrable systems and their
  deformations},'' {\em ArXiv Mathematics e-prints} (May, 2003) ,
  \href{http://arxiv.org/abs/math/0305216}{{\ttfamily math/0305216}}.

\bibitem{Litvinov:2013}
A.~{Litvinov}, S.~{Lukyanov}, N.~{Nekrasov}, and A.~{Zamolodchikov},
  ``{Classical conformal blocks and Painlev\'e VI},''
  \href{http://arxiv.org/abs/arXiv:1309.4700}{{\ttfamily arXiv:arXiv:1309.4700
  [hep-th]}}.

\bibitem{Kuniba:2002}
A.~Kuniba, M.~Okado, J.~Suzuki, and Y.~Yamada, ``Difference {$L$} operators
  related to {$q$}-characters,''
  \href{http://dx.doi.org/10.1088/0305-4470/35/6/307}{{\em J. Phys. A}
  {\bfseries 35} no.~6, (2002) 1415--1435}.
  \url{http://dx.doi.org/10.1088/0305-4470/35/6/307}.

\bibitem{MR1062425}
M.~Jimbo, \href{http://dx.doi.org/10.1142/9789812798350_0005}{``Introduction to
  the {Y}ang-{B}axter equation,''} in {\em Braid group, knot theory and
  statistical mechanics}, vol.~9 of {\em Adv. Ser. Math. Phys.}, pp.~111--134.
\newblock World Sci. Publ., Teaneck, NJ, 1989.
\newblock \url{http://dx.doi.org/10.1142/9789812798350_0005}.

\bibitem{Enriquez:1997}
B.~Enriquez, ``Quantum currents realization of the elliptic quantum groups
  {$E_{\tau,\eta}(\mathfrak{sl}_2)$},'' in {\em Calogero-{M}oser-{S}utherland
  models ({M}ontr\'eal, {QC}, 1997)}, CRM Ser. Math. Phys., pp.~161--176.
\newblock Springer, New York, 2000.

\bibitem{Etingof:1998}
P.~I. Etingof, I.~B. Frenkel, and A.~A. Kirillov, Jr., {\em Lectures on
  representation theory and {K}nizhnik-{Z}amolodchikov equations}, vol.~58 of
  {\em Mathematical Surveys and Monographs}.
\newblock American Mathematical Society, Providence, RI, 1998.

\bibitem{Tanisaki_1991}
T.~Tanisaki, ``Killing forms, {H}arish-{C}handra isomorphisms, and universal
  {$R$}-matrices for quantum algebras,''.

\bibitem{Khoroshkin_1992b}
V.~N. Tolstoy and S.~M. Khoroshkin, ``The universal $r$-matrix for quantum
  untwisted affine lie algebras,''
  \href{http://dx.doi.org/10.1007/BF01077085}{{\em Funct. Anal. Appl.}
  {\bfseries 26} (1992) 69--71}. \url{http://dx.doi.org/10.1007/BF01077085}.

\bibitem{Khoroshkin_1993}
S.~M. Khoroshkin and V.~N. Tolstoy, ``On {D}rinfel\cprime d's realization of
  quantum affine algebras,''
  \href{http://dx.doi.org/10.1016/0393-0440(93)90070-U}{{\em J. Geom. Phys.}
  {\bfseries 11} no.~1-4, (1993) 445--452}.
  \url{http://dx.doi.org/10.1016/0393-0440(93)90070-U}. Infinite-dimensional
  geometry in physics (Karpacz, 1992).

\bibitem{Khoroshkin:1994uj}
S.~Khoroshkin and V.~Tolstoi, ``{Twisting of quantum (super)algebras:
  Connection of Drinfeld's and Cartan-Weyl realizations for quantum affine
  algebras},''
\href{http://arxiv.org/abs/hep-th/9404036}{{\ttfamily arXiv:hep-th/9404036
  [hep-th]}}.

\end{thebibliography}\endgroup
\end{document}